\documentclass[10pt,journal,compsoc]{IEEEtran}
\usepackage{url}
\usepackage[utf8]{inputenc}
\usepackage{xcolor}
\usepackage{amsmath}
\usepackage{amssymb}
\usepackage{xspace}
\usepackage{epsfig}
\usepackage{balance}

\usepackage[acronyms,nonumberlist,nopostdot,nomain,nogroupskip,acronymlists={hidden}]{glossaries}
\newglossary[algh]{hidden}{acrh}{acnh}{Hidden Acronyms}

\usepackage{booktabs}
\usepackage{tabularx}

\usepackage{algorithm}
\usepackage{algorithmic}

\usepackage{tikz}
\usepackage{pgfplots}
\pgfplotsset{compat=newest}
\pgfplotsset{plot coordinates/math parser=false}
\newlength\fheight
\newlength\fwidth
\usetikzlibrary{plotmarks,patterns,decorations.pathreplacing,backgrounds,calc,arrows,arrows.meta,spy,matrix,scopes}
\usepgfplotslibrary{patchplots,groupplots}
\usepackage{tikzscale}
\usepackage[pdfstartview=XYZ,bookmarks=true,colorlinks=true,linkcolor=blue,urlcolor=blue,citecolor=blue,pdftex,bookmarks=true,linktocpage=true,hyperindex=true]{hyperref}

\newcommand{\platform}{ns-O-RAN\xspace}

\graphicspath{{figures/}}
\newif\ifexttikz
\exttikzfalse

\ifexttikz
	\usetikzlibrary{external}
	\usepackage{fontspec}
\fi

\usepackage{multirow}

\usepackage[font=footnotesize]{caption}
\usepackage{subcaption}
\usepackage{mathtools}
\usepackage[numbers,sort&compress]{natbib}

\usepackage{soul}

\newacronym{3gpp}{3GPP}{3rd Generation Partnership Project}
\newacronym{4g}{4G}{4th generation}
\newacronym{5g}{5G}{5th generation}
\newacronym{6g}{6G}{6th generation}
\newacronym{5gc}{5GC}{5G Core}
\newacronym{adc}{ADC}{Analog to Digital Converter}
\newacronym{aerpaw}{AERPAW}{Aerial Experimentation and Research Platform for Advanced Wireless}
\newacronym{ai}{AI}{Artificial Intelligence}
\newacronym{aimd}{AIMD}{Additive Increase Multiplicative Decrease}
\newacronym{am}{AM}{Acknowledged Mode}
\newacronym{amc}{AMC}{Adaptive Modulation and Coding}
\newacronym{amf}{AMF}{Access and Mobility Management Function}
\newacronym{aops}{AOPS}{Adaptive Order Prediction Scheduling}
\newacronym{api}{API}{Application Programming Interface}
\newacronym{apn}{APN}{Access Point Name}
\newacronym{ap}{AP}{Application Protocol}
\newacronym{aqm}{AQM}{Active Queue Management}
\newacronym{asn1}{ASN.1}{Abstract Syntax Notation One}
\newacronym{ausf}{AUSF}{Authentication Server Function}
\newacronym{avc}{AVC}{Advanced Video Coding}
\newacronym{awgn}{AGWN}{Additive White Gaussian Noise}
\newacronym{balia}{BALIA}{Balanced Link Adaptation Algorithm}
\newacronym{bbu}{BBU}{Base Band Unit}
\newacronym{bdp}{BDP}{Bandwidth-Delay Product}
\newacronym{ber}{BER}{Bit Error Rate}
\newacronym{bf}{BF}{Beamforming}
\newacronym{bler}{BLER}{Block Error Rate}
\newacronym{brr}{BRR}{Bayesian Ridge Regressor}
\newacronym{bs}{BS}{Base Station}
\newacronym{bsr}{BSR}{Buffer Status Report}
\newacronym{bss}{BSS}{Business Support System}
\newacronym{ca}{CA}{Carrier Aggregation}
\newacronym{caas}{CaaS}{Connectivity-as-a-Service}
\newacronym{cb}{CB}{Code Block}
\newacronym{cc}{CC}{Congestion Control}
\newacronym{compc}{CC}{Component Carrier}
\newacronym{ccid}{CCID}{Congestion Control ID}
\newacronym{cco}{CC}{Carrier Component}
\newacronym{cdd}{CDD}{Cyclic Delay Diversity}
\newacronym{cdf}{CDF}{Cumulative Distribution Function}
\newacronym{cdn}{CDN}{Content Distribution Network}
\newacronym{cn}{CN}{Core Network}
\newacronym{cnn}{CNN}{Convolutional Neural Network}
\newacronym{codel}{CoDel}{Controlled Delay Management}
\newacronym{comac}{COMAC}{Converged Multi-Access and Core}
\newacronym{cord}{CORD}{Central Office Re-architected as a Datacenter}
\newacronym{cornet}{CORNET}{COgnitive Radio NETwork}
\newacronym{cosmos}{COSMOS}{Cloud Enhanced Open Software Defined Mobile Wireless Testbed for City-Scale Deployment}
\newacronym{cots}{COTS}{Commercial Off-the-Shelf}
\newacronym{cp}{CP}{Control Plane}
\newacronym{cpu}{CPU}{Central Processing Unit}
\newacronym{cqi}{CQI}{Channel Quality Information}
\newacronym{cr}{CR}{Cognitive Radio}
\newacronym{cql}{CQL}{Conservative Q-Learning}
\newacronym{cran}{CRAN}{Cloud \gls{ran}}
\newacronym{crs}{CRS}{Cell Reference Signal}
\newacronym{csi}{CSI}{Channel State Information}
\newacronym{csirs}{CSI-RS}{Channel State Information - Reference Signal}
\newacronym{cu}{CU}{Centralized Unit}
\newacronym{cucp}{CU-CP}{Centralized Unit - Control Plane}
\newacronym{cuup}{CU-UP}{Centralized Unit - User Plane}
\newacronym{d2tcp}{D$^2$TCP}{Deadline-aware Data center TCP}
\newacronym{d3}{D$^3$}{Deadline-Driven Delivery}
\newacronym{dac}{DAC}{Digital to Analog Converter}
\newacronym{dag}{DAG}{Directed Acyclic Graph}
\newacronym{das}{DAS}{Distributed Antenna System}
\newacronym{dash}{DASH}{Dynamic Adaptive Streaming over HTTP}
\newacronym{dc}{DC}{Dual Connectivity}
\newacronym{dccp}{DCCP}{Datagram Congestion Control Protocol}
\newacronym{dce}{DCE}{Direct Code Execution}
\newacronym{dci}{DCI}{Downlink Control Information}
\newacronym{dctcp}{DCTCP}{Data Center TCP}
\newacronym{dl}{DL}{Downlink}
\newacronym{dmr}{DMR}{Deadline Miss Ratio}
\newacronym{dmrs}{DMRS}{DeModulation Reference Signal}
\newacronym{drlcc}{DRL-CC}{Deep Reinforcement Learning Congestion Control}
\newacronym{drb}{DRB}{Data Radio Bearer}
\newacronym{drs}{DRS}{Discovery Reference Signal}
\newacronym{dnn}{DNN}{Deep Neural Network}
\newacronym{dqn}{DQN}{Deep Q-Network}
\newacronym{du}{DU}{Distributed Unit}
\newacronym{e2e}{E2E}{end-to-end}
\newacronym{e2ap}{E2AP}{E2 Application Protocol}
\newacronym{e2sm}{E2SM}{E2 Service Model}
\newacronym{ecaas}{ECaaS}{Edge-Cloud-as-a-Service}
\newacronym{ecn}{ECN}{Explicit Congestion Notification}
\newacronym{edc}{EDC}{Edge Data Center}
\newacronym{edf}{EDF}{Earliest Deadline First}
\newacronym{embb}{eMBB}{Enhanced Mobile Broadband}
\newacronym{empower}{EMPOWER}{EMpowering transatlantic PlatfOrms for advanced WirEless Research}
\newacronym{enb}{eNB}{evolved Node Base}
\newacronym{endc}{EN-DC}{E-UTRAN-NR Dual Connectivity}
\newacronym{epc}{EPC}{Evolved Packet Core}
\newacronym{eps}{EPS}{Evolved Packet System}
\newacronym{es}{ES}{Edge Server}
\newacronym{etl}{ETL}{Extract, Transform and Load}
\newacronym{etsi}{ETSI}{European Telecommunications Standards Institute}
\newacronym[firstplural=Estimated Times of Arrival (ETAs)]{eta}{ETA}{Estimated Time of Arrival}
\newacronym{eutran}{E-UTRAN}{Evolved Universal Terrestrial Access Network}
\newacronym{faas}{FaaS}{Function-as-a-Service}
\newacronym{fapi}{FAPI}{Functional Application Platform Interface}
\newacronym{fcaps}{FCAPS}{Fault, Configuration, Accounting, Performance and Security}
\newacronym{fdd}{FDD}{Frequency Division Duplexing}
\newacronym{fdm}{FDM}{Frequency Division Multiplexing}
\newacronym{fdma}{FDMA}{Frequency Division Multiple Access}
\newacronym{fed4fire}{FED4FIRE+}{Federation 4 Future Internet Research and Experimentation Plus}
\newacronym{fir}{FIR}{Finite Impulse Response}
\newacronym{fit}{FIT}{Future \acrlong{iot}}
\newacronym{fpga}{FPGA}{Field Programmable Gate Array}
\newacronym{fr1}{FR1}{Frequency Range 1}
\newacronym{fr2}{FR2}{Frequency Range 2}
\newacronym{fs}{FS}{Fast Switching}
\newacronym{fscc}{FSCC}{Flow Sharing Congestion Control}
\newacronym{ftp}{FTP}{File Transfer Protocol}
\newacronym{fw}{FW}{Flow Window}
\newacronym{gbr}{GBR}{Guaranteed Bit Rate}
\newacronym{ge}{GE}{Gaussian Elimination}
\newacronym{gnb}{gNB}{Next Generation Node Base}
\newacronym{gop}{GOP}{Group of Pictures}
\newacronym{gpr}{GPR}{Gaussian Process Regressor}
\newacronym{gpu}{GPU}{Graphics Processing Unit}
\newacronym{gtp}{GTP}{GPRS Tunneling Protocol}
\newacronym{gtpc}{GTP-C}{GPRS Tunnelling Protocol Control Plane}
\newacronym{gtpu}{GTP-U}{GPRS Tunnelling Protocol User Plane}
\newacronym{gtpv2c}{GTPv2-C}{\gls{gtp} v2 - Control}
\newacronym{gw}{GW}{Gateway}
\newacronym{harq}{HARQ}{Hybrid Automatic Repeat reQuest}
\newacronym{hetnet}{HetNet}{Heterogeneous Network}
\newacronym{hh}{HH}{Hard Handover}
\newacronym{ho}{HO}{Handover}
\newacronym{hol}{HOL}{Head-of-Line}
\newacronym{hqf}{HQF}{Highest-quality-first}
\newacronym{hss}{HSS}{Home Subscription Server}
\newacronym{http}{HTTP}{HyperText Transfer Protocol}
\newacronym{ia}{IA}{Initial Access}
\newacronym{iab}{IAB}{Integrated Access and Backhaul}
\newacronym{ic}{IC}{Incident Command}
\newacronym{ietf}{IETF}{Internet Engineering Task Force}
\newacronym{imsi}{IMSI}{International Mobile Subscriber Identity}
\newacronym{imt}{IMT}{International Mobile Telecommunication}
\newacronym{iot}{IoT}{Internet of Things}
\newacronym{ip}{IP}{Internet Protocol}
\newacronym{itu}{ITU}{International Telecommunication Union}
\newacronym{kpi}{KPI}{Key Performance Indicator}
\newacronym{kpm}{KPM}{Key Performance Measurement}
\newacronym{kvm}{KVM}{Kernel-based Virtual Machine}
\newacronym{los}{LOS}{Line-of-Sight}
\newacronym{ldc}{LDC}{Local Data Center}
\newacronym{lsm}{LSM}{Link-to-System Mapping}
\newacronym{lstm}{LSTM}{Long Short Term Memory}
\newacronym{lte}{LTE}{Long Term Evolution}
\newacronym{lxc}{LXC}{Linux Container}
\newacronym{m2m}{M2M}{Machine to Machine}
\newacronym{mac}{MAC}{Medium Access Control}
\newacronym{manet}{MANET}{Mobile Ad Hoc Network}
\newacronym{mano}{MANO}{Management and Orchestration}
\newacronym{mbr}{MBR}{Maximum Bit Rate}
\newacronym{mc}{MC}{Multi-Connectivity}
\newacronym{mcc}{MCC}{Mobile Cloud Computing}
\newacronym{mchem}{MCHEM}{Massive Channel Emulator}
\newacronym{mcs}{MCS}{Modulation and Coding Scheme}
\newacronym{mdp}{MDP}{Markov Decision Process}
\newacronym{mec}{MEC}{Multi-access Edge Computing}
\newacronym{mec2}{MEC}{Mobile Edge Cloud}
\newacronym{mfc}{MFC}{Mobile Fog Computing}
\newacronym{mgen}{MGEN}{Multi-Generator}
\newacronym{mi}{MI}{Mutual Information}
\newacronym{mib}{MIB}{Master Information Block}
\newacronym{miesm}{MIESM}{Mutual Information Based Effective SINR}
\newacronym{mimo}{MIMO}{Multiple Input, Multiple Output}
\newacronym{ml}{ML}{Machine Learning}
\newacronym{mlr}{MLR}{Maximum-local-rate}
\newacronym[plural=\gls{mme}s,firstplural=Mobility Management Entities (MMEs)]{mme}{MME}{Mobility Management Entity}
\newacronym{mmtc}{mMTC}{Massive Machine-Type Communications}
\newacronym{mmwave}{mmWave}{millimeter wave}
\newacronym{mpdccp}{MP-DCCP}{Multipath Datagram Congestion Control Protocol}
\newacronym{mptcp}{MPTCP}{Multipath TCP}
\newacronym{mr}{MR}{Maximum Rate}
\newacronym{mrdc}{MR-DC}{Multi \gls{rat} \gls{dc}}
\newacronym{mse}{MSE}{Mean Square Error}
\newacronym{mss}{MSS}{Maximum Segment Size}
\newacronym{mt}{MT}{Mobile Termination}
\newacronym{mtd}{MTD}{Machine-Type Device}
\newacronym{mtu}{MTU}{Maximum Transmission Unit}
\newacronym{mumimo}{MU-MIMO}{Multi-user \gls{mimo}}
\newacronym{mvno}{MVNO}{Mobile Virtual Network Operator}
\newacronym{nalu}{NALU}{Network Abstraction Layer Unit}
\newacronym{nas}{NAS}{Non-Access Stratum}
\newacronym{ngran}{NG-RAN}{Next Generation - \gls{ran}}
\newacronym{ns3}{ns-3}{Network Simulator 3}
\newacronym{nbiot}{NB-IoT}{Narrow Band IoT}
\newacronym{nf}{NF}{Network Function}
\newacronym{nfv}{NFV}{Network Function Virtualization}
\newacronym{nfvi}{NFVI}{Network Function Virtualization Infrastructure}
\newacronym{nic}{NIC}{Network Interface Card}
\newacronym{nlos}{NLOS}{Non-Line-of-Sight}
\newacronym{now}{NOW}{Non Overlapping Window}
\newacronym{nsm}{NSM}{Network Service Mesh}
\newacronym[type=hidden]{nr}{NR}{New Radio}
\newacronym{nrf}{NRF}{Network Repository Function}
\newacronym{nsa}{NSA}{Non Stand Alone}
\newacronym{nse}{NSE}{Network Slicing Engine}
\newacronym{nssf}{NSSF}{Network Slice Selection Function}
\newacronym{o2i}{O2I}{Outdoor to Indoor}
\newacronym{oai}{OAI}{OpenAirInterface}
\newacronym{oaicn}{OAI-CN}{\gls{oai} \acrlong{cn}}
\newacronym{oairan}{OAI-RAN}{\acrlong{oai} \acrlong{ran}}
\newacronym{oam}{OAM}{Operations, Administration and Maintenance}
\newacronym{ofdm}{OFDM}{Orthogonal Frequency Division Multiplexing}
\newacronym{olia}{OLIA}{Opportunistic Linked Increase Algorithm}
\newacronym{omec}{OMEC}{Open Mobile Evolved Core}
\newacronym{onap}{ONAP}{Open Network Automation Platform}
\newacronym{onf}{ONF}{Open Networking Foundation}
\newacronym{onos}{ONOS}{Open Networking Operating System}
\newacronym{oom}{OOM}{\gls{onap} Operations Manager}
\newacronym{opnfv}{OPNFV}{Open Platform for \gls{nfv}}
\newacronym[type=hidden]{oran}{O-RAN}{Open \gls{ran}}
\newacronym{orbit}{ORBIT}{Open-Access Research Testbed for Next-Generation Wireless Networks}
\newacronym{os}{OS}{Operating System}
\newacronym{oss}{OSS}{Operations Support System}
\newacronym{pa}{PA}{Position-aware}
\newacronym{pase}{PASE}{Prioritization, Arbitration, and Self-adjusting Endpoints}
\newacronym{pawr}{PAWR}{Platforms for Advanced Wireless Research}
\newacronym{pbch}{PBCH}{Physical Broadcast Channel}
\newacronym{pcell}{PCell}{Primary Cell}
\newacronym{pcef}{PCEF}{Policy and Charging Enforcement Function}
\newacronym{pcfich}{PCFICH}{Physical Control Format Indicator Channel}
\newacronym{pcrf}{PCRF}{Policy and Charging Rules Function}
\newacronym{pdcch}{PDCCH}{Physical Downlink Control Channel}
\newacronym{pdcp}{PDCP}{Packet Data Convergence Protocol}
\newacronym{pdcpc}{PDCP-C}{Packet Data Convergence Protocol - Control Plane}
\newacronym{pdcpu}{PDCP-U}{Packet Data Convergence Protocol - User Plane}
\newacronym{pdsch}{PDSCH}{Physical Downlink Shared Channel}
\newacronym{pdu}{PDU}{Packet Data Unit}
\newacronym{pf}{PF}{Proportional Fair}
\newacronym{pgw}{PGW}{Packet Gateway}
\newacronym{phich}{PHICH}{Physical Hybrid ARQ Indicator Channel}
\newacronym{phy}{PHY}{Physical}
\newacronym{phyu}{PHY-U}{Upper Physical}
\newacronym{phyl}{PHY-L}{Lower Physical}
\newacronym{pmch}{PMCH}{Physical Multicast Channel}
\newacronym{pmi}{PMI}{Precoding Matrix Indicators}
\newacronym{powder}{POWDER}{Platform for Open Wireless Data-driven Experimental Research}
\newacronym{ppo}{PPO}{Proximal Policy Optimization}
\newacronym{ppp}{PPP}{Poisson Point Process}
\newacronym{prach}{PRACH}{Physical Random Access Channel}
\newacronym{prb}{PRB}{Physical Resource Block}
\newacronym{pscell}{PSCell}{Primary cell of the Secondary Node}
\newacronym{psnr}{PSNR}{Peak Signal to Noise Ratio}
\newacronym{pss}{PSS}{Primary Synchronization Signal}
\newacronym{pucch}{PUCCH}{Physical Uplink Control Channel}
\newacronym{pusch}{PUSCH}{Physical Uplink Shared Channel}
\newacronym{qam}{QAM}{Quadrature Amplitude Modulation}
\newacronym{qci}{QCI}{\gls{qos} Class Identifier}
\newacronym{qoe}{QoE}{Quality of Experience}
\newacronym{qos}{QoS}{Quality of Service}
\newacronym{quic}{QUIC}{Quick UDP Internet Connections}
\newacronym{rach}{RACH}{Random Access Channel}
\newacronym{ran}{RAN}{Radio Access Network}
\newacronym[firstplural=Radio Access Technologies (RATs)]{rat}{RAT}{Radio Access Technology}
\newacronym{rcn}{RCN}{Research Coordination Network}
\newacronym{rec}{REC}{Radio Edge Cloud}
\newacronym{red}{RED}{Random Early Detection}
\newacronym{rem}{REM}{Random Ensemble Mixture}
\newacronym{renew}{RENEW}{Reconfigurable Eco-system for Next-generation End-to-end Wireless}
\newacronym{rf}{RF}{Radio Frequency}
\newacronym{rfc}{RFC}{Request for Comments}
\newacronym{rfr}{RFR}{Random Forest Regressor}
\newacronym{ric}{RIC}{\gls{ran} Intelligent Controller}
\newacronym{rlc}{RLC}{Radio Link Control}
\newacronym{rl}{RL}{Reinforcement Learning}
\newacronym{rlf}{RLF}{Radio Link Failure}
\newacronym{rlnc}{RLNC}{Random Linear Network Coding}
\newacronym{rmr}{RMR}{RIC Message Routing}
\newacronym{rmse}{RMSE}{Root Mean Squared Error}
\newacronym{rnis}{RNIS}{Radio Network Information Service}
\newacronym{rnib}{RNIB}{Radio Network Information Base}
\newacronym{rr}{RR}{Round Robin}
\newacronym{rrc}{RRC}{Radio Resource Control}
\newacronym{rrm}{RRM}{Radio Resource Management}
\newacronym{rru}{RRU}{Remote Radio Unit}
\newacronym{rs}{RS}{Remote Server}
\newacronym{rsrp}{RSRP}{Reference Signal Received Power}
\newacronym{rsrq}{RSRQ}{Reference Signal Received Quality}
\newacronym{rss}{RSS}{Received Signal Strength}
\newacronym{rssi}{RSSI}{Received Signal Strength Indicator}
\newacronym{rtt}{RTT}{Round Trip Time}
\newacronym{ru}{RU}{Radio Unit}
\newacronym{rw}{RW}{Receive Window}
\newacronym{rx}{RX}{Receiver}
\newacronym{s1ap}{S1AP}{S1 Application Protocol}
\newacronym{sa}{SA}{standalone}
\newacronym{sack}{SACK}{Selective Acknowledgment}
\newacronym{sap}{SAP}{Service Access Point}
\newacronym{sc2}{SC2}{Spectrum Collaboration Challenge}
\newacronym{scef}{SCEF}{Service Capability Exposure Function}
\newacronym{sch}{SCH}{Secondary Cell Handover}
\newacronym{scoot}{SCOOT}{Split Cycle Offset Optimization Technique}
\newacronym{sctp}{SCTP}{Stream Control Transmission Protocol}
\newacronym{sdap}{SDAP}{Service Data Adaptation Protocol}
\newacronym{sdk}{SDK}{Software Development Kit}
\newacronym{sdm}{SDM}{Space Division Multiplexing}
\newacronym{sdma}{SDMA}{Spatial Division Multiple Access}
\newacronym{sdn}{SDN}{Software-defined Networking}
\newacronym{sdr}{SDR}{Software-defined Radio}
\newacronym{seba}{SEBA}{SDN-Enabled Broadband Access}
\newacronym{sgsn}{SGSN}{Serving GPRS Support Node}
\newacronym{sgw}{SGW}{Serving Gateway}
\newacronym{si}{SI}{Study Item}
\newacronym{sib}{SIB}{Secondary Information Block}
\newacronym{sinr}{SINR}{Signal to Interference plus Noise Ratio}
\newacronym{sip}{SIP}{Session Initiation Protocol}
\newacronym{siso}{SISO}{Single Input, Single Output}
\newacronym{sla}{SLA}{Service Level Agreement}
\newacronym{sm}{SM}{Service Model}
\newacronym{smf}{SMF}{Session Management Function}
\newacronym{smo}{SMO}{Service Management and Orchestration}
\newacronym{sms}{SMS}{Short Message Service}
\newacronym{smsgmsc}{SMS-GMSC}{\gls{sms}-Gateway}
\newacronym{snr}{SNR}{Signal-to-Noise-Ratio}
\newacronym{son}{SON}{Self-Organizing Network}
\newacronym{sptcp}{SPTCP}{Single Path TCP}
\newacronym{srb}{SRB}{Service Radio Bearer}
\newacronym{srn}{SRN}{Standard Radio Node}
\newacronym{srs}{SRS}{Sounding Reference Signal}
\newacronym{ss}{SS}{Synchronization Signal}
\newacronym{sss}{SSS}{Secondary Synchronization Signal}
\newacronym{st}{ST}{Spanning Tree}
\newacronym{svc}{SVC}{Scalable Video Coding}
\newacronym{tb}{TB}{Transport Block}
\newacronym{tcp}{TCP}{Transmission Control Protocol}
\newacronym{tdd}{TDD}{Time Division Duplexing}
\newacronym{tdm}{TDM}{Time Division Multiplexing}
\newacronym{tdma}{TDMA}{Time Division Multiple Access}
\newacronym{tfl}{TfL}{Transport for London}
\newacronym{tfrc}{TFRC}{TCP-Friendly Rate Control}
\newacronym{tft}{TFT}{Traffic Flow Template}
\newacronym{tgen}{TGEN}{Traffic Generator}
\newacronym{tip}{TIP}{Telecom Infra Project}
\newacronym{tm}{TM}{Transparent Mode}
\newacronym{to}{TO}{Telco Operator}
\newacronym{tr}{TR}{Technical Report}
\newacronym{trp}{TRP}{Transmitter Receiver Pair}
\newacronym{ts}{TS}{Traffic Steering}
\newacronym{tti}{TTI}{Transmission Time Interval}
\newacronym{ttt}{TTT}{Time-to-Trigger}
\newacronym{tx}{TX}{Transmitter}
\newacronym{uas}{UAS}{Unmanned Aerial System}
\newacronym{uav}{UAV}{Unmanned Aerial Vehicle}
\newacronym{udm}{UDM}{Unified Data Management}
\newacronym{udp}{UDP}{User Datagram Protocol}
\newacronym{udr}{UDR}{Unified Data Repository}
\newacronym{ue}{UE}{User Equipment}
\newacronym{uhd}{UHD}{\gls{usrp} Hardware Driver}
\newacronym{ul}{UL}{Uplink}
\newacronym{um}{UM}{Unacknowledged Mode}
\newacronym{uml}{UML}{Unified Modeling Language}
\newacronym{upa}{UPA}{Uniform Planar Array}
\newacronym{upf}{UPF}{User Plane Function}
\newacronym{urllc}{URLLC}{Ultra Reliable and Low Latency Communications}
\newacronym{usa}{U.S.}{United States}
\newacronym{usim}{USIM}{Universal Subscriber Identity Module}
\newacronym{usrp}{USRP}{Universal Software Radio Peripheral}
\newacronym{utc}{UTC}{Urban Traffic Control}
\newacronym{vim}{VIM}{Virtualization Infrastructure Manager}
\newacronym{vm}{VM}{Virtual Machine}
\newacronym{vnf}{vNF}{Virtualized Network Function}
\newacronym{volte}{VoLTE}{Voice over \gls{lte}}
\newacronym{voltha}{VOLTHA}{Virtual OLT HArdware Abstraction}
\newacronym{vr}{VR}{Virtual Reality}
\newacronym{vran}{vRAN}{Virtualized \gls{ran}}
\newacronym{vss}{VSS}{Video Streaming Server}
\newacronym{v2x}{V2X}{Vehicle-to-everything}
\newacronym{wbf}{WBF}{Wired Bias Function}
\newacronym{wf}{WF}{Waterfilling}
\newacronym{wlan}{WLAN}{Wireless Local Area Network}
\newacronym{osm}{OSM}{Open Source \gls{nfv} Management and Orchestration}
\newacronym{pnf}{PNF}{Physical Network Function}
\newacronym{drl}{DRL}{Deep Reinforcement Learning}
\newacronym{mtc}{MTC}{Machine-type Communications}
\newacronym{osc}{OSC}{O-RAN Software Community}
\newacronym{rc}{RC}{RAN Control}
\newacronym{ar}{AR}{Augmented Reality}
\newacronym{daps}{DAPS}{Dual Active Protocol Stack}
\newacronym{nib}{NIB}{Network Information Base}
\tikzstyle{startstop} = [rectangle, rounded corners, minimum width=2cm, minimum height=0.5cm,text centered, draw=black]
\tikzstyle{io} = [trapezium, trapezium left angle=70, trapezium right angle=110, minimum width=3cm, minimum height=1cm, text centered, draw=black]
\tikzstyle{process} = [rectangle, minimum width=2cm, minimum height=0.5cm, text centered, draw=black, alignb=center]
\tikzstyle{decision} = [ellipse, minimum width=2cm, minimum height=1cm, text centered, draw=black]
\tikzstyle{arrow} = [thick,<->,>=stealth]
\tikzstyle{line} = [thick,>=stealth]
\tikzstyle{darrow} = [thick,<->,>=stealth,dashed]
\tikzstyle{sarrow} = [thick,->,>=stealth]
\tikzstyle{larrow} = [line width=0.3mm,dashdotted,->,>=stealth]
\tikzstyle{llarrow} = [line width=0.1mm,->,>=stealth]

\makeatletter
\def\grd@save@target#1{%
  \def\grd@target{#1}}
\def\grd@save@start#1{%
  \def\grd@start{#1}}
\tikzset{
  grid with coordinates/.style={
    to path={%
      \pgfextra{%
        \edef\grd@@target{(\tikztotarget)}%
        \tikz@scan@one@point\grd@save@target\grd@@target\relax
        \edef\grd@@start{(\tikztostart)}%
        \tikz@scan@one@point\grd@save@start\grd@@start\relax
        \draw[minor help lines] (\tikztostart) grid (\tikztotarget);
        \draw[major help lines] (\tikztostart) grid (\tikztotarget);
        \grd@start
        \pgfmathsetmacro{\grd@xa}{\the\pgf@x/1cm}
        \pgfmathsetmacro{\grd@ya}{\the\pgf@y/1cm}
        \grd@target
        \pgfmathsetmacro{\grd@xb}{\the\pgf@x/1cm}
        \pgfmathsetmacro{\grd@yb}{\the\pgf@y/1cm}
        \pgfmathsetmacro{\grd@xc}{\grd@xa + \pgfkeysvalueof{/tikz/grid with coordinates/major step x}}
        \pgfmathsetmacro{\grd@yc}{\grd@ya + \pgfkeysvalueof{/tikz/grid with coordinates/major step y}}
        \foreach \x in {\grd@xa,\grd@xc,...,\grd@xb}
        \node[anchor=north] at (\x,\grd@ya) {\pgfmathprintnumber{\x}};
        \foreach \y in {\grd@ya,\grd@yc,...,\grd@yb}
        \node[anchor=east] at (\grd@xa,\y) {\pgfmathprintnumber{\y}};
      }
    }
  },
  minor help lines/.style={
    help lines,
    gray,
    line cap =round,
    xstep=\pgfkeysvalueof{/tikz/grid with coordinates/minor step x},
    ystep=\pgfkeysvalueof{/tikz/grid with coordinates/minor step y}
  },
  major help lines/.style={
    help lines,
    line cap =round,
    line width=\pgfkeysvalueof{/tikz/grid with coordinates/major line width},
    xstep=\pgfkeysvalueof{/tikz/grid with coordinates/major step x},
    ystep=\pgfkeysvalueof{/tikz/grid with coordinates/major step y}
  },
  grid with coordinates/.cd,
  minor step x/.initial=.5,
  minor step y/.initial=.2,
  major step x/.initial=1,
  major step y/.initial=1,
  major line width/.initial=1pt,
}
\makeatother

\definecolor{desireRed}{RGB}{230,57,60}%
\definecolor{darkPurple}{RGB}{59,31,43}%
\definecolor{springGreen}{RGB}{37,223,145}%
\definecolor{queenBlue}{RGB}{69,123,157}%
\definecolor{spaceCadet}{RGB}{29,53,87}%

\usepackage{dblfloatfix}

\begin{document}
\bstctlcite{BSTcontrol}  

\title{Programmable and Customized Intelligence for Traffic Steering in 5G Networks Using\\Open RAN Architectures}
%
%

\author{Andrea~Lacava,~\IEEEmembership{Student~Member,~IEEE,} 
		Michele Polese,~\IEEEmembership{Member,~IEEE,}\\
		Rajarajan Sivaraj,~\IEEEmembership{Senior~Member,~IEEE},
        ~Rahul Soundrarajan,
        ~Bhawani Shanker Bhati,
        ~Tarunjeet Singh,\\
        ~Tommaso Zugno,
        ~Francesca Cuomo,~\IEEEmembership{Senior~Member,~IEEE}
        and~Tommaso Melodia~\IEEEmembership{Fellow,~IEEE}
\thanks{A. Lacava, M. Polese, and T. Melodia are with the Institute for the Wireless Internet of Things at Northeastern University, Boston, MA, 02115 USA. Tommaso Zugno was affiliated with Northeastern University and University of Padova at the time the research was conducted.}
\thanks{R. Sivaraj, R. Soundrarajan, B. S. Bhati and T. Singh are with Mavenir, Richardson, TX, USA.}
\thanks{A. Lacava and F. Cuomo are with Sapienza, University of Rome, IT 00185.}%
\thanks{The corresponding author is A. Lacava, email: lacava.a@northeastern.edu.}
\thanks{M.~Polese and R.~Sivaraj equally contributed to the paper.}%
\thanks{This work was partially supported by Mavenir and by the U.S.\ National Science Foundation under Grants CNS-1923789 and CNS-2112471.} 
\vspace{-.8cm}}

%
%
%
%



\maketitle

\begin{abstract}
5G and beyond mobile networks will support heterogeneous use cases at an unprecedented scale, thus demanding automated control and optimization of network functionalities customized to the needs of individual users. 
Such fine-grained control of the \gls{ran} is not possible with the current cellular architecture. To fill this gap, the Open \gls{ran} paradigm and its specification introduce an ``open'' architecture with abstractions that enable closed-loop control and provide data-driven, and intelligent optimization of the \gls{ran} at the user-level. 
This is obtained through custom \gls{ran} control applications (i.e., xApps) deployed on near-real-time \gls{ran} Intelligent Controller (near-RT RIC) at the edge of the network.
Despite these premises, as of today the research community lacks a sandbox to build data-driven xApps, and create large-scale datasets for effective \gls{ai} training. 
In this paper, we address this by introducing {\em \platform}, a software framework that integrates a real-world, production-grade near-RT RIC with a 3GPP-based simulated environment on ns-3, enabling at the same time the development of xApps and automated large-scale data collection and testing of Deep Reinforcement Learning (DRL)-driven control policies for the optimization at the user-level.
In addition, we propose the first user-specific O-\gls{ran} \gls{ts} intelligent handover framework. It uses Random Ensemble Mixture (REM), a  Conservative $Q$-learning (CQL) algorithm, combined with a state-of-the-art Convolutional Neural Network (CNN) architecture, to optimally assign a serving base station to each user in the network.
Our \gls{ts} xApp, trained with more than 40 million data points collected by \platform, runs on the near-RT RIC and controls the \platform base stations.
We evaluate the performance on a large-scale deployment with up to 126 users with 8 base stations, showing that the xApp-based handover improves throughput and spectral efficiency by an average of 50\% over traditional handover heuristics, with less mobility overhead.
\end{abstract}

\vspace{-.3cm}


\vspace{-.8cm}

\begin{picture}(0,0)(10,-565)
\put(0,0){
\put(0,0){\footnotesize This work has been submitted to the IEEE for possible publication.}
\put(0,-10){
\footnotesize Copyright may be transferred without notice, after which this version may no longer be accessible.}}
\end{picture}

%
\IEEEpeerreviewmaketitle
\glsresetall

\section{Introduction}
\label{sec:intro}

\IEEEPARstart{F}{ifth} generation (5G) cellular networks and beyond shall provide improved wireless communications and networking capabilities, enabling heterogeneous use cases such as \gls{urllc}, \gls{embb}, and massive machine-type communications, ranging from industrial \gls{iot} to metaverse, telepresence and remote telesurgery.
The use-case requirements and deployment scenarios keep changing with evolving radio access technologies.
As a consequence, 5G and beyond \glspl{ran} are expected to be complex systems, deployed at a scale that is unforeseen in commercial networks so far~\cite{letaief2019roadmap}.

This complexity and the evolving use-case requirements have prompted research, development, and standardization efforts in novel \gls{ran} architectures, and, notably, in the \gls{oran} paradigm.
Nowadays, classic \glspl{ran} are deployed with monolithic network functions (e.g., base stations) on black-box hardware.
Such architecture is considered static and hard to reconfigure on-demand without any manual on-site intervention.
The \gls{oran} architecture disrupts the classical approach by adopting the principles of  {\em Disaggregation}, {\em Openness}, {\em Virtualization}, and {\em Programmability}.
In \gls{oran}, the classic base station is \textit{disaggregated}, i.e divided across multiple \gls{ran} nodes.
The interfaces between the different nodes are \textit{open} and standardized, to achieve multi-vendor interoperability.
Network functions that implement the classic \gls{ran} operations are virtualized and software-based and deployed on white box hardware~\cite{polese2022understanding}. Software enables algorithmic and programmatic control based on the current network status, enabling the dynamic configuration of the infrastructure.

The combination of these principles introduces complex, virtualized architectures with \glspl{ric} that (i) have a centralized abstraction of the network; and (ii) host applications performing closed-loop control of the \gls{ran}. This custom logic leverages the centralized aggregation of analytics on multiple network functions to run advanced data-driven \gls{ai} and \gls{ml} techniques. For example, the near-real-time (near-RT) \gls{ric} hosts third-party applications called xApps that interact with the \gls{ran} through the E2 interface and take \gls{rrm} decisions at a time scale between 10 ms and 1 second~\cite{abdalla2021generation}.
Such architecture can efficiently learn complex cross-layer interactions across nodes, going beyond traditional control heuristics toward optimal \gls{rrm}~\cite{challita2020when,chinchali2018cellular}. 

Unlocking the intelligence in the networks is a crucial aspect of \gls{oran}.
Specifically, the xApps integrate custom logic and \gls{ai}/\gls{ml} algorithms for the \gls{ran}~\cite{bonati2021intelligence,mollahasani2021dynamic}, paving the way to an enhanced network control with an \gls{ue}-level granularity that would not be possible with the classical \gls{ran} architectures.
%
Indeed, the availability of data and analytics on the network in a centralized location (i.e., the \gls{ric}) enables new new approaches to traditional network management problems.
One of this use cases is the \gls{ts}, i.e., the management of the mobility management of individual \glspl{ue} served by the RAN~\cite{oran-wg1-use-cases}. 
\gls{ts} involves key \gls{ran} procedures, such as handover management, dual connectivity, and carrier aggregation, among others.
While handover management is a classic problem, the requirements and deployment scenarios for optimizing handovers keep changing with evolving radio access technologies and use-cases, posing newer challenges and requiring newer optimization strategies~\cite{8812724}.
As an example, the handover optimization requirements for broadband access, e.g., \gls{embb} \glspl{ue} streaming high-quality video, are different from those of an autonomous car, i.e., an \gls{urllc} \glspl{ue}.

In this context, traditional \gls{rrm} solutions, largely based on heuristics only involving channel quality and load thresholds, are not primed to handle \gls{ue}-centric handover decisions for new use-cases, and are often based on local, and thus limited, information.
Data-driven solutions at the \gls{ric} can leverage a centralized point of view to learn complex  inter-dependencies between RAN parameters and target the optimization to the \gls{qos} requirements of each \gls{ue}. 

Despite the promising architectural enablers, how to design and test effective and intelligent \gls{ran} control solutions to embed into xApps is still a challenge~\cite{polese2021coloran}. 
First, any \gls{ml} solution needs to be properly trained.
This requires data collection on large-scale setups, with a massive amount of data points to collect to properly represent the state of the system and allow the agent to learn an accurate representation of the system.
Then, when it comes to closed-loop control through \gls{drl}, the \gls{ml} infrastructure requires an isolated environment for testing and online exploration, to avoid impacting the performance of production \glspl{ran}.
The \gls{ts} mentioned before, for instance, includes the optimization of handover across multiple base stations through data-driven xApps. 
The data collection would need to cover large scale deployments, with different network configurations, operating frequencies, combinations of source traffic, and user mobility. 
At the same time, testing poorly trained solutions or performing online exploration on a large scale, commercial deployment may cause users to unexpectedly lose connectivity or experience a degraded service due to sub-optimal handover decisions~\cite{app12010426}.

\textbf{Contributions --- }
In this paper, we study the \gls{ts} use case and especially the handover management of a \gls{nsa} \gls{ran} deployment with an \gls{oran}-based approach.
We first build an \gls{oran}-compliant near-RT \gls{ric} platform. 
Then, we propose the first \gls{ts} xApp based on \gls{drl} to optimally control mobility procedures at a UE level, using the \gls{ric} and advanced \gls{rl} techniques that select the optimal target cells for handover of individual \glspl{ue}.
We also propose {\em \platform}, a software module to connect the near-RT \gls{ric} to \gls{ns3} to study the \gls{ts} use case over different simulation setups.
%
%
Specifically, the contributions of this paper are as follows.

\noindent $\bullet$ {\em System Design}: We design and build a standard-compliant near-RT \gls{ric} platform with \gls{oran}-defined open interfaces and service models (i.e., standardized mechanisms to interact with \gls{ran} nodes). The relevant system design details are discussed in Section \ref{sec:sys}.

\noindent $\bullet$ {\em Integration}: We build {\em \platform}, a virtualized and simulated environment for \gls{oran}, which bridges large scale 5G simulations in the open-source \gls{ns3} with a real-world near-RT \gls{ric}.
\platform combines the scale and flexibility of a simulated \gls{ran} with any real-world, E2-compliant \gls{ric}.
In this context, simulation based on realistic channel and protocol stack models contributes to the collection of data for the ML-based xApps without the need of large scale deployments.
\platform extends the \gls{ns3} 5G \gls{ran} module~\cite{mezzavilla2018end} by adding an \gls{oran} compliant E2 implementation, including the protocol capabilities and advanced service models.
\platform enables the \gls{ran} to stream events and data to the near-RT \gls{ric}, and the \gls{ric} to send control actions to the \gls{ran} over the E2 interface.
These control actions are reflected in the call processing of the \gls{ran} functions and the updated data are streamed to the RIC.
Thus, \platform enables xApps development without relying on \gls{ran} baseband and radio units; the same xApps can subsequently be tested on a real \gls{ran}, without additional development effort.
The relevant details are discussed in Section \ref{sec:sys}.
We pledge to release \platform as open-source in the \gls{osc}\footnote{The code is available at https://gerrit.o-ran-sc.org/r/gitweb?p=sim/ns3-o-ran-e2.git.} and the OpenRAN Gym platform~\cite{bonati2022openrangym}.

\noindent $\bullet$ {\em TS Optimization}: We build a data-driven \gls{ai}-powered \gls{ts} xApp in the near-RT \gls{ric} to maximize the \gls{ue} throughput utility through handover control.
We used \platform to collect data for, design, and test the \gls{ts} xApp. 
We formulate the problem as a \gls{mdp} and solve it using \gls{rl} techniques.
In particular, we use novel variants of the Deep-Q Network algorithm, namely \gls{cql} and~\gls{rem} to model the Q-function and the loss function, along with a custom \gls{cnn} design to maximize the expected reward. 
Our design enables multi-\gls{ue} control with a multi-dimensional state space using a single \gls{rl} agent.
The problem formulation and optimization details are discussed in Section \ref{sec:ts}.

\noindent $\bullet$ {\em Performance Evaluation}: We extensively evaluate the xApp using different \glspl{kpi}, such as \gls{ue} throughput, spectral efficiency, and mobility overhead on a large-scale of \gls{ran} network created by \platform.
Leveraging the fine-grained \gls{ue}-level intelligence and optimization at the near-RT \gls{ric}, we demonstrate significant performance improvements ranging from 30\% to 50\% for the above \glspl{kpi} in Section \ref{sec:results}.  


\section{Background}
\label{sec:soa}
In this section, we review the state of the art on \gls{oran}, \gls{ts} and ns-3.

\subsection{O-RAN Cellular Architecture}

\begin{figure}
    \centering
     \includegraphics[width=.9\columnwidth]{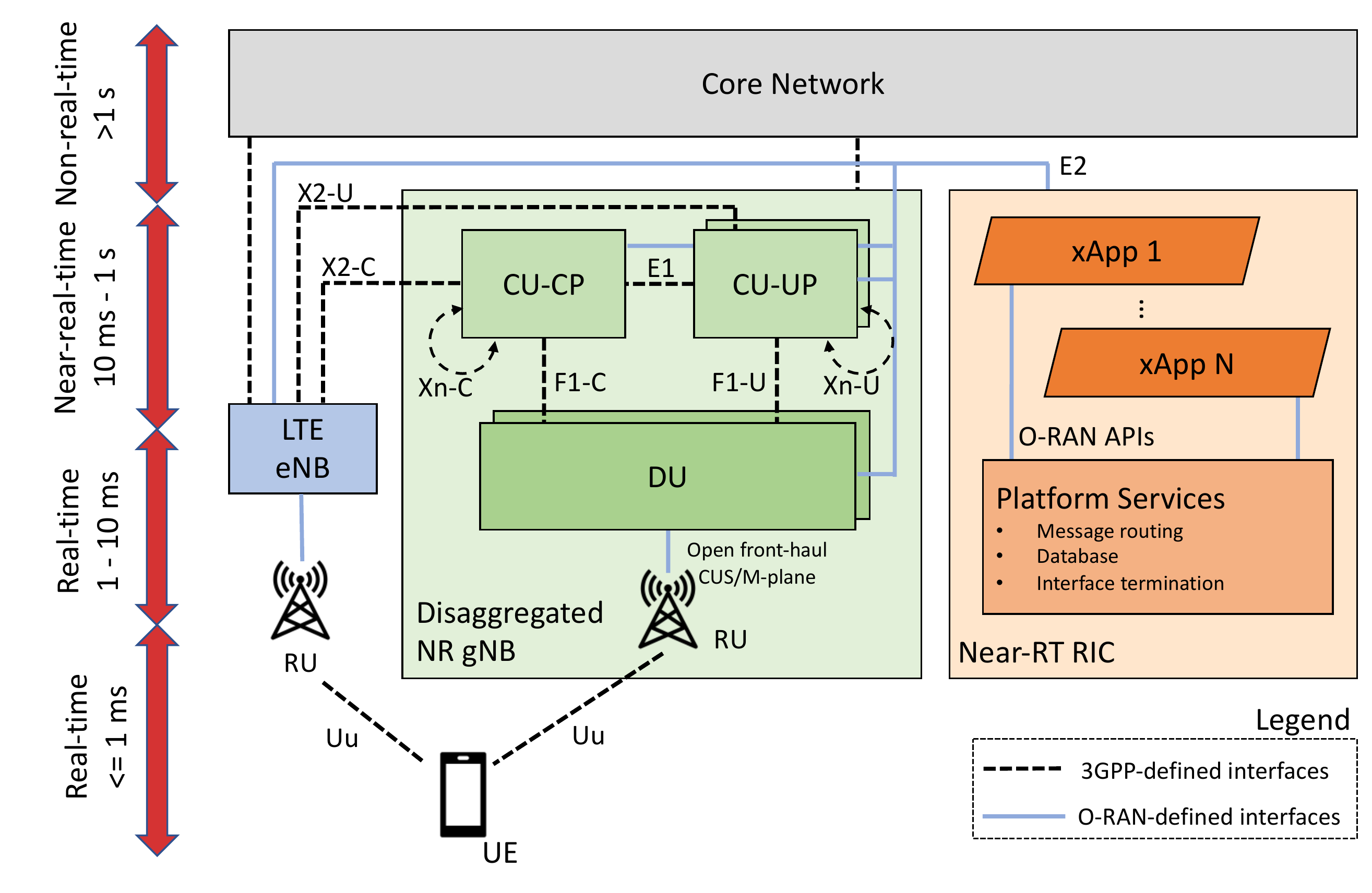}

    \caption{O-RAN architecture with the near-RT \gls{ric} functions, aside packet core.
    }
    \label{fig:oran_architecture}
\end{figure}

\textbf{RAN Protocol Stack ---} Figure~\ref{fig:oran_architecture} provides a high-level overview of the O-RAN architecture. 
In this, the baseband unit of the NR base station, also called the \gls{gnb}, is logically split into \gls{ran} functional nodes - \gls{cucp}, \gls{cuup}, \gls{du}, and \glspl{ru}.
These functions are are deployed as logical nodes in the \gls{ran} and connected through standardized \gls{oran} and \gls{3gpp}-defined interfaces.
In particular, the \gls{cucp} features the \gls{rrc} and \gls{pdcpc} layers, and manages the connectivity and mobility for the \glspl{ue}.
The \gls{cuup} handles the \gls{sdap} and \gls{pdcpu} layers, dealing with \glspl{drb} that carry user traffic.
The \gls{du} features the \gls{rlc}, \gls{mac} and \gls{phyu} layers, for buffer management, radio resource allocation, and physical layer functionalities, such as operating the NR cells.
For what it regards the \gls{lte}, all the layers are managed in a single function called \gls{enb}.
Finally, the \gls{ru} is responsible for \gls{phyl} layer, dealing with transmission and beamforming.

\textbf{RAN Intelligent Controllers ---} The near-RT \gls{ric} is typically deployed as a network function in a virtualized cloud platform at the edge of the RAN.
It onboards extensible applications (xApps), apart from \gls{oran} standardized platform framework functions, to optimize \gls{rrm} decisions for dedicated \gls{ran} functionalities using low-latency control loops at near-RT granularity (from 10 ms to 1 second).
The near-RT \gls{ric} connects through the E2 interface to the \gls{cucp}, \gls{cuup}, \gls{du} and \gls{enb}, collectively referred to as the E2 nodes.

{\bf E2 interface --- } E2 is a bi-directional interface that splits the \gls{rrm} between the E2 nodes and the near-RT \gls{ric}. 
With this architecture, the call processing and signaling procedures are implemented in the E2 nodes, but the \gls{rrm} decisions for these procedures are controlled by the \gls{ric} through xApps. 
For example, the handover procedures for a \gls{ue} are executed by the E2 node, but the \gls{ue}'s target cell for handover is decided and controlled by the \gls{ric}.

The procedures and messages exchanged over the E2 interface are standardized by \gls{e2ap}~\cite{polese2022understanding}. 
Using \gls{e2ap}, the E2 nodes can send reports to the near-RT \gls{ric} with \gls{ran} data or \gls{ue} context information.
In addition, the near-RT \gls{ric} can send control actions containing \gls{rrm} decisions and policies to the E2 node.
The xApps in the near-RT \gls{ric} encode and decode the payload of the \gls{e2ap} messages containing \gls{rrm}-specific information, as defined by the \glspl{e2sm}~\cite{oran-wg3-e2-sm}. 
The service models define the information model and semantics of \gls{rrm} operations over E2. 
Two \glspl{e2sm} of interest in this paper are {\em \gls{e2sm}-\gls{kpm}}~\cite{oran-wg3-e2-sm-kpm}, which allows E2 nodes to send \gls{ran} performance data to the \gls{ric}, with granularity down to the \gls{ue}-level, and {\em \gls{e2sm}-\gls{rc}}~\cite{oran-wg3-e2-sm-rc}, which allows the \gls{ric} to send back control based on \gls{rrm} decisions from xApps~\cite{polese2022understanding}.

\subsection{Intelligence in the RIC}

As already discussed, the disaggregation of the \gls{ran} functions enables the generation of large datasets that can be leveraged to study data driven approaches to the classical \gls{rrm} problems.
To support this trend, the \gls{oran} alliance has defined specifications for life cycle management of \gls{ml}-driven \gls{ran} control.
In fact, in \gls{oran} any \gls{ml} model shall be trained offline and deployed as xApps for online inference and \gls{rrm} control in the \gls{ric}.
One of the most promising approaches is the \gls{rl}, which teaches an {\em agent} how to choose an {\em action} from its action space, within a particular environment, to maximize {\em rewards} over time.
The goal of the \gls{rl} agent is then to compute a policy, which is a mapping between the environment {\em states} and actions so as to maximize a long term reward. 
\gls{rl} problems are particularly of interest to \gls{ric}, because of their natural {\em closed-loop} form.

The \gls{rl} model of interest to this paper is \gls{dqn}, which is a model-free, off-policy, value-based \gls{rl}.
 %
 %
Our \gls{rl} algorithm uses a $Q$-value that measures the expected reward for taking a particular action at a given state. 
\glspl{dqn} can be trained offline with an online refinement of the learned policy, thus they can subsequently keep getting deployed in the inference host towards generating optimal actions, as the agent receives live data streams from the environment~\cite{levine2020offline}.

\subsection{Dual connectivity and traffic steering}

Dual connectivity is a mode of 5G \gls{ran} deployment, where the \gls{ue} is jointly connected to more than one base station. 
One of them is designated as the master node, which is responsible for control plane procedures of a \gls{ue}, and the other is considered as a secondary node and it is responsible for data transfer for the \gls{ue} along with the master node. 
A prevalent 5G deployment in North America and globally is \gls{endc} \gls{nsa} mode 3X, where the \gls{lte} \gls{enb} is the master node, and NR \gls{gnb} is the secondary node. 

Traffic Steering is a \gls{ran} functionality of the \gls{rrc} layer for managing connectivity and mobility decisions of \glspl{ue} in the \gls{ran}. More specifically, \gls{ts} handles (on a \gls{ue} basis): (i) Primary cell selection and handover, (ii) selection and change of master and secondary nodes for dual connectivity, (iii) selection and handover of the secondary cell.


As previously discussed, the handover problem has been widely studied and optimized in the literature. Without \gls{oran}, this mechanism can be implemented by using different approaches~\cite{8812724}.
Generally, it is common practice to perform handover based on channel quality hysteresis, and/or to advance handovers from overloaded to less loaded ones for load balancing.
More recent approaches exploit \gls{rl} to select the target node for the handover.

In the literature, there are several examples of \gls{ai}-based handover procedures.
One of the possible approaches is represented by the use of a centralized \gls{rl} agent with handover control using Q-learning~\cite{8406993} and subtractive clustering techniques~\cite{liu2021intelligent} to optimize the \glspl{ue} mobility. Other work considers distributed Q-Learning approaches or cooperative multi-agents~\cite{6692634,guo2020joint} to optimize the handover process on \glspl{son}.
Another area of interest is represented by the use of the \gls{dnn} in both the online training mode on the \glspl{ue}~\cite{wang2018handover} or via offline schemes~\cite{MOLLEL2020101133,chinchali2018cellular,wang2018deep}. \cite{wang2018handover} uses \gls{dnn} with supervised learning to transfer knowledge based on traditional handover mechanisms and avoid negative effects of random exploration for an untrained agent.  
Other examples of similar works are represented by~\cite{9052936,wang2018handover}.
~\cite{MUNOZ2015112} proposes a unified self-management mechanism based on fuzzy logic and \gls{rl} to tune handover parameters of the adjacent cells.
More examples can be found in~\cite{8466370}, which discusses the state of the art and the challenges of intelligent \gls{rrm}.

These works, however, generally do not optimize the performance of individual \glspl{ue} and do not fully satisfy the need for per-\gls{ue} control and optimization. Indeed, existing cellular networks implement procedures which are mostly cell-centric, even if there are usually high variations across the performance, requirements, and channel state of different \glspl{ue} in the same cell~\cite{oran-wg1-use-cases}. Improved performance thus can be achieved with the \gls{ue}-based approach we propose, enabled by the \gls{oran} architecture.
In~\cite{s21248173}, the authors present a general overview on \gls{oran} and its potentialities by exploring the \gls{ts} use case, showing the flexibility of this novel architecture and its orchestration capabilities.
However, their work is more tailored to the O-\gls{ran} capabilities rather than their xApp performances. In~\cite{yajnanarayana20205g}, the authors consider a contextual multi-armed bandit problem to model handover across 5G cells, and considers per-UE \gls{rsrp} metrics as input. In this paper, we consider a more complete set of \gls{ran} \glspl{kpm} as input, thanks to the support of the O-RAN E2 interface between the \gls{ran} and the near-RT \gls{ric}. This makes it possible to improve the overall \gls{ran} performance, as discussed in Section~\ref{sec:results}.

Finally, none of the works presented in this Section proposes approaches that can be practically applied in \gls{oran} and \gls{3gpp} networks.
An attempt to use \gls{oran} is~\cite{9815658}, where the authors study the closed-loop power adjustment of the transmitters in a 5G cellular orientation using two different xApps that act as a simulation of the network in the \gls{ric}.
This simulator, however, does not rely on \gls{3gpp} stochastic channels and thus the results may not be plausible once the model is deployed on a real network.
In this paper, we implement an \gls{oran} compliant near-RT \gls{ric} and use xApps with standard-compliant service models that can be deployed on a real network. 
In addition, we test the performance of the xApp combining the real-world \gls{ric} with a large scale \gls{ran} deployment based on end-to-end, full-stack, \gls{3gpp}-based simulations in ns-3. 

\begin{figure*}[!t]
\centering
    \begin{subfigure}[t]{0.4\textwidth}
    \centering
    \includegraphics[width=.87\textwidth]{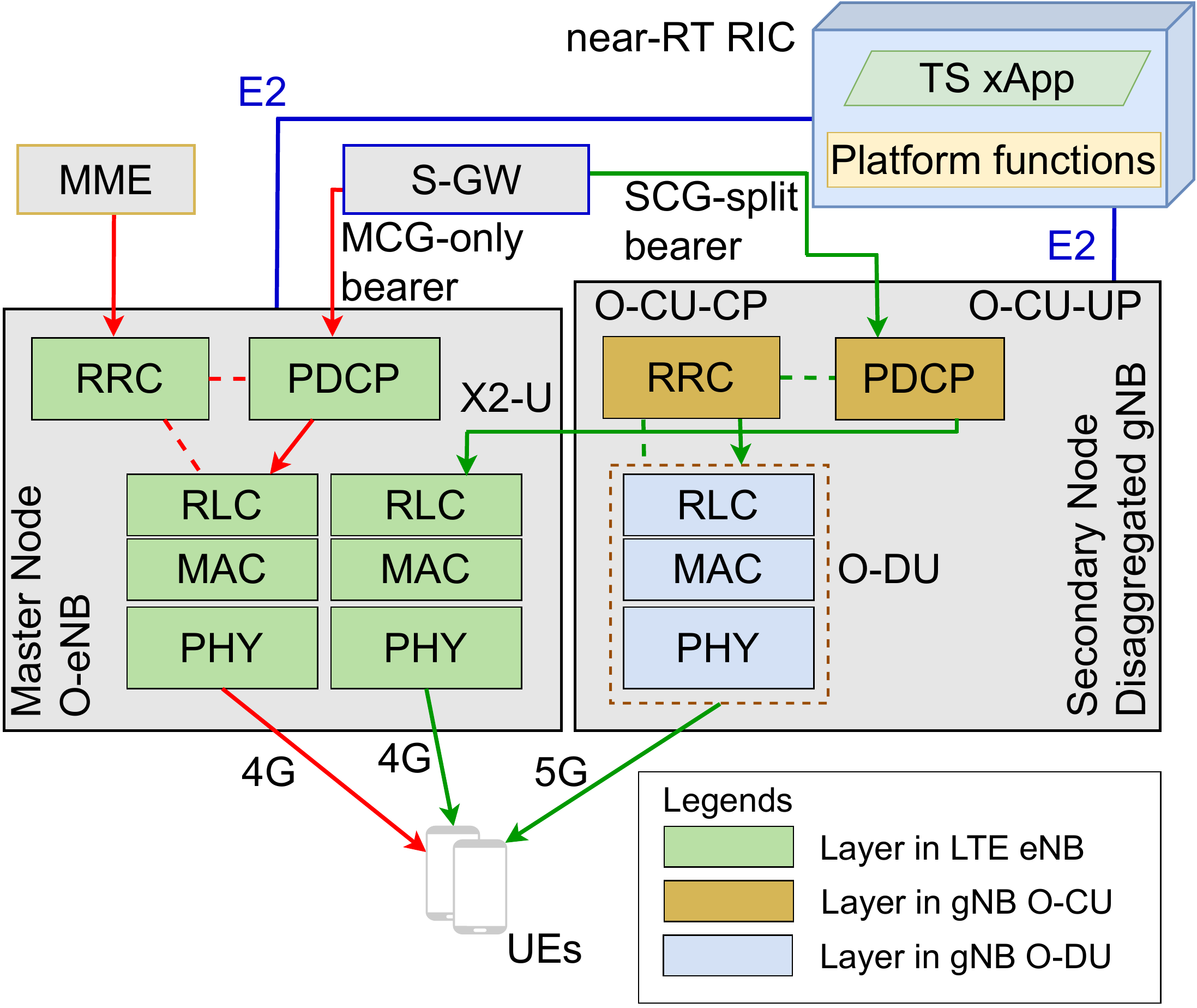}
    \caption{System model.}
    \label{fig:sys_model}
    \end{subfigure}%
    \hfill%
    \begin{subfigure}[t]{0.6\textwidth}
    \centering
    \includegraphics[width=.92\textwidth]{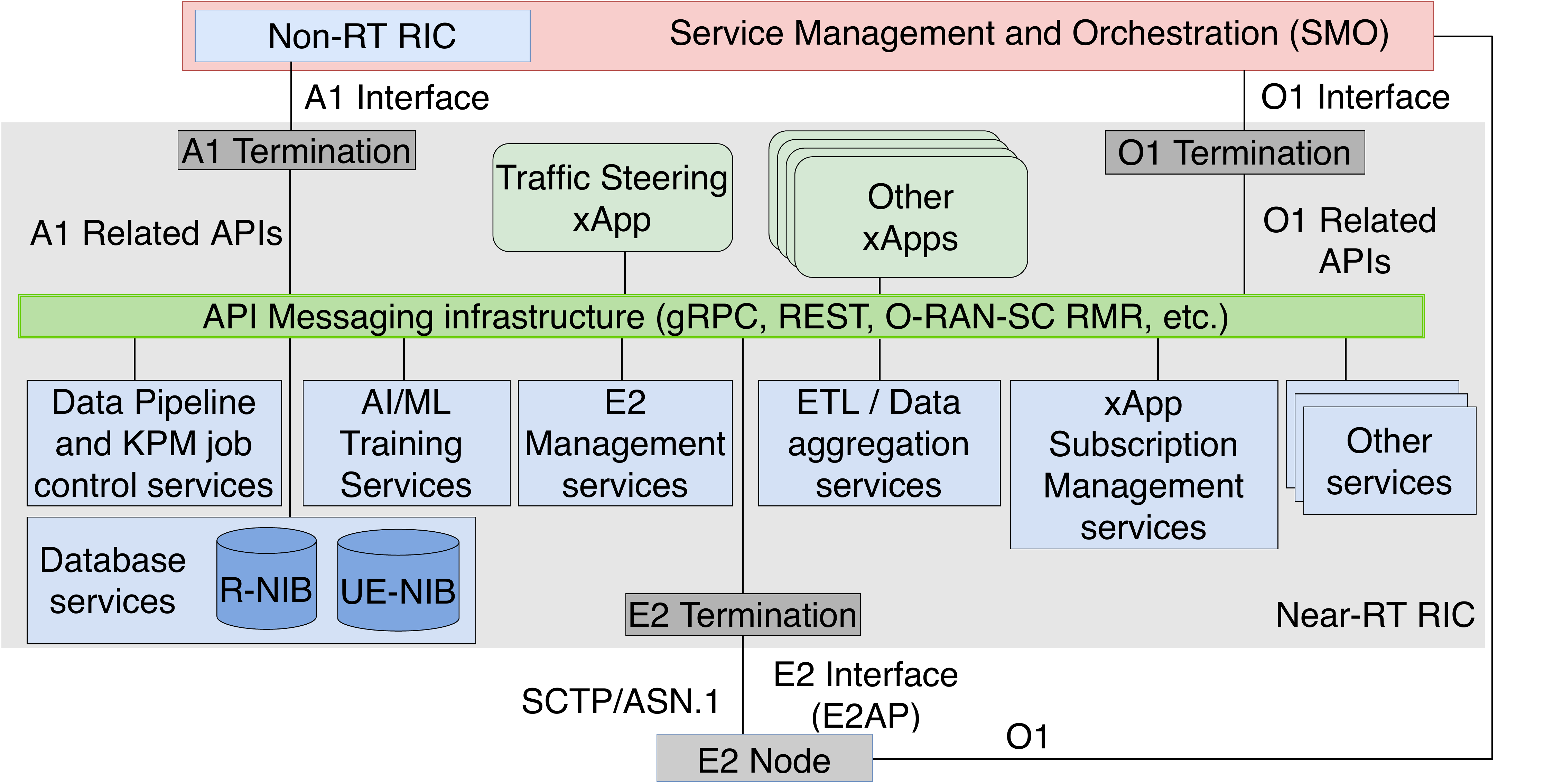}
    \caption{Near-RT \gls{ric} Software Architecture.}
    \label{fig:software_arch}
    \end{subfigure}
\caption{System Model and Software Architecture}
\label{fig:system-model}
\end{figure*}

\subsection{5G and AI in ns-3}
ns-3~\cite{riley2010ns} is a discrete-event time network simulator targeted for research and educational use.
It is considered the de facto standard for network simulators because of the variety of protocols used, its wide deployment and the widespread support of the scientific community.
The discrete event approach allows researchers to simulate the interactions in form of time-based events, allowing the modeling of each aspect of a communication, from the application layer to the the physical layer.
Such layers are the fundamental building blocks of a wireless network simulation and they can be combined in ns-3 to study particular case that otherwise would be very hard with a real deployment.
Indeed, one of the major reason of using ns-3 in this work is because of the very accurate \gls{3gpp} stochastic models~\cite{zugno2020implementation} available within the simulator combined with the possibility of simulate large scale deployments with no telecommunication hardware required.
These peculiarities are also enhanced by the possibility of integrate real world features, such as buildings and obstacles and mobility models for the wireless nodes, to create realistic scenarios.
In this work, we use the 5G and mmWave ns-3 module~\cite{mezzavilla2018end}, which extends the ns-3 \gls{lte} module with new detailed modeling of the mmWave channel that can capture spatial clusters, path dynamics, antenna patterns and beamforming algorithms.

The ns-3 simulator is also an optimal tool for implementing better \gls{ai} solutions for the networks.
In recent years, different works have extended the normal capabilities of ns-3 to combine its potentialities with some well-known \gls{ml} development software.
In~\cite{ns3gym}, the authors propose ns3-gym, a framework that integrates both OpenAI Gym and ns-3 in order to encourage usage of \gls{rl} in networking research.
Following the same principles of ns3-gym, ns3-ai~\cite{ns3ai} provides a high-efficiency solution to enable the data interaction between ns-3 and other python based \gls{ai} frameworks.
However both of these tools cannot be used as a framework for the development of \gls{oran} xApps that can be used directly in a production environment, unlike the \platform framework proposed in this paper.

\section{System Design and Architecture}
\label{sec:sys}
Here, we discuss the system model assumption, the near-RT \gls{ric} software architecture and the \platform design.


\subsection{System Model}
\noindent

The system architecture is shown in Figure~\ref{fig:sys_model}. 
We consider a network with $M$ \gls{lte} cells, and E2 nodes of $N$ NR cells, and a set $U$ of 5G \glspl{ue}.
The infrastructure is deployed as a 5G \gls{nsa} network with \gls{endc} \gls{ran} and option 3X for dual connectivity~\cite{3gpp.37.340}. 
With this, a 5G \gls{ue} is jointly connected to an \gls{lte} \gls{enb} (master node) and the E2 nodes of a 5G \gls{gnb} (secondary node).
Each \gls{ue} is jointly served by the primary cell of its master node and the secondary cell of its secondary node in \gls{endc}. 
The \glspl{ue} subscribe to heterogeneous types of data traffic (as detailed in Section \ref{sec:results}). 
In the \gls{ran}, each \gls{ue}-subscribed data traffic flow is split at the \gls{pdcpu} layer of the \gls{gnb} \gls{cuup}. 
Each packet is sent to the lower \gls{rlc} layer at either the \gls{gnb} \gls{du} (over the F1 interface) or the \gls{lte} \gls{enb} over the X2-U interface for subsequent transmission to the \gls{ue} via the NR or \gls{lte} radio, respectively. 
In addition, we consider a near-RT \gls{ric} connected to each \gls{lte} and NR cell through the E2 interface. 
The near-RT \gls{ric} is deployed at the edge of the \gls{ran} and features the \gls{ts} xApp to optimize \gls{ue} handover.
The delivery of the \gls{kpm} data between the E2 nodes and the \gls{ric} allows the exchange of network data and handover control actions at near-RT periodicity and is enabled through the use of the \gls{e2sm}-\gls{kpm} service model.
We use the \gls{e2sm}-\gls{rc} service model to generate control actions from the \gls{ric} to the E2 node for handover of specific \glspl{ue} from their current serving cells to the target cells identified by the \gls{ts} xApp.
Additionally, \gls{e2sm}-\gls{rc} is used to report \gls{ue}-specific L3 \gls{rrc} measurements (such as \gls{rsrp}, or \gls{sinr} with respect to its serving and neighbor cells) from the E2 node to the \gls{ric} periodically and during mobility events.

\subsection{Near-RT RIC Software Architecture}

We implement a near-RT \gls{ric} platform~\cite{9376232} with the components shown in Figure~\ref{fig:software_arch}.
In general, the near-RT \gls{ric} has two sets of applications, namely the xApps (for the control of the \gls{rrm} of dedicated \gls{ran} functionalities) and O-RAN-standardized platform services~\cite{polese2022understanding}. 
The latter manage integration of xApps, interfacing with E2 nodes, and the overall functioning of the \gls{ric}. 
In particular, they include the \textit{E2 Termination} service, which routes the \gls{e2ap} messages between the platform services and the E2 nodes over the E2 interface based on SCTP transport protocol. The service also performs ASN.1 encoding/decoding and manages data exposed by E2 nodes.  
The \textit{xApp Subscription Management} service maintains, manages, validates, and sends/receives xApp subscriptions toward E2 nodes.
The data collection and aggregation for the xApps is managed by two additional platform services.
The \textit{Data Pipeline and KPM job control} makes sure that xApps do not duplicate \gls{kpm} requests to the \gls{ran} by interacting with the Subscription Management service and filtering duplicated subscription requests on behalf of the xApps.
The \gls{kpm} data received by the \gls{ran} is aggregated, processed, and presented to the xApps by the \textit{\gls{etl}, data aggregation and ingestion} service.
In our implementation, the \gls{ts} xApp leverages the services of the \gls{ric} platform to (i) collect \glspl{kpm} on the status of the network; (ii) process them and perform online inference to decide if one or more \glspl{ue} should perform a handover to a different cell; and, eventually, (iii) send the handover control action to the \gls{ran}.
The \gls{ts} xApp triggers an E2 node \gls{kpm} subscription specifying the parameters for the data collection, i.e., the list of \glspl{kpm} and serving-cell and neighbor-cell L3 \gls{rrc} measurements, and the periodicity at which these values need to be reported by the E2 nodes.
The \gls{ts} xApp and the simulated \gls{ran} implemented with \platform (described in Section \ref{sec:oran-ns3}) collectively support streaming 40 \gls{ue}-level, cell-level, and node-level \glspl{kpm} from E2 nodes.

The E2 nodes accepts the subscription and starts streaming \glspl{kpm} and L3 \gls{rrc} measurements. 
The raw streamed KPM data is stored by Data Pipeline and KPM job control service.
The \gls{etl}, data aggregation and ingestion service retrieves relevant measurements stored in this data repository, and correlates and aggregates in time series the \gls{ue} level \gls{kpm} information and L3 \gls{rrc} measurements.
The \gls{ts} xApp can then fetch and process the data to perform inference with the algorithm described in Section \ref{sec:ts}. If a handover needs to be performed,  
the \gls{ts} xApp communicates with the \textit{E2 termination} service to send the control action to the \gls{ran}.




\subsection{Connecting O-RAN with ns-3: \platform}
\label{sec:oran-ns3}




One key contribution of this paper is represented by \platform, the first \gls{oran} integration for \gls{ns3}.
\platform is an ns-3 module that connects a real-world near-RT \gls{ric} with \gls{ns3}, enabling large scale (i) collection of \gls{ran} \glspl{kpm} and (ii) testing of closed-loop control of simulated cellular networks.
We use the term ``real-world'' to indicate that the \gls{ric} used in this framework is a standard compliant \gls{oran} near-RT \gls{ric} that is also capable of communicating with real hardware equipment.
This aspect allows \platform to be a powerful tool for the development of the xApp that can be then activated on real world \glspl{ran}.
Indeed, thanks to the flexibility of \gls{ns3}, such integration eases the design, development, and testing of xApps across different \gls{ran} setups with no infrastructure deployment cost. 
As already introduced in Section \ref{sec:soa}, ns-3 provides realistic modeling capabilities for large-scale wireless scenarios. 
It features a channel model with propagation and fading compliant with \gls{3gpp} specifications~\cite{3gpp.38.901}, and a full-stack 5G model for \gls{endc} \gls{ran}~\cite{mezzavilla2018end}, besides the TCP/IP stack, multiple applications, and mobility models.

\begin{figure}[t]
    \centering
    \includegraphics[width=.95\columnwidth]{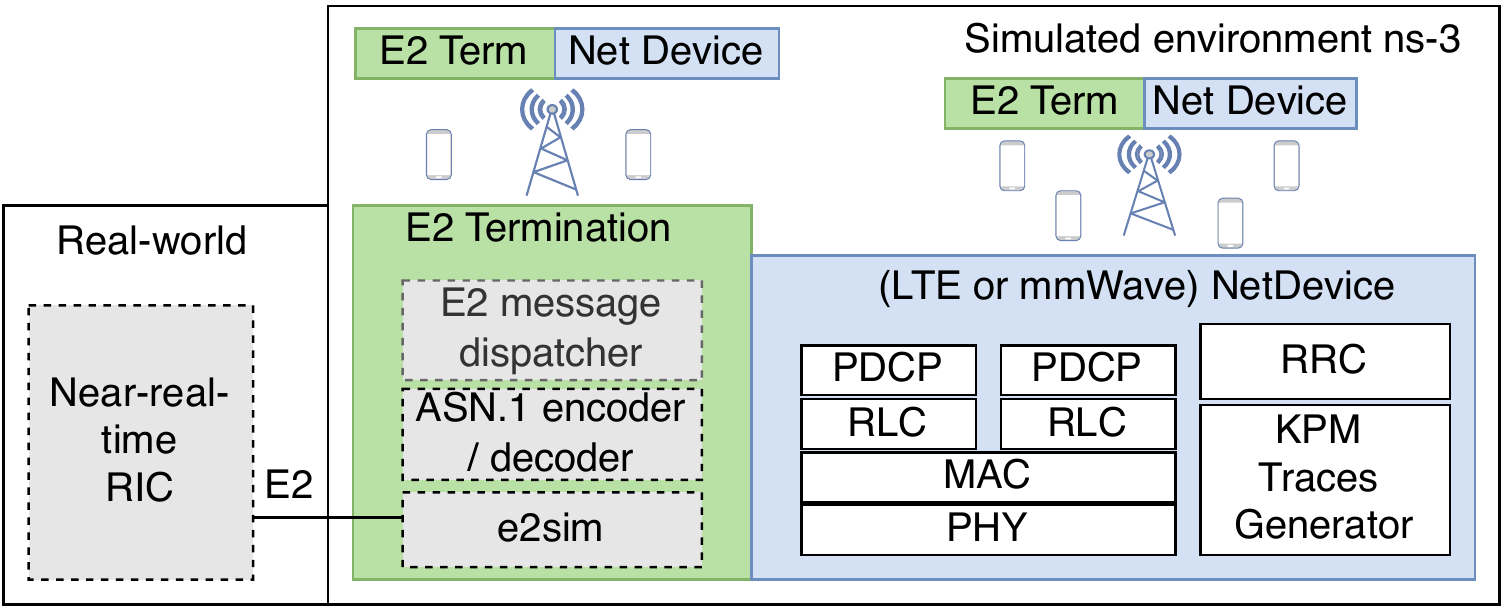}
    \caption{\platform Architecture.}
    \label{fig:architecture}
\end{figure}


\platform bridges \gls{ns3} to our real-world, O-RAN-compliant \gls{ric} to enable production code to be developed and tested against simulated \glspl{ran}.
%

\noindent $\bullet$ {\bf e2sim ---} We connect the E2 termination of the \gls{ric} to a set of E2 endpoints in ns-3, which are responsible for handling all the E2 messages from and to the simulated environment.
This connection was developed by extending the E2 simulator, namely \texttt{e2sim}~\cite{e2sim}, and wrapping it into an ad hoc module for \gls{ns3}. 
\texttt{e2sim} is a project in the O-RAN Software Community that provides basic E2 functionalities to perform integration testing of the near-RT \gls{ric}. \platform leverages and extends \texttt{e2sim} as the E2 termination on the ns-3 side. It can decode, digest, and provide feedback for all the messages coming from the \gls{ric}, and streams \gls{ran} telemetry based on simulation data to the \gls{ric}.

\noindent $\bullet$ {\bf Message dispatching ---} The design of \platform addresses several challenges that would otherwise prevent communications between the simulated and real-world environments.
Firstly, as discussed in Section \ref{sec:soa}, the near-RT \gls{ric} expects to interface with a number of disaggregated and distinct endpoints, i.e., multiple \glspl{du} and \glspl{cucp}/\glspl{cuup}, which are  usually identified by different IP addresses and/or ports. 
Instead, all the \gls{ns3} simulated \gls{ran} functions are handled by a single process.
\texttt{e2sim} itself was not designed to handle multiple hosts at once, while the E2 protocol specifications, which rely on the SCTP protocol for \gls{e2ap}, do not pose any limitation in this sense.
To address this, we extended the \texttt{e2sim} library to support multiple endpoints at the same time and created independent entities (i.e., C++ objects) in the simulated environment to represent different \gls{ran}-side E2 terminations.
Each \gls{ran} function is bound to just one E2 interface, as depicted in Figure~\ref{fig:architecture}, and has its own socket address.
\platform can successfully establish connectivity between multiple \gls{ran} nodes and the near-RT \gls{ric} even if a single IP is associated to the simulation process, as it can filter \gls{e2ap} messages through unique ports, ensuring the independence of data flow with the near-RT \gls{ric} from the others.
In this way, it is possible to assign the same IP to every base station created in the simulation, but each with a different and unique port and is possible to enable the data flow from multiple independent e2 terminations in the simulation and the \gls{ric}.
Moreover, we extended ns-3 to instantiate independent threads for each E2 termination and use callbacks that can be triggered when data is received or transmitted over E2. 

\noindent $\bullet$ {\bf Time Synchronization ---} Finally, there is also a gap in timing between the real-world \gls{ric} and the ns-3 simulator, which is a discrete-event framework that can execute faster or slower than the wall clock time. 
This may potentially lead to inconsistencies between the simulated environment and the near-RT \gls{ric} expecting the real-world timing.
To synchronize the two system, at the beginning of the simulation \gls{ns3} stores the current Unix time in milliseconds and uses it as baseline timestamp.
Whenever an E2 message is sent to the \gls{ric}, the simulator will sum the simulation time elapsed and the baseline timestamp, ensuring consistency on both sides of the happened-before relationship.

\section{Traffic Steering Optimization}
\label{sec:ts}

In this section, we formulate the optimization problem for the traffic steering xApp and discuss the algorithm design to determine the optimal target cells for handover of \glspl{ue}.

To the best of our knowledge, this is the first paper to develop a data-driven \gls{ue}-based traffic steering/handover optimization technique based on Conservative Q-learning.

Firstly, we formulate the general optimization problem in Eqs.~\ref{eqn:problem_formulation} and~\ref{eqn:problem_formulation_lagrangian} of Section~\ref{sec:problem_formulation}, showing that there is no closed forms for the optimization objective as function of the 5G \glspl{kpi} considered in production networks.
Secondly, in Section~\ref{sec:algo}, we embed the aforementioned equations in a data-driven RL framework to study the \gls{ue}-based optimization and apply the of Conservative Q-Learning algorithm which lead to a performance gain of up to 50\% in throughput and spectral efficiency that will be discussed in Section~\ref{sec:results}.

\subsection{Problem Formulation}
\label{sec:problem_formulation}

We consider as objective function the weighted cumulative sum of the logarithmic throughput of all the \glspl{ue} across time, as a function of their instantaneous target \gls{pscell}.
The optimization goal is to maximize the objective function by optimizing the choice of the target \glspl{pscell} for all \glspl{ue}.
At the same time, we want to avoid frequent handovers for individual \glspl{ue}, since they increase network overhead and deteriorate their performance.
Thus, we associate a cost function for every \gls{ue}-specific handover and model it as an exponential decay function of the linear difference in time since the previous handover for that \gls{ue}.
This means that smaller the difference in time, higher is the cost, and vice-versa.
We add this cost function as a constraint to make sure that the cost does not exceed a predefined cost threshold.

Let $\beta_u$ be a weight associated with any UE $u\in U$. $R_{u,t}$ is the throughput at any discrete window of time $t$, which depends on $c_{u,t}$, i.e., the \gls{pscell} assigned to $u$ during $t$, and on \gls{ran} performance parameters ${b_1, b_2,\dotsc b_B}$.

These metrics are available during the time window $t$ at the near-RT \gls{ric}, where the optimization is solved, thanks to the \gls{kpm} reports from the E2 nodes 
${C^{\text{NR}}}$ is the universe of all the $N$ NR cells.
The cost associated with handover for \gls{ue} $u$ at time $t$ is given by $K_{u,t}$, the initial cost is $K_0$ (where $K_0>0$), the decay constant is ${\delta}$ (where ${0<\delta< 1}$), $t_u^{\prime}$ is the time when the previous handover was executed for $u$, $X_{u,t}$ is a $0/1$ decision variable which yields a value 1, if $u$ was subject to handover at time $t$, and 0, otherwise. 
$W$ is a predefined cost threshold, which represents a maximum value that cannot be exceeded by the cost function. 
We consider any time window $t$ for an infinite time horizon ranging from $t_0$ to $\infty$. 
The constrained optimization problem is formulated as follows:  

\begin{equation}
\begin{split}
\begin{aligned}
& \mathop{Maximize}_{c_{u,t}\in C^{\text{NR}}}
& &\sum\limits_{t=t_0}^{\infty}\sum\limits_{u\in U}\beta_u\log R_{u,t}(c_{u,t},b_1,b_2,\dotsc b_B)\\
& \text{subject to}
& & K_{u,t} \cdot X_{u,t}\leq W,\\
& & & X_{u,t}\in[0,1],\\
\end{aligned}
\end{split}
\label{eqn:problem_formulation}
\end{equation}
where $K_{u,t}=K_0e^{-\delta \cdot (t-t_u^{\prime})}$, ${K_0>0}$ and ${0<\delta<1}$. Applying Lagrangian multiplier $\lambda$ to the constrained optimization problem in Eq.~\ref{eqn:problem_formulation}, the constrained optimization problem becomes as follows:
\begin{equation}
\begin{split}
\begin{aligned}
& \mathop{Maximize}_{c_{u,t}\in C^{\text{NR}}}
& &\sum\limits_{t=t_0}^{\infty}\sum\limits_{u\in U}\beta_u\log R_u(c_{u,t},b_1,b_2,\dotsc b_B)\\
&&&-K^{\prime} e^{-\delta \cdot (t-t_u^{\prime})} X_{u,t}+W^{\prime}\\
& \text{subject to}
& & X_{u,t}\in [0,1] \text{ and } \lambda\geq 0\\
\end{aligned}
\end{split}
\label{eqn:problem_formulation_lagrangian}
\end{equation}
where ${K^{\prime}=\lambda K_0}$ and ${W^{\prime}=\lambda W}$.

\subsection{Algorithm Design}
\label{sec:algo}

{\bf \gls{mdp} and \gls{rl} --- } We use a data-driven approach (specifically, \gls{rl}) to model and learn ${R_{u,t}}$ as a function of $\{{c_{u,t},b_1,b_2,\dotsc b_B}\}$, due to the lack of a deterministic closed-form equation for ${R_{u,t}}$ as a function of the parameters, and its relationship with cost ${K_{u,t}}$ and the handover decision variable ${X_{u,t}}$.
We consider the infinite time horizon \gls{mdp} to model the system, where the environment is represented by \platform, and a single \gls{rl} agent is deployed in the near-RT \gls{ric} containing the \gls{ts} xApp. 
The system is modeled as an \gls{mdp} because the \gls{ts} xApp in the \gls{ric} controls the target \gls{pscell} for the \glspl{ue} handover, while the resulting state (including the \gls{ran} performance parameters and the user throughput) is stochastic.
The \gls{mdp} is defined by the tuple ${\langle \mathcal{S},\mathcal{A},\mathcal{P},\mathcal{R},\mathcal{\gamma},\mathcal{I} \rangle}$, where:

\noindent $\bullet$ ${\mathcal{S}}$ is the state space, comprising of per-\gls{ue} \gls{e2sm}-\gls{kpm} periodic data and per-\gls{ue} \gls{e2sm}-\gls{rc} periodic/event-driven data. Let ${C^{\prime}_{u,t}\subseteq C^{\text{NR}}}$ be the set of serving \gls{pscell} and neighboring cells for any UE $u$ at time $t$.
The state vector for $u$ at time $t$ from the environment (${\vec{s}_{u,t}}$) includes the \gls{ue} identifier for $u$ and the set of parameters ${b_1,b_2,\dotsc b_B}$. 
The latter includes 
(i) the \gls{ue}-specific L3 \gls{rrc} measurements (obtained from the E2 node \gls{cucp}) such as ${\text{sinr}_{u,c,t}}$ for any cell $c\in C^{\prime}_{u,t}$ for the UE $u$; 
(ii) ${\text{PRB}_{c,t}}$, the cell-specific \gls{prb} utilization for $c$ at time $t$ obtained from the E2 node \gls{du}; 
(iii) ${Z_{c,t}}$, the cell-specific number of active \gls{ue}s in the cell $c$ with active \gls{tti} transmission at $t$ obtained from \gls{du}; 
(iv) $P_{c,t}$, the total number of \gls{mac}-layer transport blocks transmitted by cell $c$ across all UEs served by $c$ at time $t$ (obtained from the E2 node \gls{du}); 
(v) ${p^{\text{QPSK}}_{c,t}}$, ${p^{\text{16QAM}}_{c,t}}$, ${p^{\text{64QAM}}_{c,t}}$, the cell-specific number of successfully-transmitted transport blocks with QPSK, 16QAM and 64QAM modulation rates from the cell $c$ to all \glspl{ue} served by the $c$ at time $t$ normalized by $P_{c,t}$.
(vi) Finally, the set of parameters includes 
the cost the \gls{ue} $u$ would incur, if handed over to $c_{u,t}$ at $t$ (i.e., where $c_{u,t}\neq c_{u,t-1}$), given by:

\begin{equation*}
 k(c_{u,t}) = K_0 e^{-\delta \cdot (t-t_u^{\prime})} x(c_{u,t});
\end{equation*}
\begin{equation*}
 \text{where }x(c_{u,t})= 
  \begin{cases} 
   1 & \text{if } c_{u,t} \neq c_{u,t-1} \\
   0       & \text{otherwise}
  \end{cases}
\end{equation*}
Note that the cost ${k(c_{u,t})}$ is zero if there is no handover, i.e., ${c_{u,t} = c_{u,t-1}}$. 
The above state information are aggregated across all the serving and neighbor cells of $u$, i.e., ${\forall c\in C^{\prime}_{u,t}\subseteq C^{\text{NR}}}$, along with the cell identifier for $c$, during the reporting window $t$ to generate a consolidated record for $u$ for $t$. 
This aggregated state information for $u$ is fed as input feature to the \gls{rl} agent on the TS xApp. 
This is done for all \glspl{ue} in $U$, whose aggregated state information is fed to the same \gls{rl} agent. 
If any of the parameters in the state information from the environment for any \gls{ue} $u$ is missing, the \gls{ric} \gls{etl} service uses a configurable small window ${\epsilon}$ to look back into recent history (tens to few hundred of ms) and fetch those historical parameters for the missing ones.

\noindent $\bullet$ ${\mathcal{A}}$ is the action space, given by: 

\[\mathcal{A}=\{\text{HO}(c_1),\text{HO}(c_2),\dotsc\text{HO}(c_{N}),\overline{\text{HO}}\}\]
where, ${c_1,c_2,\dotsc c_{N}\in C^{\text{NR}}}$. 
Here, ${a_{u,t}=\text{HO}(c)}$, where ${a_{u,t}\in\mathcal{A}}$, indicates that the \gls{rl} agent is recommending a handover action for $u$ to any cell $c$ at $t$, and ${a_{u,t}=\overline{\text{HO}}}$ indicates no handover action for $u$ at $t$, meaning that the \gls{ue} shall continue being served by its current primary serving cell.

\noindent $\bullet$ ${\mathcal{P}(\vec{s}_{u,t+1}|\vec{s}_{u,t},a_{u,t})}$ is the state transition probability of \gls{ue} $u$ from state $\vec{s}_{u,t}$ at $t$ to $\vec{s}_{u,t+1}$ at $t+1$ caused by action ${a_{u,t}\in\mathcal{A}}$.

\noindent $\bullet$ $\mathcal{R}:\mathcal{S}\times\mathcal{A}\rightarrow \mathbb{R}$ is the reward function for \gls{ue} $u$ at $t+1$, as a result of action $a_{u,t}$, given by the following:

\begin{equation*}
    \mathcal{R}_{u,t+1}=\beta_u \cdot (\log R_{u,t+1}(c_{u,t+1}) -\log R_{u,t}(c_{u,t}))
\end{equation*}
\begin{equation}   
    -k(c_{u,t+1})
\label{eqn:reward}
\end{equation}
The reward for \gls{ue} $u$ is the improvement in the logarithmic throughput $R_{u,t}$ due to the transition from ${\vec{s}_{u,t}}$ to ${\vec{s}_{u,t+1}}$ caused by action ${a_{u,t}}$ taken at $t$, minus the cost factor.
The reward is positive, if the improvement in ${\log}$ throughput is higher than the cost, and negative, otherwise. $R_{u,t}$ is obtained from \gls{cuup} using \gls{e2sm}-\gls{kpm}.  

\noindent $\bullet$ $\gamma\in[0,1]$ is the discount factor for future rewards. The value function ${V^{\pi}(s)}$ is the net return given by the expected cumulative discounted sum reward from step $t$ onwards due to policy ${\pi}$, provided as follows:

\begin{equation}
  V^{\pi}(s)=\mathbb{E}\displaystyle\left[\sum\limits_{u\in U}\sum\limits_{i=0}^{\infty}\gamma^i \mathcal{R}_{u,t+i}|\vec{s}_{u,t}=s,\pi(a|s)\right]
\label{RL_return}
\end{equation}

\noindent $\bullet$ $\mathcal{I}$ is the initial distribution of the \gls{ue} states.

\noindent $\bullet$ We consider two policies: (i) a target policy ${\pi(a|s)}$, to learn the optimal handover action $a$ for any state $s=\vec{s}_{u,t}$; and (ii) a behavior policy ${\mu(a|s)}$, to generate the handover actions which result in state transition and a new state data from the environment.

\textbf{Q-function and Deep-Q Network --- } We use {\em Q-learning}, a model-free, off-policy, value-based \gls{rl} approach. 
We compute the $Q$ function, an action-value function which measures the expected discounted reward upon taking any action $a$ on any given state $s$ based on any policy $\pi$. 
The value returned by the $Q$-function is referred to as the $Q$-value, i.e.,

\begin{equation}
\begin{split}
    Q^{\pi}(s,a)&=\mathbb{E}\displaystyle\left[\sum\limits_{u\in U}\sum\limits_{i=0}^{\infty}\gamma^i \mathcal{R}_{u,t+i}|\vec{s}_{u,t}=s,a_{u,t}=a,\pi(a|s)\right]\\
    &=r(s,a)+\gamma\mathbb{E}_{\mathcal{P}(s'|s,a)}\displaystyle\left[Q^{\pi}(s',a')|s,a,\pi\right]
\end{split}
\label{Q_estimate}    
\end{equation}
Here, ${r(s,a)=\mathbb{E}\displaystyle\left[\sum\limits_{u\in U}\mathcal{R}_u|\vec{s}_{u,t}=s,a_{u,t}=a,\pi(a|s)\right]}$.
From ~(\ref{Q_estimate}) and~(\ref{RL_return}), we have
\begin{equation}
V^{\pi}(s)=\displaystyle\sum\limits_a\pi(a|s) Q^{\pi}(s,a).
\label{return_qvalue_relation}
\end{equation}
The optimal policy ${\pi^{\star}}$ is the one that maximizes the expected discounted return, and the optimal $Q$ function ${Q^{\star}(s,a)}$ is the action-value function for ${\pi^{\star}}$ given by the Bellman equation as follows:
\begin{equation}
\begin{split}
    \pi^{\star}(a|s)&=\arg\max\limits_{\pi}Q^{\pi}(s,a)\\
    Q^{\star}(s,a)&=r(s,a)+\\
    &+\gamma\mathbb{E}_{\mathcal{P}(s'|s,a)}\displaystyle\left[\max\limits_{a^{\prime}}Q^{\star}(s',a')|\vec{s}_{u,t}=s,a_{u,t}=a,\pi^{\star}\right]
\end{split}
\end{equation}
We use the $Q$-learning algorithm to iteratively update the $Q$-values for each state-action pair using the Bellman equation, as seen in~\ref{eqn:value_iteration}, until the $Q$ function converges to ${Q^{\star}}$. 
The value iteration by the \gls{rl} agent leverages the {\em exploration-exploitation} trade-off to update the target policy ${\pi}$. 
It {\em explores} the state space of the environment by taking random handover control actions and learning the $Q$-function for the resulting state-action pair, and {\em exploits} its learning to choose the optimal control action maximizing the $Q$-value, i.e.,
\begin{equation}
    Q_{i+1}^{\pi}(s,a)=r(s,a)+\gamma\mathbb{E}\displaystyle\left[\max_{a^{\prime}}Q_i^{\pi}(s',a'|s,a,\pi)\right]
    \label{eqn:value_iteration}.
\end{equation}
Such value iteration algorithms converge to the optimal action-value function, i.e., ${Q^{\star}:=\lim\limits_{i\rightarrow\infty}Q_i^{\pi}}$. 
The Bellman error $\Delta$, as in ~(\ref{eqn:bellman_error}), is the update to the expected return of state $s$, when we observe the next state $s'$.
$Q$-learning repeatedly adjusts the $Q$-function to minimize the Bellman error, as shown in ~(\ref{eqn:bellman_error}) as in 
\begin{equation}
\begin{split}
    \Delta_{i+1}&=\displaystyle\left[r(s,a)+\gamma\max\limits_{a'} Q^{\pi}_i(s',a')\right]-Q^{\pi}_{i+1}(s,a)\\
    Q^{\pi}_{i+1}(s,a)&\leftarrow(1-\omega)Q^{\pi}_{i+1}(s,a)+\\
    &+\omega\displaystyle\left[r(s,a)+\gamma\max\limits_{a'}Q^{\pi}_i(s',a')\right].
\end{split}
\label{eqn:bellman_error}
\end{equation}
\begin{figure}[b]
    \centering
     \includegraphics[width=0.5\textwidth]{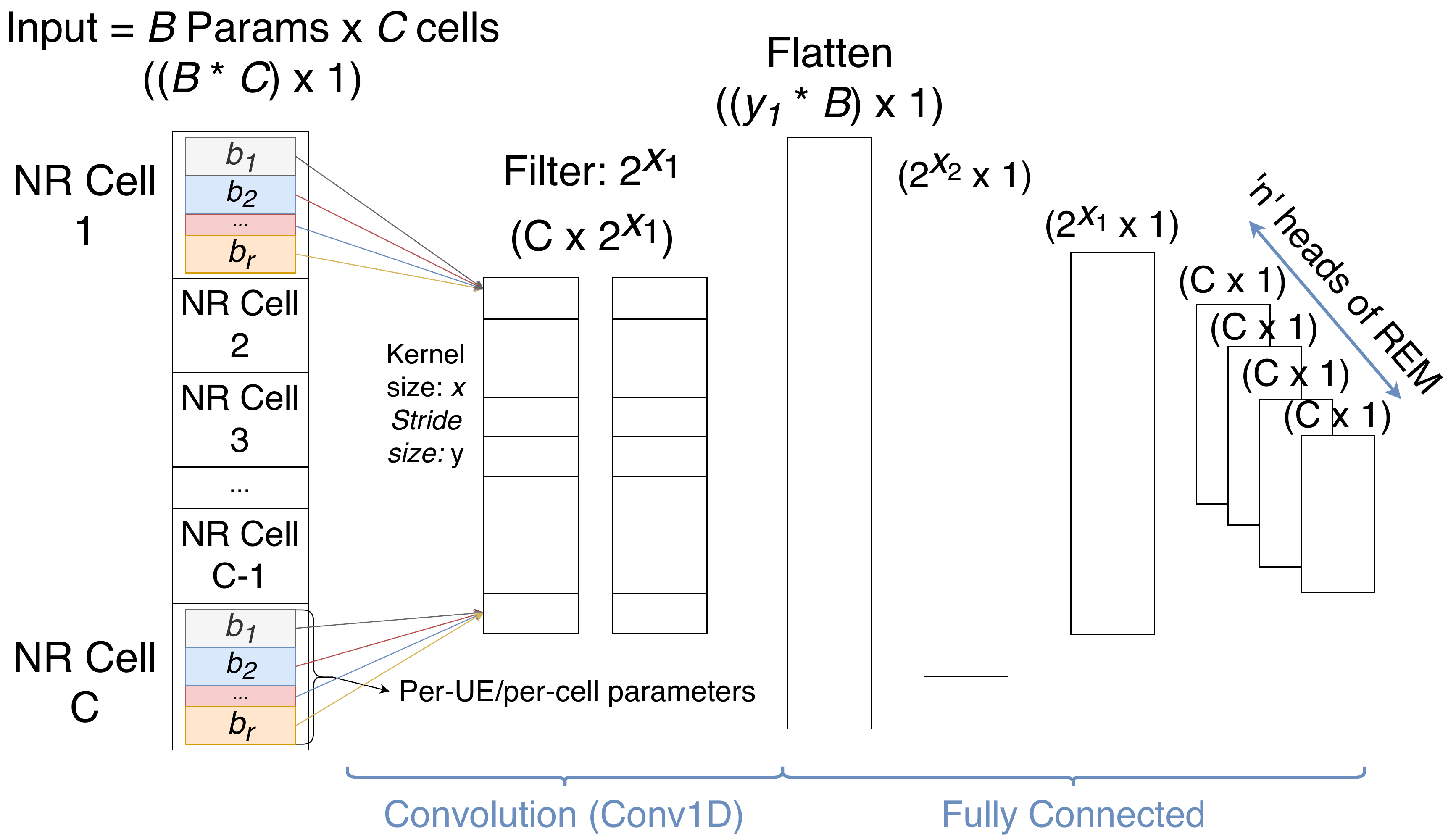}
    \caption{Our CNN architecture design}
    \label{fig:cnn_arch_design}
\end{figure}
This approach of ${\lim\limits_{i\rightarrow\infty}Q_i^{\pi}\rightarrow Q^{\star}}$ has practical constraints, as discussed in~\cite{dqn}. 
To address this, we use a \gls{cnn} approximator with weights ${\theta}$ to estimate the $Q$ function ${Q(s,a;\theta)}$, and refer to it as the $Q$-network. 
Our \gls{cnn} architecture design is shown in Fig.~\ref{fig:cnn_arch_design}. 
Deep $Q$-learning comes from parameterizing $Q$-values using \glspl{cnn}.
Therefore, instead of learning a table of $Q$-values, we learn the weights of the \gls{cnn} ${\theta}$ that outputs the $Q$-value for every given state-action pair. 
The $Q$-network is trained by minimising a sequence of loss functions $L_i(\theta_i,\pi)$ for each iteration $i$. 
The optimal $Q$-value, as a result of \gls{cnn} approximator, is given by ${\overline{Q}^{\star}}$ as follows:

\begin{equation}
\begin{split}
    L_i(\theta_i,\pi)&=\\
    =&\mathbb{E}\displaystyle\left[\displaystyle\left(r(s,a)+\gamma\max\limits_{a'}Q^{\pi}(s',a';\theta_{i-1})-Q^{\pi}(s,a;\theta_i)\right)^2\right]\\
    \overline{Q}^{\pi}_i&=\arg\min\limits_{Q^{\pi}}\displaystyle\left\{E\displaystyle\left[Q^{\pi}(s,a,\theta_i)|s,a,\pi(a|s)\right]+\omega L_i(\theta_i,\pi)\right\}\\
    \overline{Q}^{\star}&:=\lim\limits_{i\rightarrow\infty}\overline{Q}^{\pi}_i
\end{split}
\label{eqn:loss_function}
\end{equation}
Here, ${\mathbb{E}_{\mathcal{P}(s'|s,a)}\displaystyle\left[r(s,a)+\gamma\max\limits_{a'}Q^{\pi}(s',a';\theta_{i-1})|s,a,\pi\right]}$ is the target for iteration $i$. 
The parameters from the previous iteration ${\theta_{i-1}}$ are fixed for optimizing the loss function ${L_i(\theta_i)}$. The gradient of the loss function is obtained by differentiating the loss function in Eq.~\ref{eqn:loss_function} with respect to ${\theta}$ and the loss can be minimized by computing its stochastic gradient descent. 

Thanks to the \gls{cnn}, we are able to train and use the same weights for different cells, i.e., one single model for all, and we can approximate a non-linear dependency between input values and the reward function with a reduced dimensionality.

We use an {\em off-policy} $Q$-learning algorithm, called \gls{dqn}~\cite{dqn} for this purpose. The \gls{dqn} algorithm leverages an experience replay buffer, where the \gls{rl} agent's experiences at each step $e_t=(s_t,a_t,r_{t},$ $s_{t+1})$ are collected using the behavior policy ${\mu}$ and stored in a replay buffer ${\mathcal{D}=\{e_1,e_2,\dotsc e_{t-1}\}}$ for the policy iterate $\pi_i$. ${\mathcal{D}}$ is pooled over many episodes, composed of samples from policy iterates ${\pi_0, \pi_1, \dotsc \pi_{i}}$, so as to train the new policy iterate ${\pi_{i+1}}$ (as ${Q^{\star}=\lim\limits_{i\rightarrow\infty}Q_i^{\pi}}$). At each time step of data collection, the transitions are added to a circular replay buffer. To compute the loss ${L_i(\theta_i)}$ and the gradient, we use a mini-batch of transitions sampled from the replay buffer, instead of using the latest transition to compute the loss and its gradient. Using an experience replay has advantages in terms of an off-policy approach, better data efficiency from re-using transitions and better stability from uncorrelated transitions~\cite{dqn}.

{\bf \gls{rem} and \gls{cql} --- } To leverage the full potential of the integrated \gls{ns3} simulation environment in \platform and harness large datasets generated from the simulator via offline data collection for data-driven \gls{rl}, we use offline $Q$-learning. This enables us to learn the \gls{cnn} weights by training the $Q$-network using the \gls{dqn} model from dataset ${\mathcal{D}}$ collected offline based on any behavior policy  (potentially unknown, using any handover algorithm) ${\pi}$ without online interactions with the environment and hence, no additional exploration by the agent beyond the experiences $e_t$ available in ${\mathcal{D}}$ via ${\mu}$. The trained model is then deployed online to interact with the environment and the $Q$-function is iteratively updated online. We use a robust offline $Q$-learning variant of the \gls{dqn} algorithm, called {\em \gls{rem}}, which enforces optimal Bellman consistency on $J$ random convex combinations of multiple $Q$-value estimates to approximate the optimal $Q$-function~\cite{rem}. This approximator is defined by mixing probabilities on a ${(J-1)}$ simplex and is trained against its corresponding target to minimize the Bellman error~\cite{rem}.    
\begin{equation}\label{eqn:rem_loss}
\begin{split}
    \hat{L}_i(\theta_i,\pi)&=\\
    =&\mathbb{E}\displaystyle\left[\displaystyle\left(r(s,a)+\gamma\max\limits_{a'}\hat{Q}^{\pi}(s',a';\theta_{i-1})-\hat{Q}^{\pi}(s,a;\theta_i)\right)^2\right]\\
    =&\mathbb{E}[(r(s,a)+\gamma\max\limits_{a'}\displaystyle\sum\limits_{j}\alpha_{j}Q^{\pi}_{j}(s',a';\theta_{i-1})-\\
    &-\displaystyle\sum\limits_{j}\alpha_{j}Q^{\pi}_{j}(s,a;\theta_i))^2]\\
    \tilde{Q}^{\pi}_{i}&=\arg\min\limits_{Q^{\pi}} \hat{L}_i(\theta_i,\pi)\\
\end{split}
\end{equation}
Here, ${\alpha_j\in\mathbb{R}^J}$, such that ${\displaystyle\sum\limits_{j=1}^{J}\alpha_j=1}$ and ${\alpha_j\geq 0, \forall j\in [1,J]}$. ${\alpha_j}$ represents the probability distribution  over the standard ${(J-1)}$-simplex. While \gls{rem} prevents effect of outliers and can effectively address imbalances in the offline dataset ${\mathcal{D}}$, offline-$Q$ learning algorithms suffer from action distribution shift caused by bias towards out-of-distribution actions with over-estimated $Q$ values~\cite{cql}. This is because the $Q$-value iteration in Bellman equation uses actions from target policy ${\pi}$ being learned, while the $Q$-function is trained on action-value pair generated from ${\mathcal{D}}$ generated using behavior policy ${\mu}$. To avoid this problem of over-estimation of $Q$-values for out-of-distribution actions, we use a conservative variant of offline \gls{dqn}, called \gls{cql} that learns a conservative, lower-bound $Q$-function by (i) minimizing $Q$-values computed using \gls{rem} under the target policy distribution ${\pi}$ and (ii) introducing a $Q$-value maximization term under the behavior policy distribution ${\mu}$~\cite{cql}. From Eq.~\ref{eqn:loss_function}, 
the iterative update for training the $Q$-function using CQL and REM is given by:
\begin{equation}
\begin{split}
\Breve{Q}^{\pi}_{i}&\leftarrow\arg\min\limits_{\hat{Q}^{\pi}}\{\underbrace{\mathbb{E}\displaystyle\left[\hat{Q}^{\pi}(s,a_{\pi};\theta_i)|s,a_{\pi},\pi(a_{\pi}|s)\right]}_{\text{minimize REM Q-value under }\pi}\\
&-\underbrace{\mathbb{E}\displaystyle\left[\hat{Q}(s,a_{\mu};\theta_i)|s,a_{\mu},\mu(a_{\mu}|s)\right]}_{\text{maximize REM Q-value under }\mu}\\
&+\omega \hat{L}_i(\theta_i,\pi)\}\\
\Breve{Q}^{\star}&:=\lim\limits_{i\rightarrow\infty}\Breve{Q}^{\pi}_i
\end{split}
\label{cql_rem}
\end{equation}
Here, ${\hat{L}_i(\theta_i,\pi)}$ and ${\hat{Q}^{\pi}(s,a;\theta_i)}$ are as defined in Eq.~\ref{eqn:rem_loss}.

To summarize, the sequence of steps is outlined below in Algorithms~\ref{alg_qlearning_offline} and~\ref{alg_qlearning}. 
The Q-learning algorithm is trained offline with Algorithm~\ref{alg_qlearning_offline} and deployed in the \gls{ts} xApp for online inference and control following Algorithm~\ref{alg_qlearning}. 

\begin{algorithm}[t]
  \footnotesize
  \caption{Offline $Q$-learning training}\label{alg_qlearning_offline}
  \begin{algorithmic}[1]
    \STATE Store offline data (generated from \gls{ns3}) using any handover algorithm and behavior policy $\mu$ into replay buffer ${\mathcal{D}}$ consisting of \gls{ue}-specific records (${\forall u\in U}$)
    \WHILE {${\mathcal{D}}$ not empty and value iteration $i$}
      \STATE Begin training step:
      \begin{ALC@g}
	      \STATE Select a batch of ${2^{x_1}}$ samples for input to the \gls{cnn}
	      \STATE Use the $Q$-function and loss function $\hat{L}$ from Eq.~\ref{cql_rem} to train the \gls{cnn} weights ${\theta_i}$ based on \gls{cql} and \gls{rem} for value iteration $i$ of target policy ${\pi}$ for ${\Breve{Q}^{\pi}_i}$
      \end{ALC@g}
      \STATE Set $i\leftarrow i+1$
    \ENDWHILE
  \end{algorithmic}
\end{algorithm}

\begin{algorithm}[t]
  \footnotesize
  \caption{Online value iteration and inference}\label{alg_qlearning}
  \begin{algorithmic}[1]
    \WHILE {Incoming experience data $e_t$ for any \gls{ue} ${u}$ from \gls{ran} environment to near-RT \gls{ric} for $t\in[t_0,\infty]$}
      \STATE Append $e_t$ to replay buffer ${\mathcal{D}^{\prime}\subseteq\mathcal{D}}$ in \gls{ai}/\gls{ml} training services with length ${D^{\prime}\leq D}$
      \STATE Begin inference step:
      \begin{ALC@g}
	      \STATE Repeat steps 4 and 5 from Algorithm~\ref{alg_qlearning_offline}
  	      \STATE Generate HO control action for $u$ from the \gls{ts} xApp over E2 to \gls{ran} environment based on ${\Breve{Q}^{\pi}_i}$
      \end{ALC@g}
      \STATE Set $i\leftarrow i+1$
    \ENDWHILE
  \end{algorithmic}
\end{algorithm}

\section{Performance Evaluation}
\label{sec:results}

In this section, we first describe the simulation scenario, the baseline handover modes considered for the comparison, and the metrics of interest. 
We then discuss the results based on a large scale evaluation in different deployment scenarios.

\subsection{Simulation and Scenario Design}

\begin{figure}[t]
    \centering
    \includegraphics[width=\columnwidth]{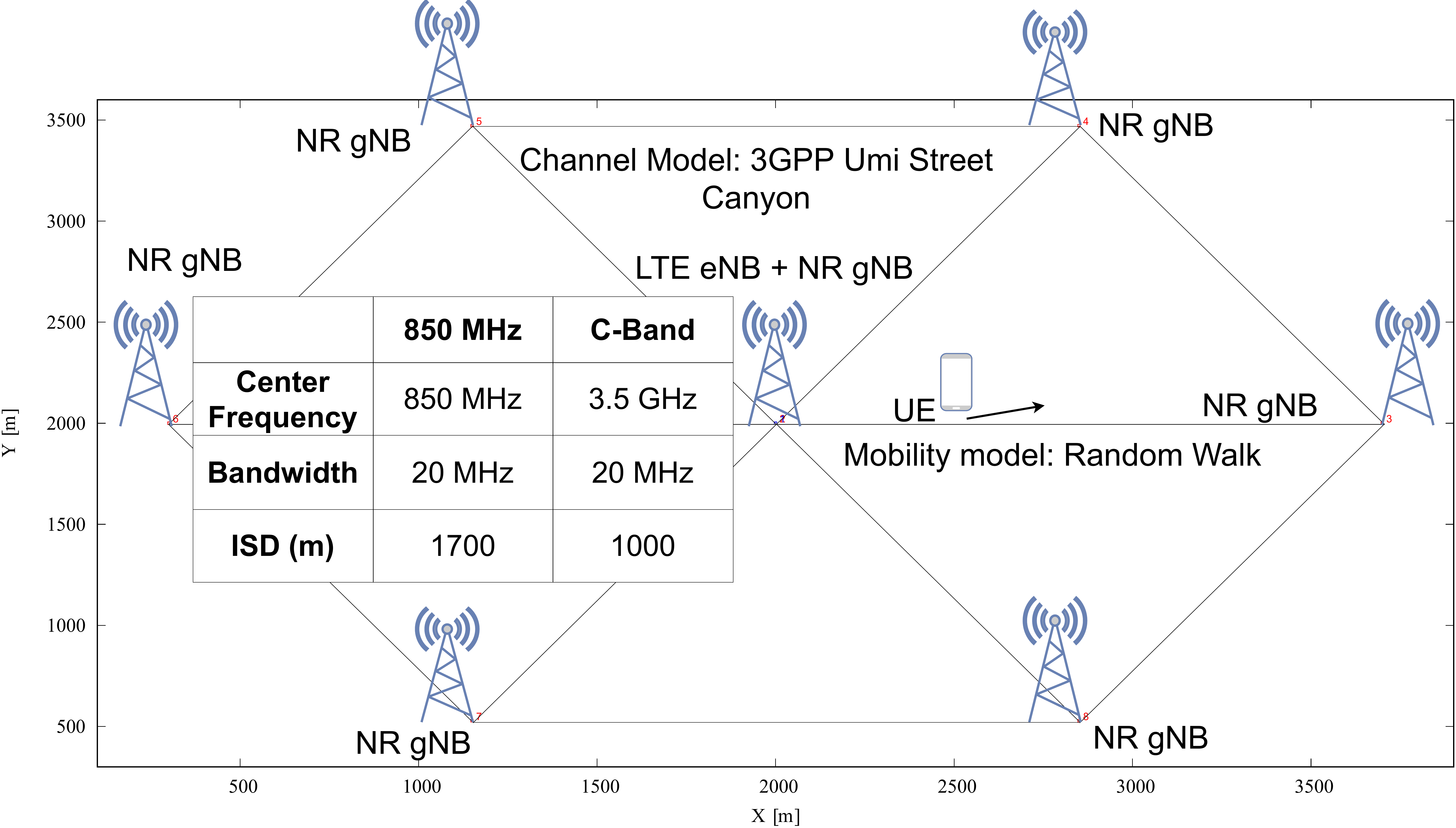}
    \caption{Simulation scenario.}
    \label{fig:simulation_scenario}
\end{figure}

{\bf Dense urban scenario --- } We model a dense urban deployment, with $M = 1$ \gls{enb} and $N = 7$ \glspl{gnb}, as shown in Figure~\ref{fig:simulation_scenario}. 
One of the \glspl{gnb} is co-located with the \gls{enb} at the center of the scenario, the others provide coverage in an hexagonal grid. 
Each node has an independent E2 termination, with reporting periodicity set to 100 ms.
We study two different configurations: (i) low band with center frequency 850 MHz and inter-site distance between the \glspl{gnb} of 1700 m; and (ii) C-band, with center frequency 3.5 GHz and inter-site distance of 1000 m.
In each configuration, the bandwidth is 10 MHz for the \gls{enb} and 20 MHz for the \glspl{gnb}.
The channel is modeled as a \gls{3gpp} Urban Macro (UMa) channel~\cite{3gpp.38.901}. 
The \gls{3gpp} NR \glspl{gnb} use numerology 2. 
$N_{\rm UE} = |U|$ dual-connected \glspl{ue} are randomly dropped in each simulation run with an uniform distribution, and move according to a random walk process with minimum speed $S_{\rm min} = 2.0$ m/s and maximum speed $S_{\rm max} = 4.0$ m/s.
This setup represents the average condition for typical 3GPP scenarios (pedestrian to slow vehicle mobility). 
We focus on the subset of UEs that are more interested by handovers, rather than, for example, static users, with a random walk model to generalize the mobility through the simulations.

{\bf Traffic model --- } The users request downlink traffic from a remote server with a mixture of four traffic models, each assigned to 25\% of the \glspl{ue}. 
The traffic models include (i) full buffer \gls{mbr} traffic, which saturates at $R_{\rm fb, max} = 20$ Mbit/s, to simulate file transfer or synchronization with cloud services; (ii) bursty traffic with an average data rate of $R_{\rm b, max} = 3$ Mbit/s, to model video streaming applications; and (iii) two bursty traffic models with an average data rate of 750 Kbit/s and 150 Kbit/s, for web browsing, instant messaging applications, and \gls{gbr} traffic (e.g., phone calls). 
The bursty traffic models feature on and off phases with a random exponential duration.


{\bf Baseline Handover Strategies --- } We consider three baseline handover models~\cite{7959177} for training the \gls{ai} agent from in Section \ref{sec:ts} and to evaluate its effectiveness.
They represent different strategies generally used for handovers in cellular networks~\cite{8812724}. 
We consider a \gls{ran} \gls{rrm} heuristic, which decides to perform a handover if a target cell has a channel quality metric (in this case, the \gls{sinr}) above a threshold (specifically, 3 dB) with respect to the current cell. 
The other algorithms use more advanced heuristics, based on a combination of a threshold and a \gls{ttt}. 
The first (called \gls{son}1 in the rest of the paper) assumes a fixed \gls{ttt}, i.e., the handover is triggered only if the target cell \gls{sinr} is above a threshold (3 dB) for a fixed amount of time (110 ms).
The second (called \gls{son}2) uses a dynamic \gls{ttt}, which is decreased proportionally to the difference between the target and current cell \gls{sinr}~\cite{7959177}.

\begin{figure}[b]
		\centering
		\setlength\fwidth{\columnwidth}
		\setlength\fheight{.5\columnwidth}
		\input{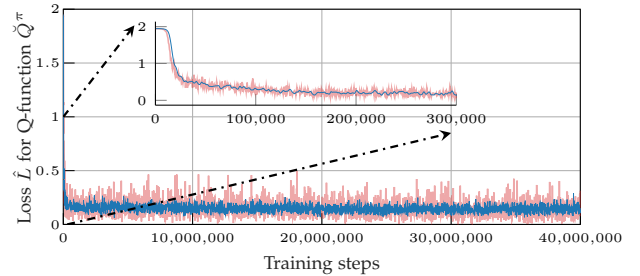}
		\caption{Loss $\hat{L}$ for the Q-function $\Breve{Q}^{\pi}$ for the offline training.}
		\label{fig:min-q-loss}
\end{figure}

\begin{figure*}[t]
\centering
	\begin{subfigure}[t]{0.5\columnwidth}
		\centering
		\setlength\fwidth{.8\textwidth}
		\setlength\fheight{.45\textwidth}
%
%
\definecolor{mycolor1}{rgb}{0.24403,0.43583,0.99883}%
\definecolor{mycolor2}{rgb}{0.00357,0.72027,0.79170}%
\definecolor{lavender}{rgb}{0.9020,0.9020,0.9804}%
\definecolor{lightskyblue}{rgb}{0.6784,0.8471,0.9020}%
\definecolor{deepskyblue}{rgb}{0,0.7490,1}%
\definecolor{steelblue}{rgb}{0.2745,0.5098,0.7059}%
\definecolor{blue}{rgb}{0,0,1}%
\definecolor{royalblue}{rgb}{0.2549,0.4118,0.8824}%

\definecolor{gainsboro}{rgb}{0.8627,0.8627,0.8627}%
\definecolor{darkslategrey}{rgb}{0.1843,0.3098,0.3098}%
\definecolor{gray}{rgb}{0.5,0.5,0.5}%

\definecolor{lightcoral}{rgb}{0.9412,0.5020,0.5020}%
\definecolor{indianred}{rgb}{0.8039,0.3608,0.3608}%
\definecolor{lightsalmon}{rgb}{1.0000,0.6275,0.4784}%
\definecolor{darksalmon}{rgb}{0.9137,0.5882,0.4784}%
\begin{tikzpicture}
\pgfplotsset{every tick label/.append style={font=\scriptsize}}

\begin{axis}[%
width=0.951\fwidth,
height=\fheight,
at={(0\fwidth,0\fheight)},
scale only axis,
xtick=data,
xmin=40,
xmax=130,
ymin=2.5,
ymax=7,
xlabel style={font=\footnotesize\color{white!15!black}},
xlabel={Number of users},
ylabel style={font=\footnotesize\color{white!15!black}},
ylabel={Throughput [Mbit/s]},
axis background/.style={fill=white},
xmajorgrids,
ymajorgrids,
legend style={font=\scriptsize,at={(2.57,1.12)},anchor=south,legend cell align=left,align=left,draw=white!15!black},
legend columns=4
]
\addplot [color=indianred, mark=asterisk, line width=1pt, mark options={solid}]
 plot [error bars/.cd, y dir = both, y explicit]
 table[row sep=crcr, y error plus expr=\thisrowno{2} - \thisrowno{1}, y error minus expr=\thisrowno{1} - \thisrowno{3}]{%
42	3.277	3.410	3.144\\
63	2.844	3.001	2.687\\
84	2.651	2.868	2.434\\
105	2.908	3.003	2.813\\
126	2.929	3.185	2.673\\
};
\addlegendentry{RAN RRM}

\addplot [color=darkslategrey, mark=x, line width=1pt, mark options={solid}]
 plot [error bars/.cd, y dir = both, y explicit]
 table[row sep=crcr, y error plus expr=\thisrowno{2} - \thisrowno{1}, y error minus expr=\thisrowno{1} - \thisrowno{3}]{%
42	3.338	3.467	3.209\\
63	2.945	3.119	2.770\\
84	2.644	2.761	2.527\\
105	3.026	3.102	2.949\\
126	2.899	3.118	2.680\\
};
\addlegendentry{SON1}

\addplot [color=darksalmon, mark=v, line width=1pt, mark options={solid}]
 plot [error bars/.cd, y dir = both, y explicit]
 table[row sep=crcr, y error plus expr=\thisrowno{2} - \thisrowno{1}, y error minus expr=\thisrowno{1} - \thisrowno{3}]{%
42	3.338	3.456	3.220\\
63	2.926	3.075	2.776\\
84	2.782	2.904	2.660\\
105	3.041	3.106	2.977\\
126	3.064	3.254	2.874\\
};
\addlegendentry{SON2}

\addplot [color=steelblue, mark=o, line width=1pt, mark options={solid}]
 plot [error bars/.cd, y dir = both, y explicit]
 table[row sep=crcr, y error plus expr=\thisrowno{2} - \thisrowno{1}, y error minus expr=\thisrowno{1} - \thisrowno{3}]{%
42	4.251	4.556	3.945\\
63	4.034	4.298	3.770\\
84	4.441	4.687	4.195\\
105	4.621	4.802	4.440\\
126	5.112	5.9	4.4\\
};
\addlegendentry{RIC RL}

\end{axis}
\end{tikzpicture}%
		\caption{Average UE throughput (\gls{lte}+NR).}
		\label{fig:avg-ue-th-850}
	\end{subfigure}%
	\hfill%
	\begin{subfigure}[t]{0.48\columnwidth}
	\centering
		\setlength\fwidth{.8\textwidth}
		\setlength\fheight{.45\textwidth}
		\input{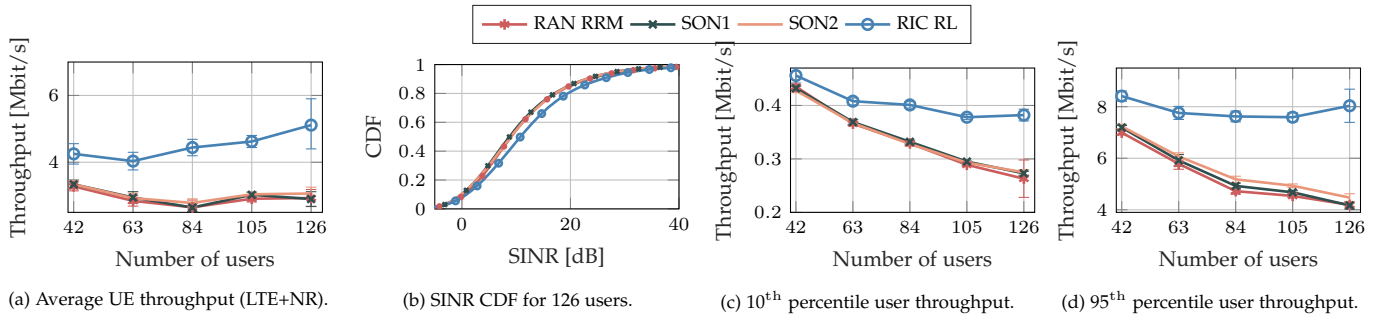}
		\vspace{-.4cm}
	\caption{SINR CDF for 126 users.}
	\label{fig:sinr-850-126}
\end{subfigure}
	\hfill%
		\begin{subfigure}[t]{0.48\columnwidth}
		\centering
		\setlength\fwidth{.8\textwidth}
		\setlength\fheight{.45\textwidth}
%
%
\definecolor{mycolor1}{rgb}{0.24403,0.43583,0.99883}%
\definecolor{mycolor2}{rgb}{0.00357,0.72027,0.79170}%
\definecolor{lavender}{rgb}{0.9020,0.9020,0.9804}%
\definecolor{lightskyblue}{rgb}{0.6784,0.8471,0.9020}%
\definecolor{deepskyblue}{rgb}{0,0.7490,1}%
\definecolor{steelblue}{rgb}{0.2745,0.5098,0.7059}%
\definecolor{blue}{rgb}{0,0,1}%
\definecolor{royalblue}{rgb}{0.2549,0.4118,0.8824}%

\definecolor{gainsboro}{rgb}{0.8627,0.8627,0.8627}%
\definecolor{darkslategrey}{rgb}{0.1843,0.3098,0.3098}%
\definecolor{gray}{rgb}{0.5,0.5,0.5}%

\definecolor{lightcoral}{rgb}{0.9412,0.5020,0.5020}%
\definecolor{indianred}{rgb}{0.8039,0.3608,0.3608}%
\definecolor{lightsalmon}{rgb}{1.0000,0.6275,0.4784}%
\definecolor{darksalmon}{rgb}{0.9137,0.5882,0.4784}%
\begin{tikzpicture}
\pgfplotsset{every tick label/.append style={font=\scriptsize}}

\begin{axis}[%
width=0.951\fwidth,
height=\fheight,
at={(0\fwidth,0\fheight)},
scale only axis,
xtick=data,
xmin=40,
xmax=130,
ymin=0.2,
ymax=0.470,
xlabel style={font=\footnotesize\color{white!15!black}},
xlabel={Number of users},
ylabel style={font=\footnotesize\color{white!15!black}},
ylabel={Throughput [Mbit/s]},
axis background/.style={fill=white},
xmajorgrids,
ymajorgrids,
legend style={font=\scriptsize,at={(1.1,1.15)},anchor=south,legend cell align=left,align=left,draw=white!15!black},
legend columns=4
]
\addplot [color=indianred, mark=asterisk, line width=1pt, mark options={solid}]
 plot [error bars/.cd, y dir = both, y explicit]
 table[row sep=crcr, y error plus expr=\thisrowno{2} - \thisrowno{1}, y error minus expr=\thisrowno{1} - \thisrowno{3}]{%
42	0.435	0.439	0.430\\
63	0.366	0.370	0.362\\
84	0.329	0.332	0.325\\
105	0.289	0.291	0.288\\
126	0.263	0.298	0.228\\
};

\addplot [color=darkslategrey, mark=x, line width=1pt, mark options={solid}]
 plot [error bars/.cd, y dir = both, y explicit]
 table[row sep=crcr, y error plus expr=\thisrowno{2} - \thisrowno{1}, y error minus expr=\thisrowno{1} - \thisrowno{3}]{%
42	0.432	0.436	0.428\\
63	0.369	0.374	0.365\\
84	0.332	0.335	0.328\\
105	0.295	0.296	0.293\\
126	0.273	0.277	0.269\\
};

\addplot [color=darksalmon, mark=v, line width=1pt, mark options={solid}]
 plot [error bars/.cd, y dir = both, y explicit]
 table[row sep=crcr, y error plus expr=\thisrowno{2} - \thisrowno{1}, y error minus expr=\thisrowno{1} - \thisrowno{3}]{%
42	0.428	0.432	0.424\\
63	0.367	0.371	0.363\\
84	0.328	0.332	0.324\\
105	0.293	0.295	0.292\\
126	0.275	0.280	0.270\\
};

\addplot [color=steelblue, mark=o, line width=1pt, mark options={solid}]
 plot [error bars/.cd, y dir = both, y explicit]
 table[row sep=crcr, y error plus expr=\thisrowno{2} - \thisrowno{1}, y error minus expr=\thisrowno{1} - \thisrowno{3}]{%
42	0.456	0.470	0.442\\
63	0.408	0.414	0.401\\
84	0.401	0.407	0.395\\
105	0.378	0.383	0.374\\
126	0.382	0.393	0.371\\
};

\end{axis}
\end{tikzpicture}%
		\caption{10$^{\rm th}$ percentile user throughput.}
		\label{fig:10-th-thp}
	\end{subfigure}%
	\hfill%
		\begin{subfigure}[t]{0.48\columnwidth}
		\centering
		\setlength\fwidth{.8\textwidth}
		\setlength\fheight{.45\textwidth}
%
%
\definecolor{mycolor1}{rgb}{0.24403,0.43583,0.99883}%
\definecolor{mycolor2}{rgb}{0.00357,0.72027,0.79170}%
\definecolor{lavender}{rgb}{0.9020,0.9020,0.9804}%
\definecolor{lightskyblue}{rgb}{0.6784,0.8471,0.9020}%
\definecolor{deepskyblue}{rgb}{0,0.7490,1}%
\definecolor{steelblue}{rgb}{0.2745,0.5098,0.7059}%
\definecolor{blue}{rgb}{0,0,1}%
\definecolor{royalblue}{rgb}{0.2549,0.4118,0.8824}%

\definecolor{gainsboro}{rgb}{0.8627,0.8627,0.8627}%
\definecolor{darkslategrey}{rgb}{0.1843,0.3098,0.3098}%
\definecolor{gray}{rgb}{0.5,0.5,0.5}%

\definecolor{lightcoral}{rgb}{0.9412,0.5020,0.5020}%
\definecolor{indianred}{rgb}{0.8039,0.3608,0.3608}%
\definecolor{lightsalmon}{rgb}{1.0000,0.6275,0.4784}%
\definecolor{darksalmon}{rgb}{0.9137,0.5882,0.4784}%
\begin{tikzpicture}
\pgfplotsset{every tick label/.append style={font=\scriptsize}}

\begin{axis}[%
width=0.951\fwidth,
height=\fheight,
at={(0\fwidth,0\fheight)},
scale only axis,
xtick=data,
xmin=40,
xmax=130,
ymin=3.9,
ymax=9.5,
xlabel style={font=\footnotesize\color{white!15!black}},
xlabel={Number of users},
ylabel style={font=\footnotesize\color{white!15!black}},
ylabel={Throughput [Mbit/s]},
axis background/.style={fill=white},
xmajorgrids,
ymajorgrids,
legend style={font=\scriptsize,at={(0.99,0.99)},anchor=north east,legend cell align=left,align=left,draw=white!15!black},
legend columns=2
]
\addplot [color=indianred, mark=asterisk, line width=1pt, mark options={solid}, forget plot]
 plot [error bars/.cd, y dir = both, y explicit]
 table[row sep=crcr, y error plus expr=\thisrowno{2} - \thisrowno{1}, y error minus expr=\thisrowno{1} - \thisrowno{3}]{%
42	6.995	7.085	6.905\\
63	5.784	5.995	5.573\\
84	4.722	4.840	4.604\\
105	4.539	4.608	4.470\\
126	4.190	4.310	4.070\\
};

\addplot [color=darkslategrey, mark=x, line width=1pt, mark options={solid}, forget plot]
 plot [error bars/.cd, y dir = both, y explicit]
 table[row sep=crcr, y error plus expr=\thisrowno{2} - \thisrowno{1}, y error minus expr=\thisrowno{1} - \thisrowno{3}]{%
42	7.193	7.269	7.117\\
63	5.917	6.141	5.693\\
84	4.927	5.039	4.815\\
105	4.678	4.740	4.616\\
126	4.170	4.314	4.026\\
};

\addplot [color=darksalmon, mark=v, line width=1pt, mark options={solid}, forget plot]
 plot [error bars/.cd, y dir = both, y explicit]
 table[row sep=crcr, y error plus expr=\thisrowno{2} - \thisrowno{1}, y error minus expr=\thisrowno{1} - \thisrowno{3}]{%
42	7.236	7.319	7.153\\
63	6.069	6.209	5.929\\
84	5.182	5.302	5.062\\
105	4.932	5.003	4.861\\
126	4.472	4.621	4.323\\
};

\addplot [color=steelblue, mark=o, line width=1pt, mark options={solid}, forget plot]
 plot [error bars/.cd, y dir = both, y explicit]
 table[row sep=crcr, y error plus expr=\thisrowno{2} - \thisrowno{1}, y error minus expr=\thisrowno{1} - \thisrowno{3}]{%
42	8.411	8.610	8.212\\
63	7.758	8.012	7.504\\
84	7.622	7.832	7.412\\
105	7.592	7.764	7.420\\
126	8.036	8.680	7.392\\
};

\end{axis}
\end{tikzpicture}%
		\caption{95$^{\rm th}$ percentile user throughput.}
		\label{fig:95-th-thp}
	\end{subfigure}%
\caption{Throughput and \gls{sinr} \gls{cdf} for the 850 MHz deployment, for the different baselines and the xApp-driven handover control. The average throughput accounts for the traffic on the \gls{lte} and NR split bearer.}
\label{fig:th-perc}
\end{figure*}

\begin{figure*}[t]
	\centering

	\hfill%
		\begin{subfigure}[t]{0.245\textwidth}
		\centering
		\setlength\fwidth{.8\textwidth}
		\setlength\fheight{.45\textwidth}
%
%
\definecolor{mycolor1}{rgb}{0.24403,0.43583,0.99883}%
\definecolor{mycolor2}{rgb}{0.00357,0.72027,0.79170}%
\definecolor{lavender}{rgb}{0.9020,0.9020,0.9804}%
\definecolor{lightskyblue}{rgb}{0.6784,0.8471,0.9020}%
\definecolor{deepskyblue}{rgb}{0,0.7490,1}%
\definecolor{steelblue}{rgb}{0.2745,0.5098,0.7059}%
\definecolor{blue}{rgb}{0,0,1}%
\definecolor{royalblue}{rgb}{0.2549,0.4118,0.8824}%

\definecolor{gainsboro}{rgb}{0.8627,0.8627,0.8627}%
\definecolor{darkslategrey}{rgb}{0.1843,0.3098,0.3098}%
\definecolor{gray}{rgb}{0.5,0.5,0.5}%

\definecolor{lightcoral}{rgb}{0.9412,0.5020,0.5020}%
\definecolor{indianred}{rgb}{0.8039,0.3608,0.3608}%
\definecolor{lightsalmon}{rgb}{1.0000,0.6275,0.4784}%
\definecolor{darksalmon}{rgb}{0.9137,0.5882,0.4784}%
\begin{tikzpicture}
\pgfplotsset{every tick label/.append style={font=\scriptsize}}

\begin{axis}[%
width=0.951\fwidth,
height=\fheight,
at={(0\fwidth,0\fheight)},
scale only axis,
xtick=data,
xmin=40,
xmax=130,
ymin=0,
ymax=1.5,
xlabel style={font=\footnotesize\color{white!15!black}},
xlabel={Number of users},
ylabel style={font=\footnotesize\color{white!15!black}},
ylabel={Spectral efficiency [bit/s/Hz]},
axis background/.style={fill=white},
xmajorgrids,
ymajorgrids,
ylabel shift = -3 pt,
yticklabel shift = -1 pt,
legend style={font=\scriptsize,at={(2.45,1.12)},anchor=south,legend cell align=left,align=left,draw=white!15!black},
legend columns=4
]
\addplot [color=indianred, mark=asterisk, line width=1pt, mark options={solid}]
 plot [error bars/.cd, y dir = both, y explicit]
 table[row sep=crcr, y error plus expr=\thisrowno{2} - \thisrowno{1}, y error minus expr=\thisrowno{1} - \thisrowno{3}]{%
42	0.482	0.508	0.456\\
63	0.568	0.622	0.513\\
84	0.642	0.707	0.573\\
105	0.804	0.848	0.759\\
126	0.857	0.941	0.767\\
};
\addlegendentry{RAN RRM}

\addplot [color=darkslategrey, mark=x, line width=1pt, mark options={solid}]
 plot [error bars/.cd, y dir = both, y explicit]
 table[row sep=crcr, y error plus expr=\thisrowno{2} - \thisrowno{1}, y error minus expr=\thisrowno{1} - \thisrowno{3}]{%
42	0.489	0.509	0.469\\
63	0.694	0.781	0.601\\
84	0.684	0.739	0.626\\
105	0.812	0.850	0.774\\
126	0.913	1.008	0.811\\
};
\addlegendentry{SON1}

\addplot [color=darksalmon, mark=v, line width=1pt, mark options={solid}]
 plot [error bars/.cd, y dir = both, y explicit]
 table[row sep=crcr, y error plus expr=\thisrowno{2} - \thisrowno{1}, y error minus expr=\thisrowno{1} - \thisrowno{3}]{%
42	0.517	0.554	0.479\\
63	0.616	0.665	0.565\\
84	0.715	0.786	0.641\\
105	0.810	0.848	0.770\\
126	0.881	0.967	0.789\\
};
\addlegendentry{SON2}

\addplot [color=steelblue, mark=o, line width=1pt, mark options={solid}]
 plot [error bars/.cd, y dir = both, y explicit]
 table[row sep=crcr, y error plus expr=\thisrowno{2} - \thisrowno{1}, y error minus expr=\thisrowno{1} - \thisrowno{3}]{%
42	0.809	0.915	0.695\\
63	0.985	1.130	0.824\\
84	1.246	1.357	1.127\\
105	1.232	1.299	1.160\\
126	1.297	1.446	1.130\\
};
\addlegendentry{RIC RL}

\end{axis}
\end{tikzpicture}%
		\caption{Average spectral efficiency, per UE.}
		\label{fig:avg-ue-se-850}
	\end{subfigure}%
	\hfill%
		\begin{subfigure}[t]{0.245\textwidth}
		\centering
		\setlength\fwidth{.8\textwidth}
		\setlength\fheight{.45\textwidth}
%
%
\definecolor{mycolor1}{rgb}{0.24403,0.43583,0.99883}%
\definecolor{mycolor2}{rgb}{0.00357,0.72027,0.79170}%
\definecolor{lavender}{rgb}{0.9020,0.9020,0.9804}%
\definecolor{lightskyblue}{rgb}{0.6784,0.8471,0.9020}%
\definecolor{deepskyblue}{rgb}{0,0.7490,1}%
\definecolor{steelblue}{rgb}{0.2745,0.5098,0.7059}%
\definecolor{blue}{rgb}{0,0,1}%
\definecolor{royalblue}{rgb}{0.2549,0.4118,0.8824}%

\definecolor{gainsboro}{rgb}{0.8627,0.8627,0.8627}%
\definecolor{darkslategrey}{rgb}{0.1843,0.3098,0.3098}%
\definecolor{gray}{rgb}{0.5,0.5,0.5}%

\definecolor{lightcoral}{rgb}{0.9412,0.5020,0.5020}%
\definecolor{indianred}{rgb}{0.8039,0.3608,0.3608}%
\definecolor{lightsalmon}{rgb}{1.0000,0.6275,0.4784}%
\definecolor{darksalmon}{rgb}{0.9137,0.5882,0.4784}%
\begin{tikzpicture}
\pgfplotsset{every tick label/.append style={font=\scriptsize}}

\begin{axis}[%
width=0.951\fwidth,
height=\fheight,
at={(0\fwidth,0\fheight)},
scale only axis,
xtick=data,
xmin=40,
xmax=130,
ymin=0,
ymax=1,
xlabel style={font=\footnotesize\color{white!15!black}},
xlabel={Number of users},
ylabel style={font=\footnotesize\color{white!15!black}},
ylabel={Spectral efficiency [bit/s/Hz]},
axis background/.style={fill=white},
xmajorgrids,
ymajorgrids,
legend style={font=\scriptsize,at={(0.99,0.01)},anchor=south east,legend cell align=left,align=left,draw=white!15!black},
legend columns=2
]
\addplot [color=indianred, mark=asterisk, line width=1pt, mark options={solid}, forget plot]
 plot [error bars/.cd, y dir = both, y explicit]
 table[row sep=crcr, y error plus expr=\thisrowno{2} - \thisrowno{1}, y error minus expr=\thisrowno{1} - \thisrowno{3}]{%
42	0.500	0.525	0.475\\
63	0.414	0.442	0.384\\
84	0.398	0.425	0.370\\
105	0.440	0.460	0.419\\
126	0.455	0.488	0.423\\
};

\addplot [color=darkslategrey, mark=x, line width=1pt, mark options={solid}, forget plot]
 plot [error bars/.cd, y dir = both, y explicit]
 table[row sep=crcr, y error plus expr=\thisrowno{2} - \thisrowno{1}, y error minus expr=\thisrowno{1} - \thisrowno{3}]{%
42	0.500	0.528	0.472\\
63	0.444	0.482	0.405\\
84	0.379	0.398	0.360\\
105	0.440	0.453	0.426\\
126	0.447	0.481	0.411\\
};

\addplot [color=darksalmon, mark=v, line width=1pt, mark options={solid}, forget plot]
 plot [error bars/.cd, y dir = both, y explicit]
 table[row sep=crcr, y error plus expr=\thisrowno{2} - \thisrowno{1}, y error minus expr=\thisrowno{1} - \thisrowno{3}]{%
42	0.488	0.509	0.466\\
63	0.422	0.448	0.396\\
84	0.401	0.423	0.379\\
105	0.438	0.451	0.425\\
126	0.469	0.509	0.428\\
};

\addplot [color=steelblue, mark=o, line width=1pt, mark options={solid}, forget plot]
 plot [error bars/.cd, y dir = both, y explicit]
 table[row sep=crcr, y error plus expr=\thisrowno{2} - \thisrowno{1}, y error minus expr=\thisrowno{1} - \thisrowno{3}]{%
42	0.650	0.714	0.582\\
63	0.536	0.580	0.491\\
84	0.587	0.619	0.554\\
105	0.601	0.627	0.574\\
126	0.776	0.862	0.688\\
};

\end{axis}
\end{tikzpicture}%
		\caption{Average spectral efficiency, per cell.}
		\label{fig:avg-cell-se-850}
	\end{subfigure}%
	\hfill%
	\begin{subfigure}[t]{0.245\textwidth}
		\centering
		\setlength\fwidth{.8\textwidth}
		\setlength\fheight{.45\textwidth}
%
%
\definecolor{mycolor1}{rgb}{0.24403,0.43583,0.99883}%
\definecolor{mycolor2}{rgb}{0.00357,0.72027,0.79170}%
\definecolor{lavender}{rgb}{0.9020,0.9020,0.9804}%
\definecolor{lightskyblue}{rgb}{0.6784,0.8471,0.9020}%
\definecolor{deepskyblue}{rgb}{0,0.7490,1}%
\definecolor{steelblue}{rgb}{0.2745,0.5098,0.7059}%
\definecolor{blue}{rgb}{0,0,1}%
\definecolor{royalblue}{rgb}{0.2549,0.4118,0.8824}%

\definecolor{gainsboro}{rgb}{0.8627,0.8627,0.8627}%
\definecolor{darkslategrey}{rgb}{0.1843,0.3098,0.3098}%
\definecolor{gray}{rgb}{0.5,0.5,0.5}%

\definecolor{lightcoral}{rgb}{0.9412,0.5020,0.5020}%
\definecolor{indianred}{rgb}{0.8039,0.3608,0.3608}%
\definecolor{lightsalmon}{rgb}{1.0000,0.6275,0.4784}%
\definecolor{darksalmon}{rgb}{0.9137,0.5882,0.4784}%
\begin{tikzpicture}
\pgfplotsset{every tick label/.append style={font=\scriptsize}}

\begin{axis}[%
width=0.951\fwidth,
height=\fheight,
at={(0\fwidth,0\fheight)},
scale only axis,
xtick=data,
xmin=40,
xmax=130,
ymin=45,
ymax=55,
xlabel style={font=\footnotesize\color{white!15!black}},
xlabel={Number of users},
ylabel style={font=\footnotesize\color{white!15!black}},
ylabel={PRB used in DL [\%]},
axis background/.style={fill=white},
xmajorgrids,
ymajorgrids,
ylabel shift = -3 pt,
yticklabel shift = -1 pt,
legend style={font=\scriptsize,at={(0.99,0.01)},anchor=south east,legend cell align=left,align=left,draw=white!15!black},
legend columns=2
]
\addplot [color=indianred, mark=asterisk, line width=1pt, mark options={solid}, forget plot]
 plot [error bars/.cd, y dir = both, y explicit]
 table[row sep=crcr, y error plus expr=\thisrowno{2} - \thisrowno{1}, y error minus expr=\thisrowno{1} - \thisrowno{3}]{%
42	48.272	48.966	47.578\\
63	50.553	51.265	49.842\\
84	49.379	49.862	48.896\\
105	48.197	48.394	48.001\\
126	46.242	46.594	45.889\\
};

\addplot [color=darkslategrey, mark=x, line width=1pt, mark options={solid}, forget plot]
 plot [error bars/.cd, y dir = both, y explicit]
 table[row sep=crcr, y error plus expr=\thisrowno{2} - \thisrowno{1}, y error minus expr=\thisrowno{1} - \thisrowno{3}]{%
42	49.990	50.618	49.363\\
63	51.654	52.381	50.926\\
84	52.211	52.641	51.781\\
105	50.129	50.310	49.949\\
126	48.309	48.712	47.907\\
};

\addplot [color=darksalmon, mark=v, line width=1pt, mark options={solid}, forget plot]
 plot [error bars/.cd, y dir = both, y explicit]
 table[row sep=crcr, y error plus expr=\thisrowno{2} - \thisrowno{1}, y error minus expr=\thisrowno{1} - \thisrowno{3}]{%
42	50.102	50.774	49.429\\
63	52.097	52.873	51.322\\
84	51.736	52.260	51.212\\
105	50.723	50.923	50.523\\
126	48.433	48.907	47.959\\
};

\addplot [color=steelblue, mark=o, line width=1pt, mark options={solid}, forget plot]
 plot [error bars/.cd, y dir = both, y explicit]
 table[row sep=crcr, y error plus expr=\thisrowno{2} - \thisrowno{1}, y error minus expr=\thisrowno{1} - \thisrowno{3}]{%
42	48.911	50.139	47.683\\
63	54.618	55.680	53.556\\
84	54.061	54.660	53.463\\
105	54.354	54.667	54.040\\
126	52.571	53.327	51.814\\
};

\end{axis}
\end{tikzpicture}%
		\caption{Average number of PRB used in downlink, per cell, percentage.}
		\label{fig:avg-prb-perc-850}
	\end{subfigure}
		\hfill%
	\begin{subfigure}[t]{0.245\textwidth}
		\centering
		\setlength\fwidth{.8\textwidth}
		\setlength\fheight{.45\textwidth}
%
%
\definecolor{mycolor1}{rgb}{0.24403,0.43583,0.99883}%
\definecolor{mycolor2}{rgb}{0.00357,0.72027,0.79170}%
\definecolor{lavender}{rgb}{0.9020,0.9020,0.9804}%
\definecolor{lightskyblue}{rgb}{0.6784,0.8471,0.9020}%
\definecolor{deepskyblue}{rgb}{0,0.7490,1}%
\definecolor{steelblue}{rgb}{0.2745,0.5098,0.7059}%
\definecolor{blue}{rgb}{0,0,1}%
\definecolor{royalblue}{rgb}{0.2549,0.4118,0.8824}%

\definecolor{gainsboro}{rgb}{0.8627,0.8627,0.8627}%
\definecolor{darkslategrey}{rgb}{0.1843,0.3098,0.3098}%
\definecolor{gray}{rgb}{0.5,0.5,0.5}%

\definecolor{lightcoral}{rgb}{0.9412,0.5020,0.5020}%
\definecolor{indianred}{rgb}{0.8039,0.3608,0.3608}%
\definecolor{lightsalmon}{rgb}{1.0000,0.6275,0.4784}%
\definecolor{darksalmon}{rgb}{0.9137,0.5882,0.4784}%
\begin{tikzpicture}
\pgfplotsset{every tick label/.append style={font=\scriptsize}}

\begin{axis}[%
width=0.951\fwidth,
height=\fheight,
at={(0\fwidth,0\fheight)},
scale only axis,
xtick=data,
xmin=40,
xmax=130,
ymin=0,
ymax=2,
xlabel style={font=\footnotesize\color{white!15!black}},
xlabel={Number of users},
ylabel style={font=\footnotesize\color{white!15!black}},
ylabel={Mobility overhead $H_u$},
axis background/.style={fill=white},
xmajorgrids,
ymajorgrids,
ylabel shift = -3 pt,
yticklabel shift = -1 pt,
legend style={font=\scriptsize,at={(0.99,0.99)},anchor=north east,legend cell align=left,align=left,draw=white!15!black},
legend columns=2
]
\addplot [color=indianred, mark=asterisk, line width=1pt, mark options={solid}, forget plot]
 plot [error bars/.cd, y dir = both, y explicit]
 table[row sep=crcr, y error plus expr=\thisrowno{2} - \thisrowno{1}, y error minus expr=\thisrowno{1} - \thisrowno{3}]{%
42	1.302	1.391	1.214\\
63	0.878	0.886	0.870\\
84	0.687	0.780	0.595\\
105	0.507	0.532	0.483\\
126	0.399	0.432	0.365\\
};

\addplot [color=darkslategrey, mark=x, line width=1pt, mark options={solid}, forget plot]
 plot [error bars/.cd, y dir = both, y explicit]
 table[row sep=crcr, y error plus expr=\thisrowno{2} - \thisrowno{1}, y error minus expr=\thisrowno{1} - \thisrowno{3}]{%
42	0.779	0.837	0.721\\
63	0.531	0.625	0.437\\
84	0.485	0.541	0.428\\
105	0.308	0.321	0.295\\
126	0.242	0.269	0.215\\
};

\addplot [color=darksalmon, mark=v, line width=1pt, mark options={solid}, forget plot]
 plot [error bars/.cd, y dir = both, y explicit]
 table[row sep=crcr, y error plus expr=\thisrowno{2} - \thisrowno{1}, y error minus expr=\thisrowno{1} - \thisrowno{3}]{%
42	0.700	0.754	0.646\\
63	0.485	0.560	0.409\\
84	0.448	0.511	0.386\\
105	0.262	0.275	0.250\\
126	0.207	0.255	0.159\\
};

\addplot [color=steelblue, mark=o, line width=1pt, mark options={solid}, forget plot]
 plot [error bars/.cd, y dir = both, y explicit]
 table[row sep=crcr, y error plus expr=\thisrowno{2} - \thisrowno{1}, y error minus expr=\thisrowno{1} - \thisrowno{3}]{%
42	1.862	2.590	1.134\\
63	0.622	0.689	0.556\\
84	0.488	0.540	0.437\\
105	0.284	0.312	0.256\\
126	0.234	0.291	0.177\\
};

\end{axis}
\end{tikzpicture}%
		\caption{UE mobility overhead $H_u$.}
		\label{fig:ho-metric-850}
	\end{subfigure}%
	\caption{Spectral efficiency and mobility overhead metrics for the 850 MHz deployment, as a function of the number of users, for the different baselines and the xApp-driven handover control.}
	\label{fig:perf-850}
\end{figure*}
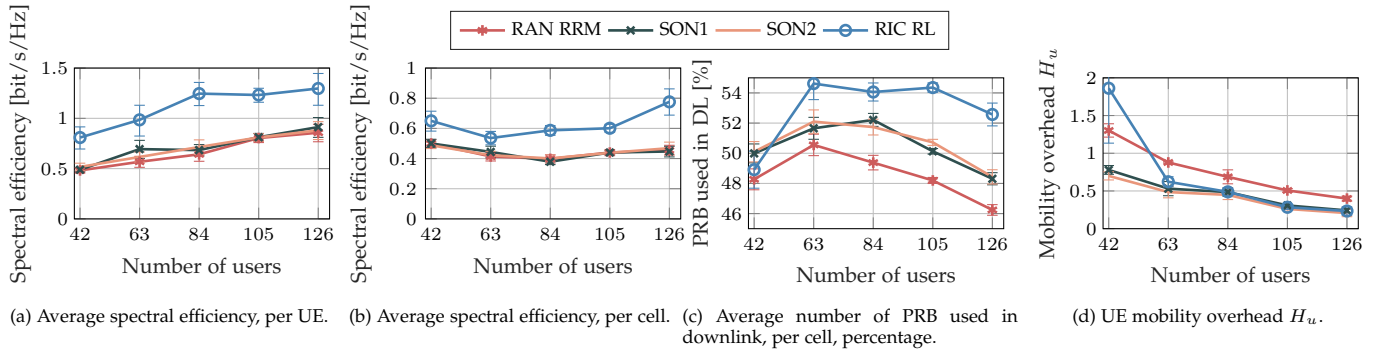


{\bf Performance metrics --- } 
For the performance evaluation of the \gls{ts} xApp we consider the metrics related to throughput, channel quality, spectral efficiency, and mobility overhead. 
For the first, we report the average \gls{ue} throughput at the \gls{pdcp} layer, i.e., including both \gls{lte} and NR split bearers, as well as the 10$^{\rm th}$ and 95$^{\rm th}$ percentiles of all the users in a simulation, averaged over multiple independent runs. 
The channel quality is represented by the \gls{sinr}. For the spectral efficiency, we analyze the average value for each \glspl{ue} and cell, as well as the 10$^{\rm th}$ percentile, and the percentage of \glspl{prb} used for downlink traffic. 
Finally, we evaluate the \gls{ue} mobility overhead $H_{u}$ as the number of handovers per unit time weighted by a throughput factor $\hat{R}_u = \mathbb{E}(R_u) / \sum_{u' \in U} \mathbb{E}(R_u')$, where $\mathbb{E}(R_u)$ is the average throughput for the user over the same unit time.

\begin{table}
\centering
  \begin{tabular}{ll}
    \toprule
    {\bf Hyperparameters} & {\bf Value} \\
    \midrule
        {\bf DQN Agent (Offline)} & \\
        \hspace{0.5cm} Target update period & 8000 \\
        \hspace{0.5cm} Batch size & 32 \\
        \hspace{0.5cm} Number of heads ($n$ heads in Fig.~\ref{fig:cnn_arch_design}) & 200 \\        
        \hspace{0.5cm} Number of actions ($N$ in Fig.~\ref{fig:cnn_arch_design})) & 7 \\        
        \hspace{0.5cm} Minimum replay history & 20000 \\        
        \hspace{0.5cm} Terminal (Episode) length & 1 \\    
        \hspace{0.5cm} Gamma & 0.99 \\        
        \hspace{0.5cm} Replay capacity & 1000000 \\                
        \hspace{0.5cm} Number of iterations & 400 \\
        \hspace{0.5cm} Training steps & 100000 \\    
        
        {\bf Optimizer} & \\
        \hspace{0.5cm} Optimizer & AdamOptimizer \\        
        \hspace{0.5cm} Learning rate & 0.00005 \\        
        {\bf Neural Network (Fig.~\ref{fig:cnn_arch_design})} & \\        
        \hspace{0.5cm} Conv1D Layer & filters=32 \\        
                                    & kernel size=8 \\        
                                    & strides=8\\        
                                    & activation=ReLu \\                             
        \hspace{0.5cm} Flatten Layer &  225 neurons \\
        \hspace{0.5cm} Dense Layer 1 & 128 neurons \\
        \hspace{0.5cm} Dense Layer 2 & 32 neurons \\
        \hspace{0.5cm} Dense Layer 3 & 1400 neurons \\

  \bottomrule
\end{tabular}
  \caption{\gls{rl} hyperparameters and their values.}
  \label{tab:RL_hyperparmeters_value}
\end{table}
{\bf Data collection and agent training --- }
The data collection is based on a total of more than 2000 simulations for the different configurations, including multiple independent simulation runs for each scenario.
Table~\ref{tab:RL_hyperparmeters_value} provides the list of \gls{rl} hyperparameters and their values considered in this paper.
In the offline training, frequency with which the target network gets updated is set to 8000 training steps.
We perform 400 iterations during the offline training, and each iteration has 100K training steps, for a total of 40 million training steps.
In a training step a batch of 32 samples (or data points) are selected randomly for input to Neural Network.
The first layer of the network is the Conv1D layer with $2^{x_1} = 32$ filters (see Fig.~\ref{fig:cnn_arch_design}).
The kernel size and strides are set to 8, as each cell has 8 input parameters (Section \ref{sec:ts}), and activation function is ReLU.
This is followed by a flattening layer which flattens the output of the Conv1D layer (with $y_1 B = 225$), and the output of the flattened layer is concatenated with the $t-t_{u'}$ parameter. 
Third, fourth and fifth layers are fully-connected layers with $2^{x_2} = 128$, $2^{x_3} = 32$ and 1400 units/neurons, respectively. 
The number of units in the last layer is given by the product of $n=200$, the number of heads of the \gls{rem}, and the number of actions $N=7$. 
We use the Adam optimizer with a learning rate of 0.00005.
Figure~\ref{fig:min-q-loss} shows the trend of the loss $\hat{L}$ for the Q-function $\Breve{Q}^{\pi}$ (as discussed in Section \ref{sec:ts}) during the training of the \gls{rl} agent, including a focus on the first $3\cdot 10^5$ iterations. 
The initial cost $K_0$ from Eq.~\ref{eqn:problem_formulation} is 1, and ${\delta}$ (decay constant) is 0.1.

The likelihood of the loss curve $\hat{L}$ is regular and its trend approaches values close to zero, showing that the weights of the \gls{cnn} are actually improving after each iteration and the information learned is a good approximation of the non-linear dependency between the actions and the reward.

\subsection{Results}

In this section, we comment on the results we obtained after the training and the online testing of the xApp described in Section \ref{sec:ts}.
The \gls{rl} agent was tested in simulations with the baselines \glspl{ho} disabled.
The experiments were repeated with different numbers of \glspl{ue}, and averaged around 600,000 records for FR1 850 MHz and around 300,000 records for FR1 C-band in online evaluation.

Figure~\ref{fig:avg-ue-th-850} shows the average \gls{ue} throughput for the 850 MHz deployment, while Fig.~\ref{fig:sinr-850-126} reports the \gls{cdf} of the \gls{sinr} with 126 \glspl{ue}.
%
The RIC RL introduces an improvement of the average throughput (averaging around 50\%) and \gls{sinr} with respect to the the baselines, meaning that the \gls{rl} agent is able to customize the HO-control for each single \gls{ue}.
This trend is also confirmed as the number of \glspl{ue} increases, proving the scalability of this approach over baselines.
The customization of the mobility for a single \gls{ue} is one of the advantages of using the xApps ecosystem, which has a centralized and abstracted view of the network status.

Moreover, by looking at the percentiles of the user throughput (Fig.~\ref{fig:th-perc}), it can be seen that our \gls{rl} agent brings consistent improvement not only on the average \glspl{ue}, but also between the worst (10-th percentile, Fig.~\ref{fig:10-th-thp}) users, showing 30\% improvements and best (95-th percentile, Fig.~\ref{fig:95-th-thp}) users, showing around 60\% improvement.
The 126 \glspl{ue} result is particularly relevant, as also testified by the improvement in \gls{sinr} shown in Fig.~\ref{fig:sinr-850-126}. 
%
Contrary to heuristic-based HOs, 
the \gls{rl} algorithm leverages \gls{ue}-level and cell-level \glspl{kpm} to take the decision to centrally handover/steer the user to an optimal NR neighbor, in terms of load and \gls{sinr}.
%
This results in an improved spectral efficiency (and thus throughput), as shown in Figs.~\ref{fig:avg-ue-se-850} and~\ref{fig:avg-cell-se-850}, demonstrating 52\% and 35\% improvements, respectively.
The same holds for the \gls{prb} utilization (Fig.~\ref{fig:avg-prb-perc-850}).
Indeed, since \gls{ric} \gls{rl} utilizes cell-level \glspl{kpm} at 100ms granularity, it is able to handover \glspl{ue} to a target cell with higher residual \glspl{prb}.

However, these improvements in the throughput could eventually come with a major cost in terms of HO management, and thus energy.
The mobility overhead $H_u$ of Fig.~\ref{fig:ho-metric-850} clearly shows that our \gls{rl} agent is not causing more HOs, but instead follows the trend of the baselines, while at the same time delivering better throughput.
The only exception is for 42~\glspl{ue}, where the \gls{rl} agent triggers more HOs than all baselines.
One of the possible reasons can be identified in the cost function described in Eq.~(\ref{eqn:problem_formulation_lagrangian}) (Section \ref{sec:ts}),
%
where the reward (logarithmic throughput gain, which is higher with fewer users)
compensates for the cost of handover thereby resulting in an increase in mobility overhead $H_u$.

\begin{figure}[t]
	\centering
	\setlength\fwidth{.8\columnwidth}
	\setlength\fheight{.4\columnwidth}
%
%
\definecolor{mycolor1}{rgb}{0.24403,0.43583,0.99883}%
\definecolor{mycolor2}{rgb}{0.00357,0.72027,0.79170}%
\definecolor{lavender}{rgb}{0.9020,0.9020,0.9804}%
\definecolor{lightskyblue}{rgb}{0.6784,0.8471,0.9020}%
\definecolor{deepskyblue}{rgb}{0,0.7490,1}%
\definecolor{steelblue}{rgb}{0.2745,0.5098,0.7059}%
\definecolor{blue}{rgb}{0,0,1}%
\definecolor{royalblue}{rgb}{0.2549,0.4118,0.8824}%

\definecolor{gainsboro}{rgb}{0.8627,0.8627,0.8627}%
\definecolor{darkslategrey}{rgb}{0.1843,0.3098,0.3098}%
\definecolor{gray}{rgb}{0.5,0.5,0.5}%

\definecolor{lightcoral}{rgb}{0.9412,0.5020,0.5020}%
\definecolor{indianred}{rgb}{0.8039,0.3608,0.3608}%
\definecolor{lightsalmon}{rgb}{1.0000,0.6275,0.4784}%
\definecolor{darksalmon}{rgb}{0.9137,0.5882,0.4784}%
\definecolor{color1}{rgb}{0.12156862745098,0.466666666666667,0.705882352941177}%
\begin{tikzpicture}
\pgfplotsset{every tick label/.append style={font=\scriptsize}}

\begin{axis}[%
width=0.951\fwidth,
height=\fheight,
at={(0\fwidth,0\fheight)},
scale only axis,
bar shift auto,
xmin=0.5,
xmax=7.5,
xlabel style={font=\scriptsize},
ylabel style={font=\scriptsize, align=center},
xlabel={Metric},
ymin=0,
ymax=2.45,
yminorticks=true,
ylabel={Ratio between RIC RL\\and SON2 performance},
xticklabel style={align=center},
axis background/.style={fill=white},
xmajorgrids,
ymajorgrids,
yminorgrids,
xtick=data,
xticklabels={{UE\\throughput}, {$H_u$}, {10$^{\rm th}$ perc.\\UE th.}, {95$^{\rm th}$ perc.\\UE th.}, {Average\\UE spectral\\eff.}, {Average\\cell\\spectral eff.}, {Average\\PRB\\utilization}},
xticklabel style={font=\tiny},
enlarge x limits=0.02,
legend style={legend cell align=left, align=left, draw=white!15!black, font=\scriptsize, at={(0.01, 0.99)}, anchor=north west},
legend columns=2,
]

\addplot [ybar, bar width=0.2, fill=darksalmon, draw=black, area legend, postaction={pattern=north east lines}]
  table[row sep=crcr]{%
1	1.541\\
2	1.4496\\
3	1.2158\\
4	1.4943\\
5 1.58\\
6 1.4183\\
7 1.0453\\
};
\addlegendentry{850 MHz}

\addplot [ybar, bar width=0.2, fill=color1, draw=black, area legend, postaction={pattern=horizontal lines}]
  table[row sep=crcr]{%
1	1.4275\\
2	1.2026\\
3	1.3047\\
4	1.1777\\
5 2.3760\\
6 1.3906\\
7 0.9475\\
};
\addlegendentry{C-Band (3.5 GHz)}

\end{axis}


\end{tikzpicture}%
	\caption{Comparison between the performance gain in the 850 MHz band and in the 3.5 GHz band (or C-Band). Each bar represents the ratio between the performance with the \gls{ric} \gls{rl} and \gls{son}2 for the corresponding metric.}
	\label{fig:fr1-c-comparison}
\end{figure}
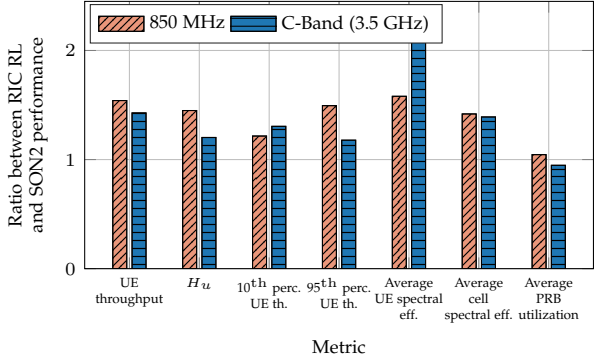

Furthermore, Fig.~\ref{fig:fr1-c-comparison} compares the already discussed results for 850 MHz with the C-Band deployment.
In this figure, we show the relative gains of the performances of the \gls{rl} agent in the two bands.
The gain of each \gls{kpm} shown in the x axis is defined as the ratio between the
performance with the \gls{ric} \gls{rl} and \gls{son}2 for the corresponding metric.
Given this definition, the \gls{rl} agent is performing better than the baseline when the ratio is greater than 1.
The analysis of the relative gains shows that while the average \gls{prb} utilization of the \gls{ric} falls below the baseline, the other \glspl{kpm} improves consistently, showing the adaptability of \gls{ric} \gls{rl} through different bands.

We also compare the performance of the proposed RIC-enabled \gls{rl} agent against the contextual multi-armed bandit \gls{rl} agent proposed in~\cite{yajnanarayana20205g}. To do so, we implemented the agent from~\cite{yajnanarayana20205g} in the xApp, and trained it on the same dataset used to train our agent. Fig.~\ref{fig:10-th-comparison} compares the performance between the two agents in terms of 10$^{\rm th}$ percentile throughput, for different numbers of users and for deployments in the 850 MHz and C-Band frequencies. The method proposed in~\cite{yajnanarayana20205g} aims at improving the \gls{rsrp} for individual \glspl{ue} through handover to different cells. However, this does not improve the cell-edge throughput as much as the RIC-enabled \gls{rl} optimization proposed in this paper, which provides consistently higher cell-edge user throughput in the two frequency bands and with different numbers of users. Compared to~\cite{yajnanarayana20205g}, the O-RAN-driven solution we introduce in this paper exploits a richer input feature set, which makes it possible to characterize the user status with higher precision, and thus to select control actions that go beyond the \gls{rsrp} improvement optimizing the user throughput itself.

\begin{figure}[t]
	\centering
	\setlength\fwidth{.8\columnwidth}
	\setlength\fheight{.4\columnwidth}
%
%
\definecolor{mycolor1}{rgb}{0.24403,0.43583,0.99883}%
\definecolor{mycolor2}{rgb}{0.00357,0.72027,0.79170}%
\definecolor{lavender}{rgb}{0.9020,0.9020,0.9804}%
\definecolor{lightskyblue}{rgb}{0.6784,0.8471,0.9020}%
\definecolor{deepskyblue}{rgb}{0,0.7490,1}%
\definecolor{steelblue}{rgb}{0.2745,0.5098,0.7059}%
\definecolor{blue}{rgb}{0,0,1}%
\definecolor{royalblue}{rgb}{0.2549,0.4118,0.8824}%

\definecolor{gainsboro}{rgb}{0.8627,0.8627,0.8627}%
\definecolor{darkslategrey}{rgb}{0.1843,0.3098,0.3098}%
\definecolor{gray}{rgb}{0.5,0.5,0.5}%

\definecolor{lightcoral}{rgb}{0.9412,0.5020,0.5020}%
\definecolor{indianred}{rgb}{0.8039,0.3608,0.3608}%
\definecolor{lightsalmon}{rgb}{1.0000,0.6275,0.4784}%
\definecolor{darksalmon}{rgb}{0.9137,0.5882,0.4784}%
\definecolor{color1}{rgb}{0.12156862745098,0.466666666666667,0.705882352941177}%
\begin{tikzpicture}
\pgfplotsset{every tick label/.append style={font=\scriptsize}}

\begin{axis}[%
width=0.951\fwidth,
height=\fheight,
at={(0\fwidth,0\fheight)},
scale only axis,
bar shift auto,
xmin=0.5,
xmax=4.5,
xlabel style={font=\scriptsize},
ylabel style={font=\scriptsize, align=center},
xlabel={Application class},
ymin=0,
ymax=0.8,
yminorticks=true,
ylabel={Spectral efficiency [bit/s/Hz]},
xticklabel style={align=center},
axis background/.style={fill=white},
xmajorgrids,
ymajorgrids,
yminorgrids,
xtick=data,
xticklabels={{Full buffer}, {Video streaming}, {Web browsing}, {Instant messaging}},
xticklabel style={font=\tiny},
enlarge x limits=0.02,
legend style={legend cell align=left, align=left, draw=white!15!black, font=\scriptsize, at={(0.99, 0.99)}, anchor=north east},
legend columns=2,
]

\addplot [ybar, bar width=0.16, fill=indianred, draw=black, area legend, postaction={pattern=north east lines}]
  table[row sep=crcr]{%
1	0.675\\
2	0.368\\
3	0.325\\
4	0.362\\
};
\addlegendentry{RAN RRM}

\addplot [ybar, bar width=0.16, fill=darkslategrey, draw=black, area legend, postaction={pattern=vertical lines}]
  table[row sep=crcr]{%
1	0.680\\
2	0.359\\
3	0.336\\
4	0.354\\
};
\addlegendentry{SON1}

\addplot [ybar, bar width=0.16, fill=darksalmon, draw=black, area legend, postaction={pattern=north west lines}]
  table[row sep=crcr]{%
1	0.676\\
2	0.356\\
3	0.341\\
4	0.352\\
};
\addlegendentry{SON2}

\addplot [ybar, bar width=0.16, fill=steelblue, draw=black, area legend, postaction={pattern=north east lines}]
  table[row sep=crcr]{%
1	0.755\\
2	0.591\\
3	0.556\\
4	0.453\\
};
\addlegendentry{RIC RL}

\end{axis}


\end{tikzpicture}%
	\caption{Average cell spectral efficiency for the different traffic types, for 105 users and the 850 MHz deployment.}
	\label{fig:850-traffic-types}
\end{figure}
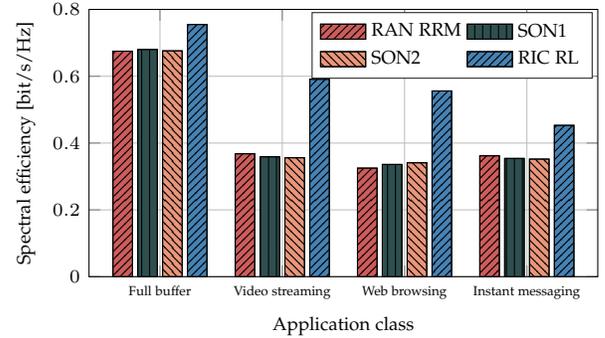

Finally, one key aspect enabled by the per-UE control made possible by our xApps design and the \gls{oran} architecture is the possibility of improving the performances of a heterogeneous \glspl{ue}, with different traffic models.
Fig.~\ref{fig:850-traffic-types} indeed shows the average cell spectral efficiency for the different traffic types, for 105 users and the 850 MHz deployment.
Thanks to the optimized handover management, the \gls{ric} \gls{rl} policy is able to improve the conditions of the all the \glspl{ue} with significant gains for traffic models such as the video streaming, the web browsing, and the instant messages, whose performance fails to be optimized by the baselines policies.  

\begin{figure}[t]
	\centering
	\setlength\fwidth{.8\columnwidth}
	\setlength\fheight{.3\columnwidth}
%
%
\definecolor{mycolor1}{rgb}{0.24403,0.43583,0.99883}%
\definecolor{mycolor2}{rgb}{0.00357,0.72027,0.79170}%
\definecolor{lavender}{rgb}{0.9020,0.9020,0.9804}%
\definecolor{lightskyblue}{rgb}{0.6784,0.8471,0.9020}%
\definecolor{deepskyblue}{rgb}{0,0.7490,1}%
\definecolor{steelblue}{rgb}{0.2745,0.5098,0.7059}%
\definecolor{blue}{rgb}{0,0,1}%
\definecolor{royalblue}{rgb}{0.2549,0.4118,0.8824}%

\definecolor{gainsboro}{rgb}{0.8627,0.8627,0.8627}%
\definecolor{darkslategrey}{rgb}{0.1843,0.3098,0.3098}%
\definecolor{gray}{rgb}{0.5,0.5,0.5}%

\definecolor{lightcoral}{rgb}{0.9412,0.5020,0.5020}%
\definecolor{indianred}{rgb}{0.8039,0.3608,0.3608}%
\definecolor{lightsalmon}{rgb}{1.0000,0.6275,0.4784}%
\definecolor{darksalmon}{rgb}{0.9137,0.5882,0.4784}%
\begin{tikzpicture}
\pgfplotsset{every tick label/.append style={font=\scriptsize}}

\begin{axis}[%
width=0.951\fwidth,
height=\fheight,
at={(0\fwidth,0\fheight)},
scale only axis,
xtick=data,
xmin=40,
xmax=130,
ymin=0.13,
ymax=0.6,
xlabel style={font=\footnotesize\color{white!15!black}},
xlabel={Number of users},
ylabel style={font=\footnotesize\color{white!15!black}},
ylabel={Throughput [Mbit/s]},
axis background/.style={fill=white},
xmajorgrids,
ymajorgrids,
legend style={font=\tiny,at={(0.99,0.99)},anchor=north east,legend cell align=left,align=left,draw=white!15!black},
legend columns=2
]

\addplot [color=steelblue, mark=o, line width=1pt, mark options={solid}]
table[row sep=crcr]{%
42	0.456\\
63	0.408\\
84	0.401\\
105	0.378\\
126	0.382\\
};
\addlegendentry{RIC RL, 850 MHz}

\addplot [color=steelblue, dashed, mark=v, line width=1pt, mark options={solid}]
table[row sep=crcr]{%
42	0.300\\
63	0.243\\
84	0.206\\
105	0.188\\
126	0.172\\
};
\addlegendentry{CMAB RL~\cite{yajnanarayana20205g}, 850 MHz}

\addplot [color=indianred, mark=o, line width=1pt, mark options={solid}]
table[row sep=crcr]{%
42	0.418\\
63	0.385\\
84	0.334\\
105	0.211\\
126	0.200\\
};
\addlegendentry{RIC RL, C-Band}

\addplot [color=indianred, dashed, mark=v, line width=1pt, mark options={solid}]
table[row sep=crcr]{%
42	0.374\\
63	0.212\\
84	0.176\\
105	0.154\\
126	0.143\\
};
\addlegendentry{CMAB RL~\cite{yajnanarayana20205g}, C-Band MHz}

\end{axis}
\end{tikzpicture}%
	\caption{Comparison between the 10$^{\rm th}$ percentile user throughput with the proposed xApp (RIC RL) and an xApp implementing the handover control logic from~\cite{yajnanarayana20205g}.}
	\label{fig:10-th-comparison}
\end{figure}
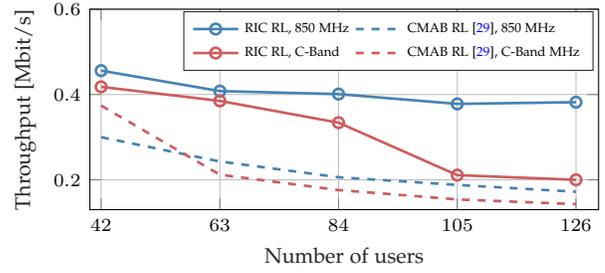

\section{Conclusions}
\label{sec:conclusions}

This paper introduced a complete, system-level, \gls{oran}-compliant framework for the optimization of \gls{ts} in \gls{3gpp} networks. 
Specifically, we focused on throughput maximization through the selection of the NR serving cell in an \gls{endc} setup.
We implemented a cloud-native near-RT \gls{ric}, which we connect through open, O-RAN interfaces to a simulated \gls{ran} environment in ns-3. 
We developed a custom xApp for the near-RT \gls{ric}, with a data-driven handover control based on \gls{rem} and \gls{cql}.
Finally, we profiled the performance of the agent on a large scale deployment in multiple frequency bands, evaluating its gain over traditional handover heuristics.
The results show that, thanks to the \gls{ue}-level control at the near-RT \gls{ric}, our solution achieves significant performance improvements ranging from 30\% to 50\% for the average throughput and spectral efficiency, demonstrating its effectiveness over different combinations of \glspl{ue}.



\ifCLASSOPTIONcaptionsoff
  \newpage
\fi



\footnotesize 
\bibliographystyle{IEEEtran}
\bibliography{IEEEabrv,biblio}

\begin{thebibliography}{10}
\providecommand{\url}[1]{#1}
\csname url@samestyle\endcsname
\providecommand{\newblock}{\relax}
\providecommand{\bibinfo}[2]{#2}
\providecommand{\BIBentrySTDinterwordspacing}{\spaceskip=0pt\relax}
\providecommand{\BIBentryALTinterwordstretchfactor}{4}
\providecommand{\BIBentryALTinterwordspacing}{\spaceskip=\fontdimen2\font plus
\BIBentryALTinterwordstretchfactor\fontdimen3\font minus
  \fontdimen4\font\relax}
\providecommand{\BIBforeignlanguage}[2]{{%
\expandafter\ifx\csname l@#1\endcsname\relax
\typeout{** WARNING: IEEEtran.bst: No hyphenation pattern has been}%
\typeout{** loaded for the language `#1'. Using the pattern for}%
\typeout{** the default language instead.}%
\else
\language=\csname l@#1\endcsname
\fi
#2}}
\providecommand{\BIBdecl}{\relax}
\BIBdecl

\bibitem{letaief2019roadmap}
K.~B. {Letaief}, W.~{Chen}, Y.~{Shi}, J.~{Zhang}, and Y.~A. {Zhang}, ``{The
  Roadmap to 6G: AI Empowered Wireless Networks},'' \emph{IEEE Communications
  Magazine}, vol.~57, no.~8, pp. 84--90, August 2019.

\bibitem{polese2022understanding}
\BIBentryALTinterwordspacing
M.~Polese, L.~Bonati, S.~D'Oro, S.~Basagni, and T.~Melodia, ``{Understanding
  O-RAN: Architecture, Interfaces, Algorithms, Security, and Research
  Challenges},'' \emph{arXiv:2202.01032 [cs.NI]}, February 2022. [Online].
  Available: \url{https://arxiv.org/abs/2202.01032}
\BIBentrySTDinterwordspacing

\bibitem{abdalla2021generation}
A.~S. Abdalla, P.~S. Upadhyaya, V.~K. Shah, and V.~Marojevic, ``{Toward Next
  Generation Open Radio Access Network--What O-RAN Can and Cannot Do!}''
  \emph{arXiv preprint arXiv:2111.13754 [cs.NI]}, November 2021.

\bibitem{challita2020when}
U.~Challita, H.~Ryden, and H.~Tullberg, ``When machine learning meets wireless
  cellular networks: Deployment, challenges, and applications,'' \emph{IEEE
  Communications Magazine}, vol.~58, no.~6, pp. 12--18, June 2020.

\bibitem{chinchali2018cellular}
S.~Chinchali \emph{et~al.}, ``Cellular network traffic scheduling with deep
  reinforcement learning,'' in \emph{Proc. of Thirty-Second AAAI Conf. on
  Artificial Intelligence}, New Orleans, LA, 2018, pp. 766--774.

\bibitem{bonati2021intelligence}
L.~Bonati, S.~D'Oro, M.~Polese, S.~Basagni, and T.~Melodia, ``{Intelligence and
  Learning in O-RAN for Data-driven NextG Cellular Networks},'' \emph{IEEE
  Communications Magazine}, vol.~59, no.~10, pp. 21--27, October 2021.

\bibitem{mollahasani2021dynamic}
S.~Mollahasani, M.~Erol-Kantarci, and R.~Wilson, ``{Dynamic CU-DU Selection for
  Resource Allocation in O-RAN Using Actor-Critic Learning},'' \emph{arXiv
  preprint arXiv:2110.00492 [cs.NI]}, October 2021.

\bibitem{oran-wg1-use-cases}
{O-RAN Working Group 1}, ``{{O-RAN} Use Cases Detailed Specification 6.0},''
  O-RAN.WG1.Use-Cases-Detailed-Specification-v06.00 Technical Specification,
  July 2021.

\bibitem{8812724}
M.~Tayyab, X.~Gelabert, and R.~Jantti, ``A survey on handover management: From
  lte to nr,'' \emph{IEEE Access}, vol.~7, pp. 118\,907--118\,930, 2019.

\bibitem{polese2021coloran}
M.~Polese, L.~Bonati, S.~D'Oro, S.~Basagni, and T.~Melodia, ``{ColO-RAN:
  Developing Machine Learning-based xApps for Open RAN Closed-loop Control on
  Programmable Experimental Platforms},'' \emph{IEEE Transactions on Mobile
  Computing}, pp. 1--14, 2022.

\bibitem{app12010426}
J.~Tanveer, A.~Haider, R.~Ali, and A.~Kim, ``{An Overview of Reinforcement
  Learning Algorithms for Handover Management in 5G Ultra-Dense Small Cell
  Networks},'' \emph{Applied Sciences}, vol.~12, no.~1, 2022.

\bibitem{mezzavilla2018end}
M.~Mezzavilla \emph{et~al.}, ``End-to-end simulation of {5G mmWave} networks,''
  \emph{IEEE Communications Surveys Tutorials}, vol.~20, no.~3, pp. 2237--2263,
  April 2018.

\bibitem{bonati2022openrangym}
L.~Bonati, M.~Polese, S.~D'Oro, S.~Basagni, and T.~Melodia, ``{OpenRAN Gym: An
  Open Toolbox for Data Collection and Experimentation with AI in O-RAN},'' in
  \emph{Proc. of IEEE WCNC Workshop on Open RAN Architecture for 5G Evolution
  and 6G}, Austin, TX, USA, April 2022.

\bibitem{oran-wg3-e2-sm}
{O-RAN Working Group 3}, ``{O-RAN} near-real-time {RAN} intelligent controller
  {E2} service model 2.00,'' ORAN-WG3.E2SM-v02.00 Technical Specification, July
  2021.

\bibitem{oran-wg3-e2-sm-kpm}
------, ``{O-RAN} near-real-time {RAN} intelligent controller {E2} service
  model {(E2SM) KPM} 2.0,'' ORAN-WG3.E2SM-KPM-v02.00 Technical Specification,
  July 2021.

\bibitem{oran-wg3-e2-sm-rc}
------, ``{O-RAN} near-real-time {RAN} intelligent controller {E2} service
  model, ran control 1.0,'' ORAN-WG3.E2SM-RC-v01.00 Technical Specification,
  July 2021.

\bibitem{levine2020offline}
\BIBentryALTinterwordspacing
S.~Levine, A.~Kumar, G.~Tucker, and J.~Fu, ``Offline reinforcement learning:
  Tutorial, review, and perspectives on open problems,'' \emph{arXiv:2005.01643
  [cs.LG]}, 2020. [Online]. Available: \url{https://arxiv.org/abs/2005.01643}
\BIBentrySTDinterwordspacing

\bibitem{8406993}
Y.~Koda, K.~Yamamoto, T.~Nishio, and M.~Morikura, ``Reinforcement learning
  based predictive handover for pedestrian-aware mmwave networks,'' in
  \emph{IEEE Conference on Computer Communications Workshops (INFOCOM WKSHPS)},
  2018, pp. 692--697.

\bibitem{liu2021intelligent}
Q.~Liu, C.~F. Kwong, S.~Wei, L.~Li, and S.~Zhang, ``Intelligent handover
  triggering mechanism in 5g ultra-dense networks via clustering-based
  reinforcement learning,'' \emph{Mobile Networks and Applications}, vol.~26,
  no.~1, pp. 27--39, 2021.

\bibitem{6692634}
S.~Mwanje and A.~Mitschele-Thiel, ``Minimizing handover performance degradation
  due to lte self organized mobility load balancing,'' in \emph{IEEE 77th
  Vehicular Technology Conference (VTC Spring)}, 2013.

\bibitem{guo2020joint}
D.~Guo, L.~Tang, X.~Zhang, and Y.~Liang, ``Joint optimization of handover
  control and power allocation based on multi-agent deep reinforcement
  learning,'' \emph{IEEE Transactions on Vehicular Technology}, vol.~69,
  no.~11, pp. 13\,124--13\,138, 2020.

\bibitem{wang2018handover}
Z.~Wang, L.~Li, Y.~Xu, H.~Tian, and S.~Cui, ``Handover control in wireless
  systems via asynchronous multiuser deep reinforcement learning,'' \emph{IEEE
  Internet Things J.}, vol.~5, no.~6, pp. 4296--4307, December 2018.

\bibitem{MOLLEL2020101133}
M.~S. Mollel \emph{et~al.}, ``Intelligent handover decision scheme using double
  deep reinforcement learning,'' \emph{Physical Communication}, vol.~42, p.
  101133, 2020.

\bibitem{wang2018deep}
S.~Wang, H.~Liu, P.~Gomes, and B.~Krishnamachari, ``Deep reinforcement learning
  for dynamic multichannel access in wireless networks,'' \emph{IEEE Trans.
  Cogn. Commun. Netw.}, vol.~4, no.~2, pp. 257--265, June 2018.

\bibitem{9052936}
M.~Sana, A.~De~Domenico, E.~Strinati, and A.~Clemente, ``Multi-agent deep
  reinforcement learning for distributed handover management in dense mmwave
  networks,'' in \emph{IEEE International Conference on Acoustics, Speech and
  Signal Processing (ICASSP)}, 2020, pp. 8976--8980.

\bibitem{MUNOZ2015112}
P.~Muñoz, R.~Barco, and I.~{de la Bandera}, ``Load balancing and handover
  joint optimization in lte networks using fuzzy logic and reinforcement
  learning,'' \emph{Computer Networks}, vol.~76, pp. 112--125, 2015.

\bibitem{8466370}
F.~D. Calabrese \emph{et~al.}, ``Learning radio resource management in rans:
  Framework, opportunities, and challenges,'' \emph{IEEE Communications
  Magazine}, vol.~56, no.~9, pp. 138--145, 2018.

\bibitem{s21248173}
M.~Dryjański, L.~Kułacz, and A.~Kliks, ``Toward modular and flexible open ran
  implementations in 6g networks: Traffic steering use case and o-ran xapps,''
  \emph{Sensors}, vol.~21, no.~24, 2021.

\bibitem{yajnanarayana20205g}
V.~Yajnanarayana, H.~Rydén, and L.~Hévizi, ``{5G Handover using Reinforcement
  Learning},'' in \emph{IEEE 3rd 5G World Forum (5GWF)}, 2020, pp. 349--354.

\bibitem{9815658}
T.~Karamplias \emph{et~al.}, ``Towards closed-loop automation in 5g open ran:
  Coupling an open-source simulator with xapps,'' in \emph{2022 Joint European
  Conference on Networks and Communications \& 6G Summit (EuCNC/6G Summit)},
  2022, pp. 232--237.

\bibitem{riley2010ns}
G.~F. Riley and T.~R. Henderson, ``The ns-3 network simulator,'' in
  \emph{Modeling and tools for network simulation}.\hskip 1em plus 0.5em minus
  0.4em\relax Springer, 2010, pp. 15--34.

\bibitem{zugno2020implementation}
T.~Zugno \emph{et~al.}, ``Implementation of a spatial channel model for ns-3,''
  in \emph{Proceedings of the 2020 Workshop on Ns-3}, ser. WNS3 2020.\hskip 1em
  plus 0.5em minus 0.4em\relax New York, NY, USA: Association for Computing
  Machinery, 2020, p. 49–56.

\bibitem{ns3gym}
\BIBentryALTinterwordspacing
P.~Gaw{\l}owicz and A.~Zubow, ``{ns-3 meets OpenAI Gym: The Playground for
  Machine Learning in Networking Research},'' in \emph{{ACM International
  Conference on Modeling, Analysis and Simulation of Wireless and Mobile
  Systems (MSWiM)}}, November 2019. [Online]. Available:
  \url{http://www.tkn.tu-berlin.de/fileadmin/fg112/Papers/2019/gawlowicz19_mswim.pdf}
\BIBentrySTDinterwordspacing

\bibitem{ns3ai}
\BIBentryALTinterwordspacing
H.~Yin \emph{et~al.}, ``Ns3-ai: Fostering artificial intelligence algorithms
  for networking research,'' in \emph{Proceedings of the 2020 Workshop on
  Ns-3}, ser. WNS3 2020.\hskip 1em plus 0.5em minus 0.4em\relax New York, NY,
  USA: Association for Computing Machinery, 2020, p. 57–64. [Online].
  Available: \url{https://doi.org/10.1145/3389400.3389404}
\BIBentrySTDinterwordspacing

\bibitem{3gpp.37.340}
3GPP, ``{NR}; multi-connectivity; overall description; stage-2,'' {3GPP},
  Technical Specification (TS) 37.340, 12 2021, version 16.8.0.

\bibitem{9376232}
B.~Balasubramanian \emph{et~al.}, ``{RIC: A RAN Intelligent Controller Platform
  for AI-Enabled Cellular Networks},'' \emph{IEEE Internet Computing}, vol.~25,
  no.~2, pp. 7--17, 2021.

\bibitem{3gpp.38.901}
3GPP, ``Study on channel model for frequencies from 0.5 to 100 ghz,'' {3GPP},
  Technical Specification (TS) 38.901, 1 2020, version 16.1.0.

\bibitem{e2sim}
{O-RAN Software Community}. (2022) {sim-e2-interface repository}.
  \url{https://github.com/o-ran-sc/sim-e2-interface}. Accessed March 2022.

\bibitem{dqn}
V.~Mnih \emph{et~al.}, ``Playing atari with deep reinforcement learning,''
  \emph{arXiv preprint arXiv:1312.5602}, 2013.

\bibitem{rem}
R.~Agarwal, D.~Schuurmans, and M.~Norouzi, ``{An optimistic perspective on
  offline reinforcement learning},'' in \emph{International Conference on
  Machine Learning}.\hskip 1em plus 0.5em minus 0.4em\relax PMLR, 2020, pp.
  104--114.

\bibitem{cql}
A.~Kumar, A.~Zhou, G.~Tucker, and S.~Levine, ``{Conservative Q-Learning for
  Offline Reinforcement Learning},'' in \emph{Advances in Neural Information
  Processing Systems}, H.~Larochelle, M.~Ranzato, R.~Hadsell, M.~F. Balcan, and
  H.~Lin, Eds., vol.~33.\hskip 1em plus 0.5em minus 0.4em\relax Curran
  Associates, Inc., 2020, pp. 1179--1191.

\bibitem{7959177}
M.~Polese, M.~Giordani, M.~Mezzavilla, S.~Rangan, and M.~Zorzi, ``Improved
  handover through dual connectivity in 5g mmwave mobile networks,'' \emph{IEEE
  Journal on Selected Areas in Communications}, vol.~35, no.~9, pp. 2069--2084,
  2017.

\end{thebibliography}

%

\begin{IEEEbiography}[{\includegraphics[width=1in,height=1.25in,clip,keepaspectratio]{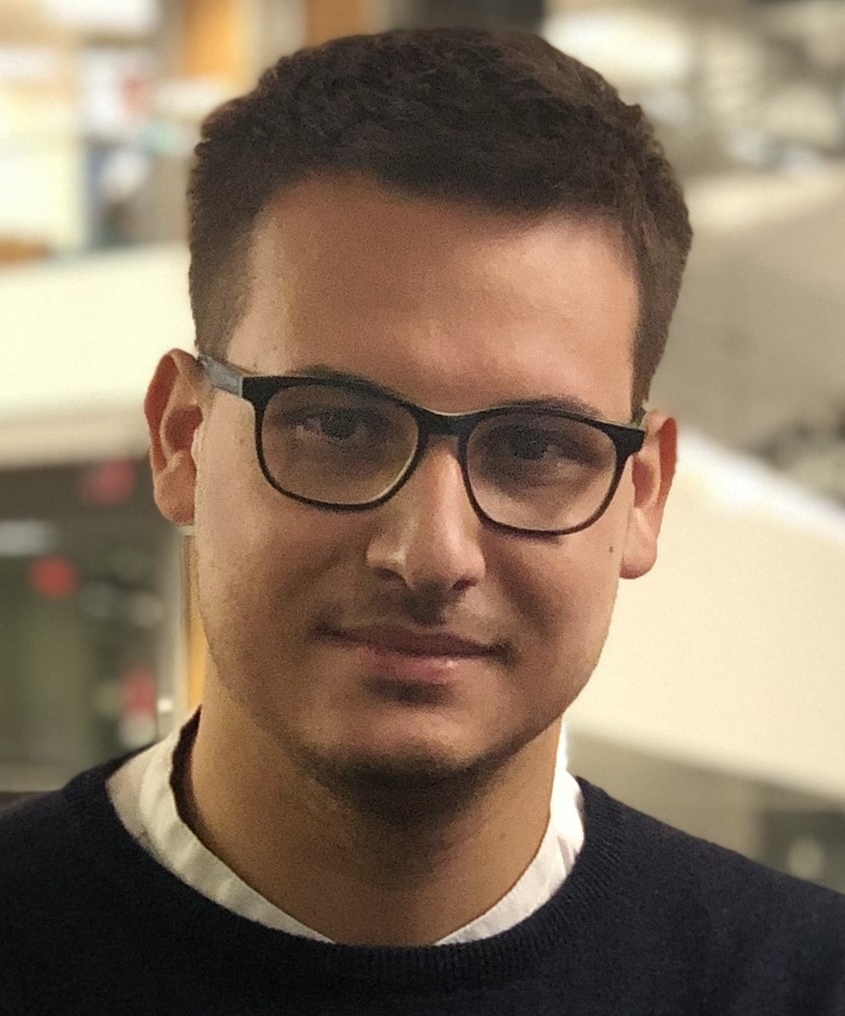}}]{Andrea Lacava} received his B.S. in Computer Engineering
and his M.S. in Cybersecurity from Sapienza, University of Rome, Italy in 2018 and 2021, respectively. He is currently pursuing a double Ph.D. degree in Computer Engineering at the Institute for the Wireless Internet of Things at Northeastern University, MA, USA and in Information and Communication Technology (ICT) at Sapienza, University of Rome, Italy. His research interests focus on the O-RAN architecture, 5G and beyond cellular networks. He is IEEE graduate student member.
\end{IEEEbiography}

\begin{IEEEbiography}
[{\includegraphics[width=1in,height=1.25in,keepaspectratio]{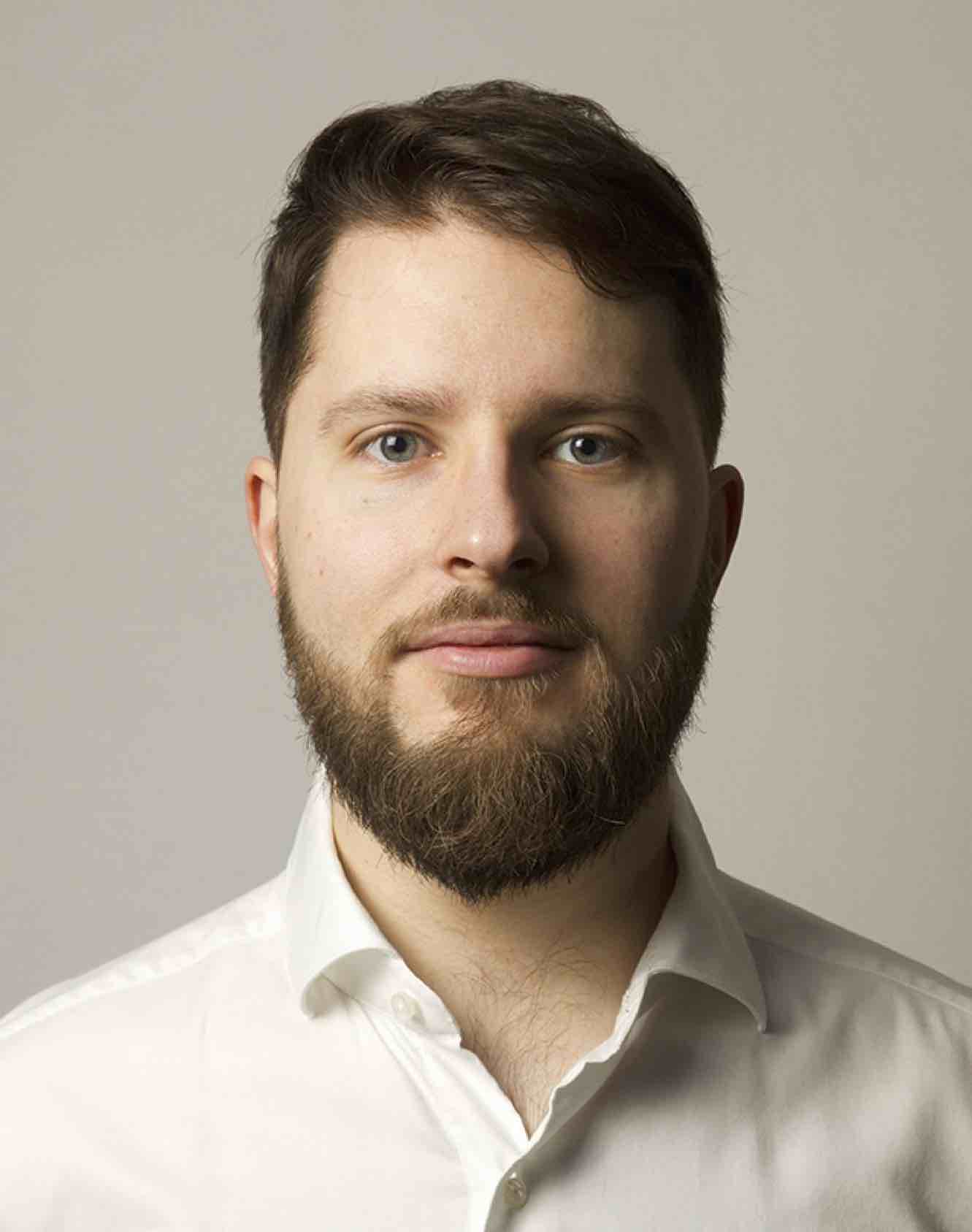}}]{Michele Polese} is a Principal Research Scientist at the Institute for the Wireless Internet of Things, Northeastern University, Boston, since March 2020. He received his Ph.D. at the Department of Information Engineering of the University of Padova in 2020. He also was an adjunct professor and postdoctoral researcher in 2019/2020 at the University of Padova, and a part-time lecturer in Fall 2020 and 2021 at Northeastern University. During his Ph.D., he visited New York University (NYU), AT\&T Labs in Bedminster, NJ, and Northeastern University.
His research interests are in the analysis and development of protocols and architectures for future generations of cellular networks (5G and beyond), in particular for millimeter-wave and terahertz networks, spectrum sharing and passive/active user coexistence, open RAN development, and the performance evaluation of end-to-end, complex networks. He has contributed to O-RAN technical specifications and submitted responses to multiple FCC and NTIA notice of inquiry and requests for comments, and is a member of the Committee on Radio Frequency Allocations of the American Meteorological Society (2022-2024). He collaborates and has collaborated with several academic and industrial research partners, including AT\&T, Mavenir, NVIDIA, InterDigital, NYU, University of Aalborg, King's College, and NIST. He was awarded with several best paper awards, is serving as TPC co-chair for WNS3 2021-2022, as an Associate Technical Editor for the IEEE Communications Magazine, and has organized the Open 5G Forum in Fall 2021. He is a Member of the IEEE.
\end{IEEEbiography}

\begin{IEEEbiography}[{\includegraphics[width=1in,height=1.25in,clip,keepaspectratio]{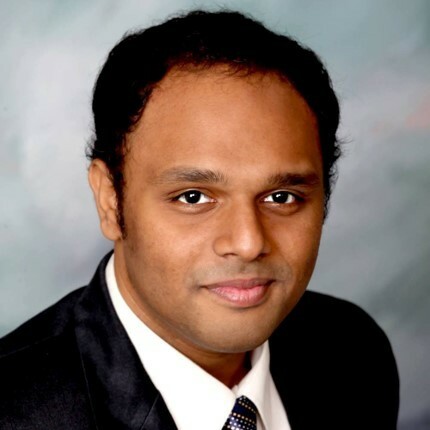}}]{Rajarajan Sivaraj} is director for RIC architecture and standards at Mavenir, where he is responsible for the standardization and productization of the Near-RT RIC, Non-RT RIC, SMO, xApps, rApps and interfaces. He holds a PhD in Computer Science from UC Davis, and has a career spanning over 12 years in cellular telecommunications with prior positions at AT\&T labs, Microsoft Research, Intel labs, NEC labs, Broadcom, Uhana (VMWare), etc. He has had numerous publications and granted patents, and his scholarly works have been highly cited. 
\end{IEEEbiography}

\begin{IEEEbiography}[{\includegraphics[width=1in,height=1.25in,clip,keepaspectratio]{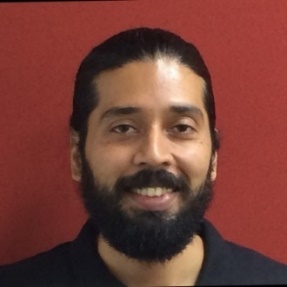}}]{Rahul Soundrarajan} is Sr. Director for RAN Analytics at Mavenir. As of his role, he is responsible for design, development and evaluation of Machine Learning Algorithms for Near-RT RIC and Non-RT RIC. His career spanning 21 years is a rich mix of Systems Architecture and Engineering of 2/3/4G RAN and applying ML algorithms for network optimization, for which he holds several patents. His prior experiences include positions at Lucent, Alcatel-Lucent, Nokia and HCL Technologies.
\end{IEEEbiography}

\begin{IEEEbiography}[{\includegraphics[width=1in,height=1.25in,clip,keepaspectratio]{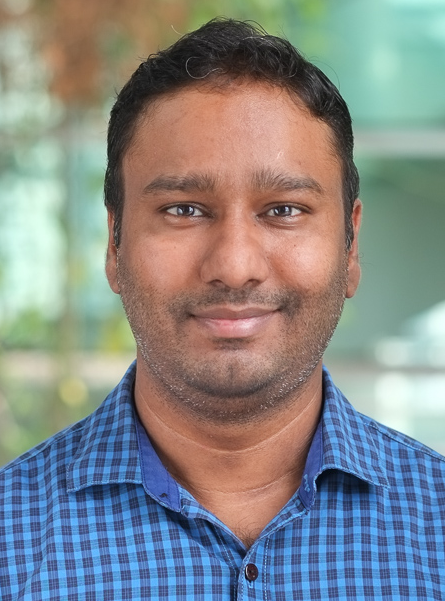}}]{Bhawani Shanker Bhati} received the Ph.D. degree in engineering from the Indian Institute of Science (IISc), Bengaluru, India, in 2018. He is currently a Senior Member of Technical Staff at Mavenir, India. His research interests include ad-hoc networks, Near-RT RIC, communication protocols, ubiquitous computing, security, and privacy in wireless networks.
\end{IEEEbiography}

\begin{IEEEbiography}[{\includegraphics[width=1in,height=1.25in,clip,keepaspectratio]{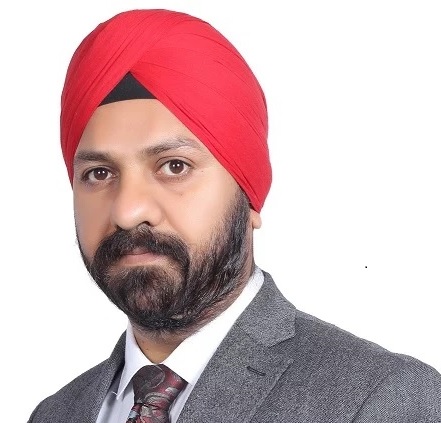}}]{Tarunjeet Singh} is Director Engineering at Mavenir Systems.
He has wide array of experience in data analytics, cloud native technologies, telecom networks, telecom VAS and customer experience and billing.
He received his Bachelor's degree in Computer Science in 2002, from the University of Delhi, India.
In his current role, he is leading R\&D for Near RT RIC platform and xApps at Mavenir Systems. He has held positions with erstwhile Alcatel-Lucent, Nokia and Bharti Airtel providing him exposure to the engineering methods of some of the key leaders and innovators in the telecom space. 
\end{IEEEbiography}

\begin{IEEEbiography}[{\includegraphics[width=1in,height=1.25in,clip,keepaspectratio]{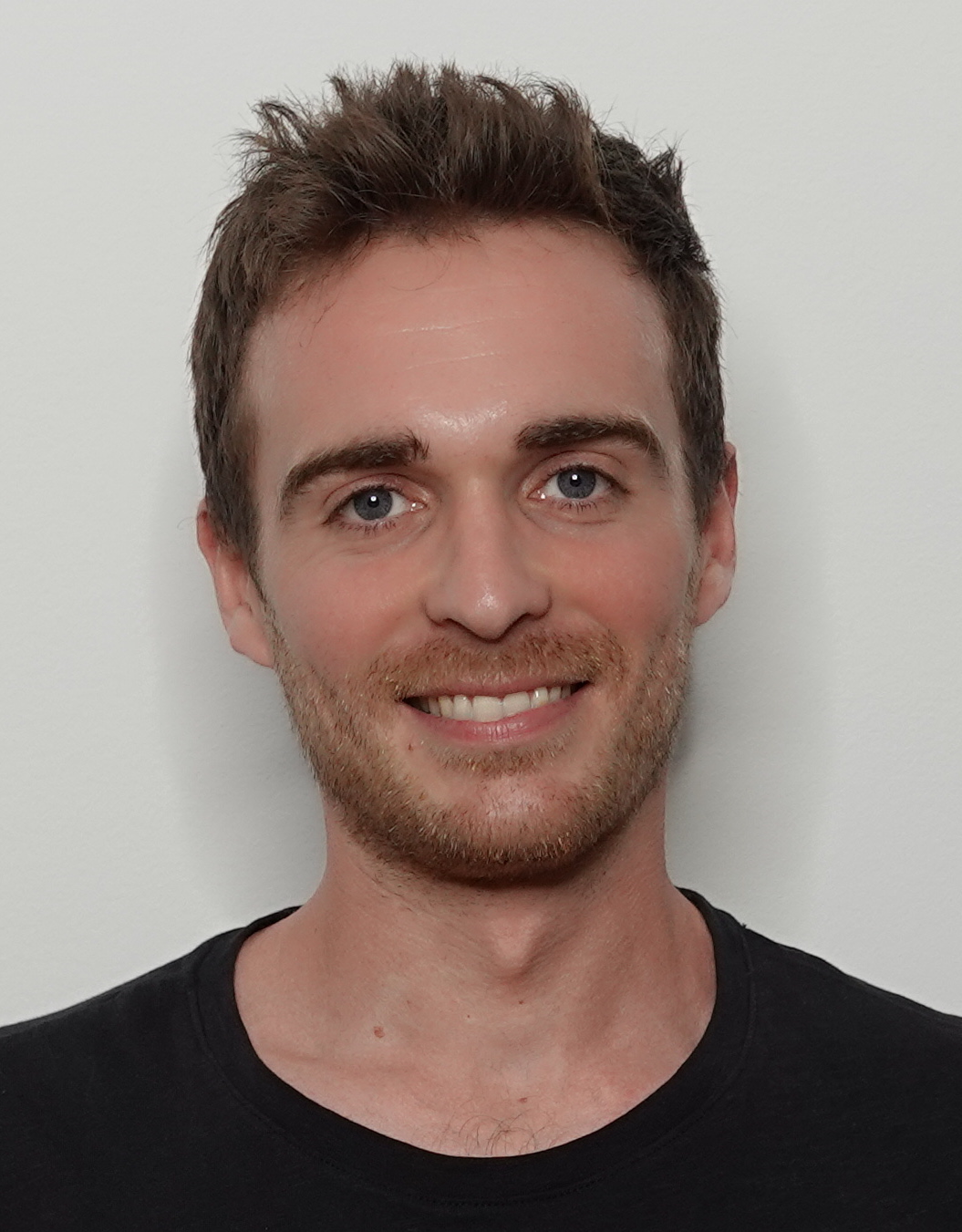}}]{Tommaso Zugno}  received the Ph.D. degree from the Department of Information Engineering, University of Padua, in 2022. From May 2018 to October 2018, he was a Postgraduate Researcher with the Department of Information Engineering, University of Padua. He is with Huawei Technologies, Munich Research Center, Germany. His research interests include the design and evaluation of algorithms and architectures for next-generation cellular networks. He was awarded the Best Paper Awards at WNS3 2020 and IEEE MedComNet 2020.
\end{IEEEbiography}

\begin{IEEEbiography}[{\includegraphics[width=1in,height=1.25in,clip,keepaspectratio]{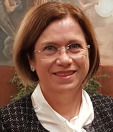}}]{Francesca Cuomo} received the Ph.D. in Information and Communications Engineering in 1998 from Sapienza University of Rome. 
From 2005 to October 2020 she was Associate Professor and from November 2020 she joined ‘‘Sapienza’’ as Full Professor teaching courses in Telecommunication and Networks. Prof. Cuomo has advised numerous master students in computer engineering, and has been the advisor of 13 PhD students in Networking.
Her current research interests focus on: Vehicular networks and Sensor networks, Low Power Wide Area
Networks and IoT, 5G Networks, Multimedia Networking,
Energy saving in the Internet and in the wireless system.
Francesca Cuomo has authored over 156 peer-reviewed papers published in prominent international journals and
conferences. Her Google Scholar h-index is 30 with over 3850 citations.
Relevant scientific international recognitions: Two Best Paper Awards. 
She has been in the editorial board of Computer Networks (Elsevier) and now is member of the editorial board of the Ad-Hoc Networks (Elsevier), IEEE
Transactions on Mobile Computing, Sensors (MDPI), Frontiers in Communications and Networks Journal. 
She has been the TPC co-chair of several editions of the ACM
PE-WASUN workshop, TPC Co-Chair of ICCCN 2016, TPC Symposium Chair of IEEE WiMob 2017, General Co-Chair of the First Workshop on Sustainable Networking through Machine Learning and Internet of Things (SMILING), in conjunction with IEEE INFOCOM 2019; Workshop Co-Chair of AmI 2019: European Conference on Ambient Intelligence 2019. 
She is IEEE senior member.
\end{IEEEbiography}

\begin{IEEEbiography}
[{\includegraphics[width=1in,height=1.25in,keepaspectratio]{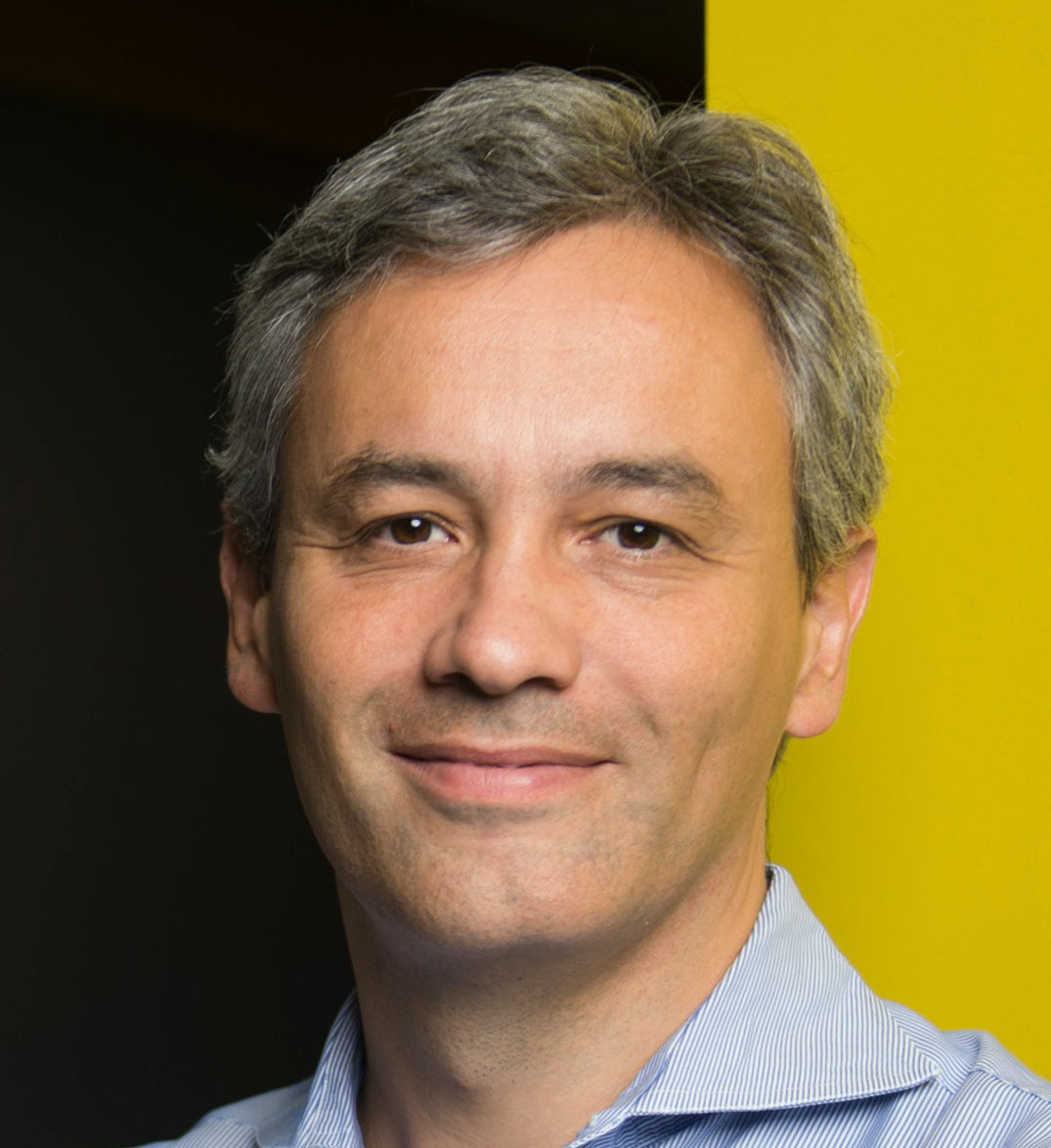}}]{Tommaso Melodia}
is the William Lincoln Smith Chair Professor with the Department of Electrical and Computer Engineering at Northeastern University in Boston. He is also the Founding Director of the Institute for the Wireless Internet of Things and the Director of Research for the PAWR Project Office. He received his Ph.D. in Electrical and Computer Engineering from the Georgia Institute of Technology in 2007. He is a recipient of the National Science Foundation CAREER award. Prof. Melodia has served as Associate Editor of IEEE Transactions on Wireless Communications, IEEE Transactions on Mobile Computing, Elsevier Computer Networks, among others. He has served as Technical Program Committee Chair for IEEE Infocom 2018, General Chair for IEEE SECON 2019, ACM Nanocom 2019, and ACM WUWnet 2014. Prof. Melodia is the Director of Research for the Platforms for Advanced Wireless Research (PAWR) Project Office, a \$100M public-private partnership to establish 4 city-scale platforms for wireless research to advance the US wireless ecosystem in years to come. Prof. Melodia's research on modeling, optimization, and experimental evaluation of Internet-of-Things and wireless networked systems has been funded by the National Science Foundation, the Air Force Research Laboratory the Office of Naval Research, DARPA, and the Army Research Laboratory. Prof. Melodia is a Fellow of the IEEE and a Senior Member of the ACM.
\end{IEEEbiography}





\end{document}